%% file: main.tex
\documentclass[preprint,prd,aps,showpacs,preprintnumbers,amsmath,amssymb,nofootinbib,superscriptaddress]{revtex4-1}

\usepackage{amsmath}
\usepackage{amssymb}
\usepackage{graphicx}
\usepackage{dcolumn}
\usepackage{bm}

\usepackage{color}

\usepackage{hyperref}

\graphicspath{{fig/}}

\def\simge{
    \mathrel{\rlap{\raise 0.511ex
        \hbox{$>$}}{\lower 0.511ex \hbox{$\sim$}}}}
\def\simle{
    \mathrel{\rlap{\raise 0.511ex
        \hbox{$<$}}{\lower 0.511ex \hbox{$\sim$}}}}

\begin{document}

\date{\today}

\preprint{KEK Preprint 2012-18}

\title{
Lattice study of conformality in twelve-flavor QCD
}


\author{Yasumichi~Aoki}
\affiliation{Kobayashi-Maskawa Institute for the Origin of Particles and the Universe (KMI), Nagoya University, Nagoya 464-8602, Japan}

\author{Tatsumi~Aoyama}
\affiliation{Kobayashi-Maskawa Institute for the Origin of Particles and the Universe (KMI), Nagoya University, Nagoya 464-8602, Japan}

\author{Masafumi~Kurachi}
\affiliation{Kobayashi-Maskawa Institute for the Origin of Particles and the Universe (KMI), Nagoya University, Nagoya 464-8602, Japan}

\author{Toshihide~Maskawa}
\affiliation{Kobayashi-Maskawa Institute for the Origin of Particles and the Universe (KMI), Nagoya University, Nagoya 464-8602, Japan}

\author{Kei-ichi~Nagai}
\affiliation{Kobayashi-Maskawa Institute for the Origin of Particles and the Universe (KMI), Nagoya University, Nagoya 464-8602, Japan}

\author{Hiroshi~Ohki}
\affiliation{Kobayashi-Maskawa Institute for the Origin of Particles and the Universe (KMI), Nagoya University, Nagoya 464-8602, Japan}

\author{Akihiro~Shibata}
\affiliation{Computing Research Center, High Energy Accelerator Research Organization (KEK), Tsukuba 305-0801, Japan}

\author{Koichi~Yamawaki}
\affiliation{Kobayashi-Maskawa Institute for the Origin of Particles and the Universe (KMI), Nagoya University, Nagoya 464-8602, Japan}

\author{Takeshi~Yamazaki}
\affiliation{Kobayashi-Maskawa Institute for the Origin of Particles and the Universe (KMI), Nagoya University, Nagoya 464-8602, Japan}

\collaboration{LatKMI Collaboration}
\noaffiliation


\input{text_sections/abstract.tex}

\maketitle


\section{Introduction}
\label{sec:introduction}
\input{text_sections/introduction.tex}

\section{The basis methods for the study of the chiral properties}
\label{sec:method}
\input{text_sections/method.tex}

\section{Numerical results and primary analysis}
\label{sec:primary_results}
\input{text_sections/results.tex}

\section{
Finite-size hyperscaling
}
\label{sec:fss}
\input{text_sections/fss.tex}

\section{
Nonuniversal corrections to finite-size hyperscaling
}
\label{sec:fshp}
\input{text_sections/fshp.tex}

\section{Chiral perturbation theory analysis}
\label{sec:chpt_summary}
\input{text_sections/chpt_summary.tex}

\section{Summary and discussion}
\label{sec:summary}
\input{text_sections/summary.tex}


\section*{Acknowledgments}
Numerical calculations have been carried out
on the cluster system ``$\varphi$'' at KMI, Nagoya University.
We thank Katsuya Hasebe for the encouragement and discussion.
We thank Luigi Del Debbio for a crucial suggestion and discussion
during his visit to KMI.
We also thank Julius Kuti and Maria Paola Lombardo for fruitful discussion.
This work is supported in part by JSPS Grants-in-Aid for Scientific
Research (S) No.~22224003, (C) No.~21540289 (Y.A.), (C) No.~23540300 (K.Y.),
(C) No.~24540252 (A.S.)
and 
Grants-in-Aid of the Japanese Ministry for Scientific Research on
Innovative 
Areas No.~23105708 (T.Y.).
%

\appendix

\section{Staggered flavor symmetry}
\label{sec:other_spectra}
\input{text_sections/other_spectra.tex}
\section{
Details of fit analysis for conformal hypothesis with corrections
}
\label{sec:conformal_fit}
\input{text_sections/conformal_fit.tex}

\section{
Details of chiral perturbation theory analysis}
\label{sec:chpt}
\input{text_sections/chpt.tex}


\clearpage

\bibliographystyle{apsrev4-1}
\bibliography{main}


\end{document}

%% file: text_sections/abstract.tex
\begin{abstract}
We study infrared conformality of the twelve-flavor QCD on the lattice.
Utilizing the highly improved staggered quarks (HISQ) type action 
which is useful to study the continuum physics, 
we analyze the lattice data of the mass and the decay constant of 
a pseudoscalar meson and the mass of a vector meson as well  
at several values of lattice spacing and fermion mass.
  Our result is consistent with the conformal hypothesis for the mass anomalous dimension
  $\gamma_m \sim 0.4-0.5$.
\end{abstract}

%% file: text_sections/introduction.tex
%
\thispagestyle{empty}

There has been a renewed interest in 
the study of 
the QCD with a large 
number of the massless fermions  in fundamental representation (``large $N_f$ QCD'') 
in the context of walking technicolor having approximate scale invariance and 
large anomalous dimension $\gamma_m \simeq 1$~\cite{Yamawaki:1985zg}
\footnote{For subsequent similar works without notion of scale invariance
and anomalous dimension see
\cite{Akiba:1985rr,Appelquist:1986an}.
For earlier work on this line based on purely numerical analysis see
\cite{Holdom:1984sk}.}.
The model was proposed to cure the fatal flavor-changing neutral current (FCNC) problem of the original technicolor
and predicted a technidilaton as a pseudo-Nambu-Goldstone (NG) boson of the 
approximate scale invariance\footnote{The technidilaton may be identified with 125-126 GeV boson 
observed recently at LHC~\cite{Matsuzaki:2012fq}.
More detailed discussions can be seen in Ref.~\cite{Matsuzaki:2012mk} which appeared after submission of this paper. },
based on the ladder Schwinger-Dyson (SD) 
equation with nonrunning (scale-invariant) gauge coupling. (See, for a review, Ref.~\cite{Hill:2002ap}.)  
Such an approximate scale-invariant dynamics may in fact be realized in the large $N_f$ QCD: 
The perturbative two-loop beta function predicts a 
nontrivial infrared (IR) fixed point $\alpha_*$ ($0<\alpha_*<\infty$) in the range of $9 \le N_f \le 16$ 
in the asymptotically free SU(3) gauge theory
~\cite{Caswell:1974gg, Banks:1981nn}, where $N_f$ is the number of massless flavor. 
The theory has an intrinsic scale $\Lambda$, analogue of
$\Lambda_{QCD}$, which characterizes scale-symmetry breaking (scale anomaly)  
in the ultraviolet (UV) region ($\mu>\Lambda$) where the coupling runs as an asymptotically free theory, $\alpha(\mu) \sim 1/\log(\mu/\Lambda)$, 
in the same way as the
 ordinary QCD. 
However, the theory in the IR region ($0<\mu<\Lambda$) governed by the IR fixed point respects approximate scale invariance
 due to the almost nonrunning (walking)  coupling (IR conformality), $\alpha(\mu) \simeq \alpha_*$.
 If such an IR fixed point survives the nonperturbative effects,  it would imply that  the theory is in the deconfined and chiral-symmetric 
 phase,  
the so-called ``conformal window'' (though the theory has no conformality in the UV region), which 
is expected to exist 
for a 
certain range of large $N_f$.

 Actually, the conformal window and phase structure of the large $N_f$ QCD was studied by the (improved) ladder 
 SD equation 
 in a way to simply replace the coupling by the two-loop running coupling 
having the IR fixed point mentioned above~\cite{Appelquist:1996dq, Miransky:1996pd}:
 For $9\le N_f \le 11$, the IR fixed-point coupling exceeds the critical coupling $\alpha_*>\alpha_{cr}$ 
 and hence triggers the spontaneous chiral symmetry breaking (S$\chi$SB)
 to give rise to the dynamical generation of the fermion mass $m_D (\ne 0)$, which actually washes out  the would-be IR fixed point,
 since the fermions, once acquiring mass, are decoupled from the beta function in the very IR region below the scale of $m_D$.
The conformal window thus is expected to be $12  \le N_f\le 16$.
If it is the case, the walking technicolor 
should be realized in a slightly broken phase  very close to the conformal window, 
$N_f \approx 12$, such that $m_D \ll \Lambda$,  with $m_D$ being on 
the electroweak scale 
order and $\Lambda$ being 
usually identified with the extended technicolor scale.  
Although the IR fixed point actually disappears at $\mu \sim m_D (\ll \Lambda)$,  
the coupling is still almost nonrunning as a remnant of the would-be IR fixed point 
for the wide IR region $m_D < \mu < \Lambda$, which is relevant to the physics 
with the critical value $\gamma_m \simeq 1$ valid all the way up to the scale $\Lambda$.

In the conformal phase with the bare fermion mass $m_f$,  
the low-energy behavior of the hadron spectra 
should obey the same scaling relation
(hyperscaling relation) near the IR fixed point, 
$M \sim
m_f^{1/(1+\gamma_*)}$, where $M$ is the hadron mass and   
$\gamma_*$ is the mass anomalous dimension $\gamma_m$ at the IR fixed point.  
For small $m_f$, we can expect that  $\gamma_m\simeq  \gamma_*$ 
in the wide region all the way up to the scale $\Lambda$  where the IR conformality is operative.   
On the other hand, if the chiral symmetry is spontaneously broken (not in the conformal phase), 
the low-energy physics could be described by the chiral perturbation theory (ChPT).

As a powerful nonperturbative study of such a problem,  one can use  
lattice QCD simulations, which can, in principle, 
determine the phase structure of the SU(3) gauge  
theory with a various number of fermions through investigating  
nonperturbative running of 
gauge coupling, 
low-energy spectra, 
chiral condensates, and so on. 
Furthermore, 
when the theory is inside the conformal window, one can also
calculate 
critical exponents such as the mass anomalous dimension. 
A similar analysis could be approximately
applied to the slightly broken phase ($m_D\ll m_f \ll \Lambda$) 
reflecting the remnant of the conformality, 
which is vital to the study of the walking technicolor.

In lattice calculation, there exist effects of the 
finite size boundary and 
finite fermion masses, which 
breaks the IR conformality explicitly and also deforms the ChPT. 
In the case of the conformal phase, it may cause 
serious
uncertainties,  since the IR scale is given by the inverse lattice size $L^{-1}$ or input 
$m_f$ but not by the intrinsic scale $\Lambda (\gg m_f, L^{-1})$ 
which behaves as the UV scale where the IR conformality breaks down,  
in sharp contrast to the ordinary QCD where 
the IR scale  is given by the intrinsic scale $\Lambda_{QCD} (\sim m_f, L^{-1})$. 
Actually, the finite mass $m_f$ as well as finite size $L$ can easily 
distort the hyperscaling relation, while finite $m_f$ can also trigger the dynamical mass $m_D$ of the fermion
since the coupling should blow up due to the decoupling of the fermion in the IR region $\mu <m_f$, thus mocking up the ChPT. 

On the other hand, 
even in the chiral-broken phase, 
there still can exist an approximate hyperscaling relation,  
as far as $m_D  \ll  m_f \ll \Lambda$, as a remnant of the approximate scale invariance, while
there exist 
finite size effects and higher-order contributions which make it  
difficult to understand the ChPT 
particularly in the large $N_f$ 
QCD.
Here, we should point out that the validity region of the ChPT is extremely restricted 
near the conformal window where 
the decay constant of $\pi$ ($f_\pi$) in the chiral limit 
is expected to be very small and, particularly, $N_f$ is large. 
For the consistency of the ChPT~\cite{Gasser:1984gg}, 
it is required that the expansion parameter~\cite{Soldate:1989fh,Chivukula:1992gi,Harada:2003jx}   
for the mass of $\pi$ ($M_\pi$) is small, 
\begin{equation}\label{eq:X}
\mathcal{X} =N_f \left( \frac{M_\pi}{4 \pi f_\pi} \right)^2 <1 
\end{equation}
which, however,
could become easily violated when simulation is made for $M_\pi$ away from chiral limit.

There are many lattice results on the large $N_f$ 
QCD in recent years.
In addition to the pioneered works~\cite{Iwasaki:1991mr, Brown:1992fz,Damgaard:1997ut} 
for the study of the phase structures, 
there has been 
a lot of progress on the lattice calculations 
such as the renormalized running coupling, hadron spectra, finite temperature transitions, 
etc.~\cite{Appelquist:2007hu,Deuzeman:2008sc,Deuzeman:2009mh,  
Fodor:2009wk,Hayakawa:2010yn,Fodor:2011tu,Aoyama:2011ry,DeGrand:2012yq,
Jin:2010vm,Hasenfratz:2010fi,Cheng:2011ic,Miura:2011mc}.      
In particular,  for  the $N_f=12$ the system has been 
widely investigated
by the lattice approach focusing on 
determining the phase 
structure:
A recent large-volume analysis for the hadron spectra gives 
the result favoring 
the chiral-broken phase~\cite{Fodor:2011tu},
while 
other groups with  
similar analyses using the same data conclude
that this model  
favors 
the conformal phase~\cite{Appelquist:2011dp,DeGrand:2011cu}.  
Thus, it is controversial at present whether this theory is in the conformal or the chiral-broken phase. 

Another concern is the value of $\beta (\equiv 6/g^2)
$ in simulations. 
It is suggested that 
there are some phase structures  
inherent to the lattice model even for a small number of flavors. 
Actually, following 
earlier suggestions~\cite{Banks:1981nn, Miransky:1996pd},
there have been 
some studies 
about the lattice phase diagram in the large 
$N_f$
QCD~\cite{Deuzeman:2009mh,Cheng:2011ic,Hasenfratz:2011xn} 
which, in fact, indicate 
a nontrivial lattice phase with a bulk transition at $\beta$ other than 
the critical value of $\beta$ which is associated with
the IR fixed point we are interested in.
Thus, it is important to  survey the $\beta$ dependence 
for the investigation of the low-energy physics.

In this work, we study the phase structure of the twelve-flavor QCD.
Preliminary results were given in Ref.~\cite{Aoki:2012kr}. 
We utilize the  highly improved staggered
quarks (HISQ) type action~\cite{Follana:2006rc, Bazavov:2010ru}  to reduce the discretization error.
This is 
 the first attempt to use the HISQ-type action for the large $N_f$ QCD. 
A salient feature of our work  is that we perform  simulations of $N_f=12$ 
in comparison with other cases of $N_f=(0), 4, 8, 16$ 
which we do simultaneously with the same systematics 
based on the same HISQ-type action~\cite{Aoki:2012ep, Aoki:2012kr}.  
We make simulations 
at two values of $\beta \, (\equiv 6/g^2$) (or two lattice-spacings), with $g$ being the bare gauge coupling,
to study the lattice spacing dependence of the data for the reason we described above. 
Our results suggest that two coupling regions are consistent to be
in the asymptotically free region,
where we discuss lattice spacing dependence of the physical quantities.

We investigate 
several bound-state masses such as the pseudoscalar meson $\pi$ (corresponding to the exact
NG boson if it is in the broken phase)  and vector meson $\rho$ as well as 
the decay
constant of $\pi$, 
by varying the fermion bare mass $m_f$.

Concerning the controversy about the previous lattice studies of $N_f=12$ QCD mentioned before, 
it 
should be pointed out 
that there is a problem in judging whether this theory is in the
conformal or the chiral-broken 
phase from the fit analysis using lattice
data of the hadron spectra with finite fermion bare mass $m_f$,
which fully depends on the fitting forms of 
the ChPT
fit in the chiral-broken phase 
as well as 
on the hyperscaling relation in the conformal hypothesis.  
Indeed, we do not know the 
systematics of the 
corrections 
due to the finiteness of the  mass $m_f$ and the volume
size $L$ in either case, which could make some confusion.

In the present work, 
we introduce 
 an alternative quantity for the scaling test of the
conformal hypothesis  with finite volume in the analyses 
of the simulations for $N_f=12$ QCD. 
Using this method, it is possible to analyze the data without any
specific function form as a fitting function to the data.
We can discuss 
possible finite mass and finite size corrections.   
We find that the results are consistent with hyperscaling 
for $\gamma_* \sim 0.4-0.5$.
This can be compared with the value 
$\gamma_\ast^{\rm SD} \simeq 0.80$, 
the value calculated through the ladder Schwinger-Dyson equation
$\gamma_\ast^{\rm SD} \simeq 1-\sqrt{1-\frac{\alpha_\ast}{\alpha_{cr}}}$~\cite{Leung:1985sn, 
Yamawaki:1985zg}
for two-loop $\alpha_\ast \simeq 0.754$ and $\alpha_{cr}=\pi/4$~\footnote{
Using three-loop (four-loop) $\alpha_\ast$ in the MS-bar scheme in this expression, 
one obtains $\gamma_\ast^{\rm SD} \simeq 0.33\ (0.37)$. 
Note that $\gamma_\ast^{\rm SD}$ can be written as 
$\gamma_\ast^{\rm SD} \simeq 1 - \sqrt{1-2\gamma_\ast^\text{1-loop}}$.
This is also compared 
with perturbative two-loop ($\gamma_\ast \simeq 0.77$), three-loop ($\gamma_\ast \simeq 0.31$), 
and four-loop ($\gamma_\ast \simeq 0.25$) calculations~\cite{Ryttov:2010iz}.
}. 
Moreover, for 
concrete understanding of possible corrections, we try to perform 
several fits
with some correction terms. 
 
We also perform the ChPT analysis of our data.
It turns out that our data are far from satisfying the consistency condition 
in Eq.~(\ref{eq:X}).

Our paper is organized as follows.
In Sec.~\ref{sec:method},  
we explain our model and simulation setup.
The results of the numerical simulations for the spectra 
are shown in Sec.~\ref{sec:primary_results}. 
Further study of dimensionless ratios 
composed of these measurements will be performed
for the study of the hyperscaling.
The consistency of the asymptotically free domain is also discussed
from the $\beta$ dependence of these observations.
Our main result is given in Sec.~\ref{sec:fss},
where we introduce a method for the scaling test for conformal hypothesis
and analyze the data using this method.
We deduce the anomalous dimension from our analysis.
In Sec.~\ref{sec:fshp} 
we perform the analysis based on the fit function which 
includes the corrections to hyperscaling as suggested by certain models.
In Sec.~\ref{sec:chpt_summary}, 
we briefly summarize the results on the analysis of the S$\chi$SB scenario.
The summary and discussion are given in Sec.~\ref{sec:summary}.
In Appendix \ref{sec:other_spectra}, we show the results for other spectra than those discussed in the text.
In Appendix \ref{sec:conformal_fit}, we show the numerical details of the fit results on 
the finite size hyperscaling.
In Appendix \ref{sec:chpt}, the details of the fit analysis based on the ChPT are shown.

%% file: text_sections/method.tex
%
\subsection{Continuum theory}

Our target is the SU(3) gauge theory with $N_f=12$ of massless
Dirac fermions in the fundamental representation. Investigating the
spectral quantities in mesonic channel of the mass-deformed theory,
we try to determine the phase and the quantities which are characteristic
to the associated phase. We use the technique of lattice gauge theory,
whose continuum counterpart is briefly described here.
The continuum Euclidean action reads,
\begin{equation}
  \label{eq:c_action}
  S= \int d^4 x    
  \left\{
    \frac{1}{2g^2}\mathrm{Tr} F_{\mu\nu}^2  + \sum_{q=1}^{12} \bar{\psi}_q
(\gamma \cdot D+m_f) \psi_q
  \right\},
\end{equation}
where $\psi_q$ denotes the 12-flavor fundamental fermions with
degenerate mass $m_f$.
The flavor nonsinglet bilinear operators 
$P^a\equiv \bar{\psi} \gamma_5 \sigma^a \psi$ for the pseudoscalar and
$\tilde{V}_i^a\equiv \bar{\psi} \gamma_4\gamma_i \sigma^a \psi$ 
for the vector channel are used to interpolate the bound state
with the particular quantum numbers under flavor and Lorentz 
symmetries. Here, $\sigma^a$ is a generator of the SU(12) flavor
symmetry group.
The ground-state mass, $M_H$ where $H=\pi$ for the pseudoscalar
($O=P^a$) or $H=\rho$ for vector 
($O=\tilde{V}_i^a$)
characterizes the asymptotic 
fall off $G_O \sim e^{-M_H t}$ of the Euclidean correlation function 
with the zero-momentum projection
\begin{equation}
  G_O(t) \equiv \int d^3\vec{x}  
  \langle 0 | O(t,\vec{x}) O^{\dagger}(0,\vec{0})| 0 \rangle,
\end{equation}
calculated using the action Eq.~(\ref{eq:c_action}).
The pseudoscalar decay constant is obtained through the matrix
element of the pseudoscalar operator,
\begin{equation}
 F_\pi= \frac{m_f}{M_\pi^2} \langle  0| P^a(0) |\pi^a ;\vec{p}=\vec{0} \rangle,
\end{equation}
using partially conserved axial current (PCAC) 
relation\footnote{We use the convention as $F_\pi=\sqrt{2}f_\pi$, 
where $f_\pi=93$[MeV] in the real-life QCD.}.

If the theory is in the conformal window, $M_H$ and $F_\pi$ obey
the conformal hyperscaling 
\begin{equation}
 M_H \propto m_f^{\frac{1}{\gamma_*+1}},\; 
  F_\pi \propto m_f^{\frac{1}{\gamma_*+1}},
  \label{eq:hyperscaling}
\end{equation}
where $\gamma_*$ denotes the mass anomalous dimension at the
IR fixed point.
This relation is the fundamental scaling appearing 
in the critical phenomena of the statistical mechanics.
In the context of the conformal window in the large $N_f$ QCD, 
see, for example, Ref.~\cite{Miransky:1998dh}.
On the other hand, if the theory is in the phase of S$\chi$SB, 
leading mass dependence will be 
\begin{equation}
 M_\pi^2 \propto m_f,\; 
  F_\pi = c_0 + c_1 m_f,
  \label{eq:chpt_leading}
\end{equation}
with $c_0\ne 0$,
and the vector meson mass does not vanish in the chiral limit.

The spectra obtained in our lattice simulation will be tested against
these two hypothesis in the following sections.

\subsection{Lattice setup}

A staggered fermion formulation is used to define the lattice version
of the $N_f=12$ SU(3) gauge theory. A theory with three degenerate
staggered fields with the mass $m_f$ has the degree of freedom of the
$N_f=12$ Dirac fermions and matches with the
theories with twelve degenerate flavors in the continuum
limit from the asymptotic free regime.
Note that there is no problem of
locality in this lattice theory,  as the rooting trick to tune
the number of flavors to $N_f\ne 4n$ theory is not used here.
At a nonzero lattice spacing, where one can perform numerical
computations, the 12-fold degeneracy does not hold, in general.
This nondegeneracy would manifest itself as the nondegeneracy of
the would-be degenerate $12\times 12 -1$ mesons in the continuum 
theory. As the number of the effective light degrees of freedom
is important to the critical phenomena, such as the IR conformality 
near the IR fixed point, this nondegeneracy
needs to be made much smaller than other characteristic 
scales of the theory. This can be achieved by taking the
continuum limit or using the lattice action which suppresses these
effects at a given lattice spacing.

We use a version of the HISQ~\cite{Follana:2006rc,Bazavov:2010ru} 
to simulate this system.  
This action has the best continuum scaling among the staggered actions 
used so far in the real QCD simulations.
For practical reasons, 
the tree-level improved Symanzik gauge action instead of the
1-loop tadpole improved one is adopted. Exactly the same setting,
called HISQ/tree, but with the rooting for 2+1 flavor simulation, has
been used for the QCD thermodynamics and has lead to an expectedly
good 
reduction of the flavor-symmetry violation in the pseudoscalar
mass splitting \cite{Bazavov:2011nk}.
We actually observe excellent results for the flavor symmetry 
in our $N_f=12$ simulations,
which is briefly shown in Appendix \ref{sec:other_spectra}.

We use three lattice volumes
$(L,T)=(18,24), (24, 32)$, and $(30, 40)$ and two lattice spacings
with $\beta=3.7$ and $4.0$,
where $L$ and $T$ are spatial and temporal length of the finite
lattice and $\beta=6/g^2$ with bare gauge coupling $g$.
Note that the aspect ratio $T/L$ is kept fixed $=4/3$
while the volume is changed.
In this way, we only have one IR scale $1/L$,
which is ideal for the finite-size scaling analysis.
For the quark mass $m_f$, we take various values: 
$m_f=0.04, 0.05, 0.080, 0.1, 0.12, 0.16, 0.2, 0.24$, 
where $0.04$ and $0.24$ are only for $\beta=3.7$ and $4.0$, respectively.
The gauge configurations are generated by the hybrid Monte Carlo
algorithm using MILC code ver.~7 with some modifications to suit
our needs. 
The boundary condition is set periodic for all except temporal boundary
for fermions, which is made antiperiodic.
We accumulate 
400-1200 thermalized trajectories for each 
ensemble.
Calculation of the mesonic correlation functions
is performed at every 2-10 trajectories,  
thus, we have 70-400 samples for each 
ensemble.
The error analysis is performed with the standard jackknife method with a
suitable bin size $10-100$ depending on the ensemble parameter.

For comparison, we also analyze the $N_f=4$ theory which is in the
phase of S$\chi$SB, 
where the same lattice
action is used. Some detail of the $N_f=4$ simulation is given in 
Ref.~\cite{Aoki:2012ep}
and the forthcoming publication.

%% file: text_sections/results.tex
%

\begin{figure}[htbp]
  \begin{center}
      \includegraphics[width=7.5cm,clip]{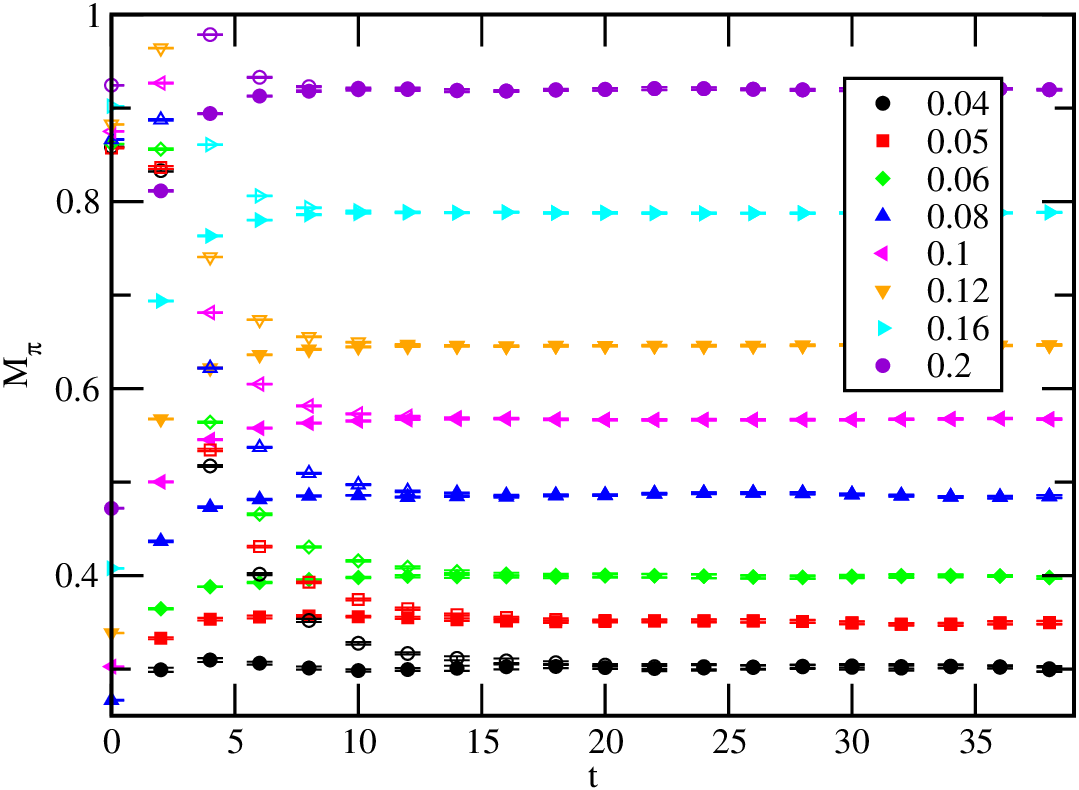} 
      \includegraphics[width=7.5cm,clip]{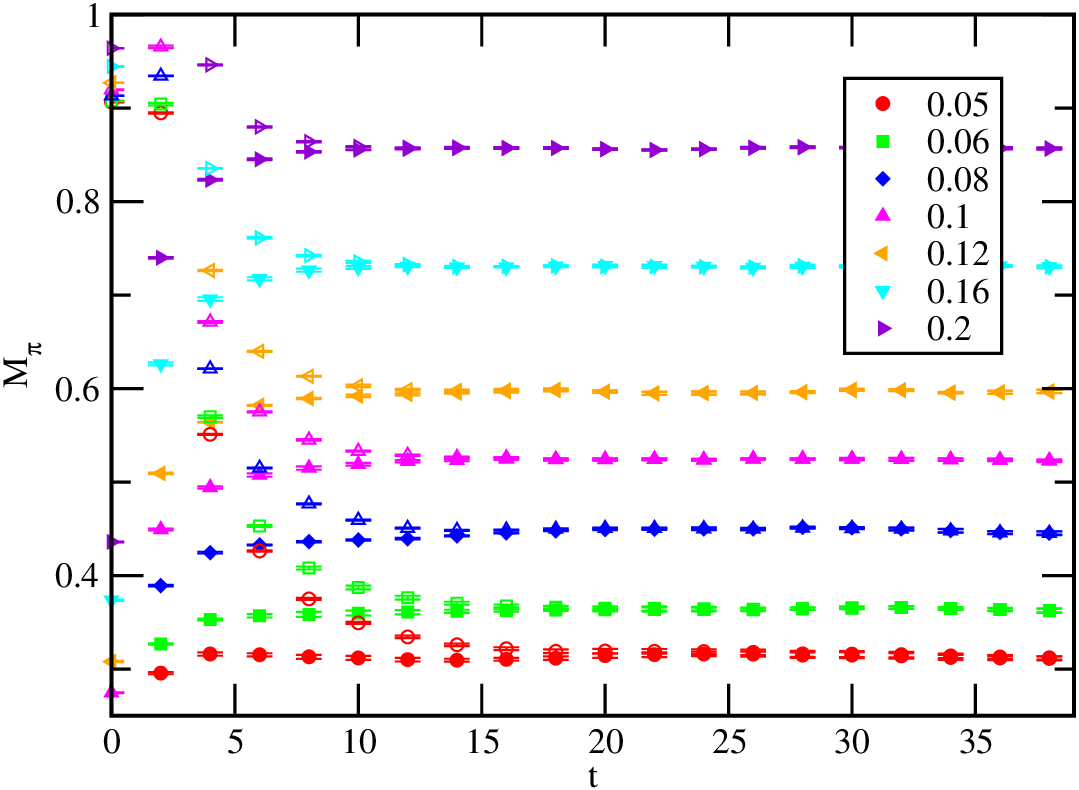} 
  \end{center}
  \caption{
 Effective mass of $\pi$ at $\beta=3.7$ (Left) and
 $\beta=4$ (Right) on $(L,T)=(30,40)$ with $\tilde{C}_{\rm PS}$ using two
 types of quark sources. See the text.
}
   \label{fig:meff}
\end{figure}

We measure the two-point correlation functions of the staggered bilinear
pseudoscalar operator which corresponds to the NG
mode associated with the exact chiral symmetry of the staggered fermions.
The corresponding spin-flavor structure is $(\gamma_5\otimes\xi_5)$,
denoted by ``PS'' in Ref.~\cite{Bowler:1986fw}.
The random wall source is used for the quark operator for the bilinear,
which becomes a noisy estimator of the point bilinear operator with
spatial sum at a given time slice $t_0$.
We combine quark propagators solved with periodic and antiperiodic
boundary conditions in the temporal direction~(see, e.g., Ref.~\cite{Blum:2001xb}).
In this well-known technique,  
the temporal size is effectively doubled, which
enables us to have sufficient range for the fitting, which only
takes into account the ground state.
Denoting such a $\pi$ correlator as $C_{\rm PS}(t)$, we use 
$\tilde{C}_{\rm PS}(2t)=C_{\rm PS}(2t)/2 +C_{\rm PS}(2t-1)/4+C_{\rm PS}(2t+1)/4$
for the analysis.
This linear combination kills the constant oscillation mode, which
could originate from the single quark line wrapping around the
antiperiodic temporal boundary. 
The results of effective mass calculated with $\tilde{C}_{\rm PS}(2t)$ 
for $L=30$ are shown in Fig.~\ref{fig:meff} (see those getting plateau from above). 
Effective mass of the same $\pi$ correlation function with
the Coulomb-gauge-fixed wall source is also plotted (ones getting
plateau from below). 
Note that the plateau starts early enough to isolate the ground state
even if the temporal size had been $T=24$ (our smallest).
The $\pi$ mass is obtained by the fit of the two-point correlators 
of $\tilde{C}_{\rm PS}$ from a random source with double period by a fit function
\begin{equation}  \label{eq:PS}
\tilde{C}_{\rm PS}(2t)=C(e^{-M_{\pi}2t}  + e^{-M_{\pi}(2T-2t)}),
\end{equation}
where $M_{\pi}$ is a mass of the $\pi$ and  the fit range is taken 
$[t_{min}, T]$. 
Similarly we can obtain $F_\pi$ from this operator.
The values of $t_{min}$ are taken as 
$t_{min}=16-22$ and $18-22$ at $\beta=3.7$ and $4$, 
respectively.

\begin{figure}[htbp]
  \begin{center}
      \includegraphics[width=7.5cm,clip]{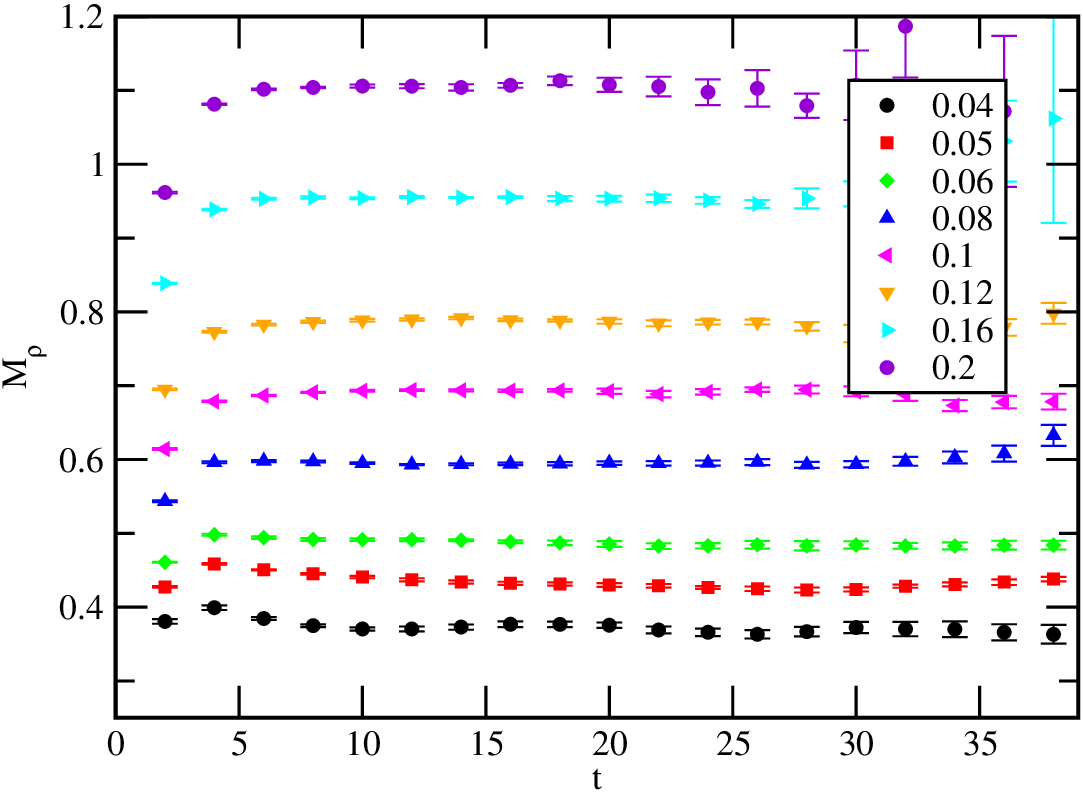} 
      \includegraphics[width=7.5cm,clip]{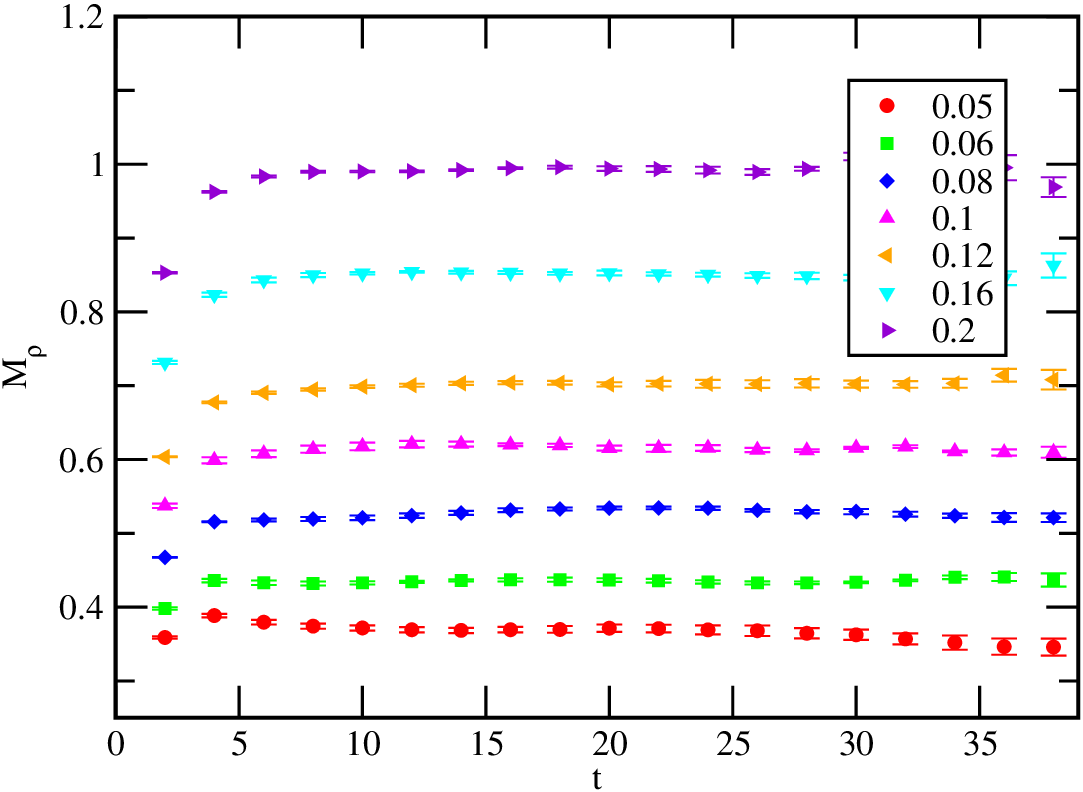} 
  \end{center}
  \caption{
Effective mass of the PV correlator ($\tilde{G}_{\rm PV}$) at $\beta=3.7$ (Left) 
and $\beta=4$ (Right) 
      on $(L,T)=(30,40)$. 
}
   \label{fig:meff_PV}
\end{figure}

We measure $M_\rho$ from the staggered vector meson operator
$(\gamma_i\gamma_4\otimes\xi_i\xi_4)$, denoted by PV in
Ref.~\cite{Bowler:1986fw}. 
The asymptotic form of the PV correlator may be written as 
\begin{equation} \label{eq:rho}
G_{\rm PV}(t)=C_1( e^{-M_\rho t} +e^{-M_\rho(2T-t)}) 
+C_2 (-1)^t ( e^{-M_{a_1} t} +e^{-M_{a_1}(2T-t)}) 
\end{equation}
where $M_{a_1}$ corresponds to the mass of the axial vector meson.
Since there exists a constant mode
due to the wrapping-around effect, we use  
$\tilde{G}_{\rm PV}(2t)=G_{\rm PV}(2t)/2 +G_{\rm PV}(2t-1)/4+G_{\rm PV}(2t+1)/4$.
Again, the results of effective mass calculated with $\tilde{G}_{\rm PV}$ 
for $L=30$ are shown in Fig.~\ref{fig:meff_PV}. 
The fitting range for the PV correlator is $[t_{min},T]$
with $t_{min}=10-12$, and $12-14$ at $\beta=3.7$ and $4$, respectively, 
which is enough to isolate the ground state, i.e. $\rho$.
We obtain $M_\rho$ by the fit of the two-point correlators of $\tilde{G}_{\rm PV}$ 
by a simple fit function: 
\begin{equation}  \label{eq:PV}
\tilde{G}_{\rm PV}(2t)=C(e^{-M_{\rho}2t}  + e^{-M_{\rho}(2T-2t)}).
\end{equation}

All the raw results of $N_f=12$ theory
which are used in the next sections 
are  shown in the Tables~\ref{tab:1}-\ref{tab:6}. 
In these tables, 
we also show the number of the thermalized trajectory for each parameter,
by which we measure the masses and the decay constant.

Besides these main channels, we study the masses of 
mesons made with local operators,
a non-NG channel $(\gamma_5\gamma_4\otimes\xi_5\xi_4)$ denoted by ``SC'', 
and a vector meson $(\gamma_i\otimes\xi_i)$ denoted by ``VT'' in Ref.~\cite{Bowler:1986fw}.
They are compared against PS and PV in Appendix \ref{sec:other_spectra}, which indicates 
the flavor-breaking effect of the HISQ is very small.

\begin{small}
\begin{table}[tbp]
\begin{tabular}{cc}
\begin{minipage}[t]{.48\textwidth}
\begin{center}
\begin{ruledtabular}
\caption{The results of the spectra on $V=18^3 \times 24$ at
$\beta=3.7$.
$N_{\rm{trj}}$ means the total number of thermalized trajectories.}
\label{tab:1}
\begin{tabular}{ccccc}
$m_f$ & $N_{\rm{trj}}$ &  $M_\pi$  & $M_{\rho}$ & $F_\pi$ \\ \hline
0.04 & 1000 & 0.3387(62) & 0.4614(85)  &  0.0633(24)  \\
0.05 & 1000 & 0.3830(38) & 0.4899(71)  &  0.0750(22)  \\
0.06 & 1200 & 0.4100(51) & 0.5144(49)  &  0.0814(12)  \\
0.08 & 1200 & 0.4922(17) & 0.6139(33)  &  0.1027(11)  \\
0.1  &  700 & 0.5717(17) & 0.7084(33)  &  0.1185(6)  \\
0.12 & 700 & 0.6485(16) & 0.7905(38)  &  0.1333(11)  \\
0.16 & 700 & 0.7878(13) & 0.9546(20)  &  0.1590(9)  \\
0.2  &  700 & 0.9209(10) & 1.1041(32)  &  0.1838(8)  
\end{tabular}
\end{ruledtabular}
\end{center}
\end{minipage}
\ \ \hfill
\begin{minipage}[t]{.48\textwidth}
\begin{center}
\begin{ruledtabular}
\caption{The results of the spectra on $V=18^3 \times 24$ at
$\beta=4$.}
\label{tab:2}
\begin{tabular}{ccccc}
$m_f$ & $N_{\rm{trj}}$ &  $M_\pi$  & $M_{\rho}$ & $F_\pi$ \\ \hline
0.05 & 1000 & 0.3993(67)  & 0.4529(88)  &  0.0573(9)  \\
0.06 & 1200 & 0.4101(41)  & 0.4687(115)  &  0.0674(11)  \\
0.08 & 1200 & 0.4551(22)  & 0.5438(44)  &  0.0878(10)  \\
0.1  &  1200 & 0.5219(43)  & 0.6133(89)  &  0.1009(19)  \\
0.12 & 1200 & 0.5960(18)  & 0.6955(40)  &  0.1149(10)  \\
0.16 & 1000 & 0.7309(21)  & 0.8564(62)  &  0.1378(11)  \\
0.2  &   700 & 0.8567(11)   & 0.9903(38)  &  0.1586(12)  \\
0.24 &  700 & 0.9760(17)  & 1.1225(29)  &  0.1785(14) 
\end{tabular}
\end{ruledtabular}
\end{center}
\end{minipage}
\end{tabular}
\end{table}
\end{small}
\begin{small}
\begin{table}[tbp]
\begin{tabular}{cc}
\begin{minipage}[t]{.48\textwidth}
\begin{center}
\begin{ruledtabular}
\caption{The results of the spectra on $V=24^3 \times 32$ at
$\beta=3.7$.}
\label{tab:3}
\begin{tabular}{ccccc}
$m_f$ & $N_{\rm{trj}}$ &  $M_\pi$  & $M_{\rho}$ & $F_\pi$ \\ \hline
0.04 & 600 & 0.3054(31) & 0.3569(86)  &  0.0621(15)  \\
0.05 & 600 & 0.3549(16) & 0.4377(44)  &  0.0750(7)  \\
0.06 & 1000 & 0.3986(22) & 0.4932(54)  &  0.0835(7)  \\
0.08 & 800 & 0.4880(8) & 0.5991(25)  &  0.1014(4)  \\
0.1  & 800 & 0.5683(17) & 0.6971(32)  &  0.1175(6)  \\
0.12 & 700 & 0.6459(5)  & 0.7891(24)  &  0.1325(4)  \\
0.16 & 700 & 0.7879(6)  & 0.9529(22)  &  0.1591(8)  \\
0.2  & 700 & 0.9193(6)  & 1.1031(21)  &  0.1818(8)  
\end{tabular}
\end{ruledtabular}
\end{center}
\end{minipage}
%
%
\begin{minipage}[t]{.48\textwidth}
\begin{center}
\begin{ruledtabular}
\caption{The results of the spectra on $V=24^3 \times 32$ at
$\beta=4$.}
\label{tab:4}
\begin{tabular}{ccccc}
$m_f$ & $N_{\rm{trj}}$ &  $M_\pi$  & $M_{\rho}$ & $F_\pi$ \\ \hline
0.05 & 1000 & 0.3290(25)  & 0.3819(81)  &  0.0615(7)  \\
0.06 & 1000 & 0.3677(13)  & 0.4353(43)  &  0.0714(7)  \\
0.08 & 1000 & 0.4459(11)  & 0.5257(31)  &  0.0875(9)  \\
0.1  & 1000 & 0.5210(7)  & 0.6165(27)  &  0.1014(4)  \\
0.12 & 1000 & 0.5946(11)  & 0.6999(21)  &  0.1147(5)  \\
0.16 &  700 & 0.7308(9)   & 0.8519(27)  &  0.1378(5)  \\
0.2  &  700 & 0.8557(5)   & 0.9893(29)  &  0.1579(4)  
\end{tabular}
\end{ruledtabular}
\end{center}
\end{minipage}
\end{tabular}
\end{table}
\end{small}

\begin{small}
\begin{table}[tbp]
\begin{tabular}{cc}
\begin{minipage}[t]{.48\textwidth}
\begin{center}
\begin{ruledtabular}
\caption{The results of the spectra on $V=30^3 \times 40$ at
$\beta=3.7$.}
\label{tab:5}
\begin{tabular}{ccccc}
$m_f$ & $N_{\rm{trj}}$ &  $M_\pi$  & $M_{\rho}$ & $F_\pi$ \\ \hline
0.04 & 800 & 0.3028(14) & 0.3713(39)  &  0.0637(6)  \\
0.05 & 700 & 0.3504(11) & 0.4302(20)  &  0.0741(5)  \\
0.06 & 600 & 0.3990(15) & 0.4864(28)  &  0.0835(5)  \\
0.08 & 500 & 0.4869(8)  & 0.5949(15)  &  0.1017(5)  \\
0.1   & 500 & 0.5670(7)  & 0.6925(16)  &  0.1167(3)  \\
0.12 & 500 & 0.6460(7)  & 0.7866(18)  &  0.1328(4)  \\
0.16 & 400 & 0.7877(6)  & 0.9542(15)  &  0.1586(5)  \\
0.2   & 400 & 0.9199(8)  & 1.1068(30)  &  0.1828(8)  
\end{tabular}
\end{ruledtabular}
\end{center}
\end{minipage}
\hfill
\begin{minipage}[t]{.45\textwidth}
\begin{center}
\begin{ruledtabular}
\caption{The results of the spectra on $V=30^3 \times 40$ at
$\beta=4$.}
\label{tab:6}
\begin{tabular}{ccccc}
$m_f$ & $N_{\rm{trj}}$ &  $M_\pi$  & $M_{\rho}$ & $F_\pi$ \\ \hline
0.05 &  600 & 0.3167(27)  & 0.3671(56)  &  0.0634(8)  \\
0.06 &  700 & 0.3648(14)  & 0.4357(17)  &  0.0732(6)  \\
0.08 &  600 & 0.4499(8)  & 0.5301(13)  &  0.0901(5)  \\
0.1  &   600 & 0.5243(7)   & 0.6150(16)  &  0.1027(6)  \\
0.12 &  600 & 0.5966(10)  & 0.7027(29)  &  0.1149(7)  \\
0.16 &  500 & 0.7308(8)   & 0.8508(21)   &  0.1380(7)  \\
0.2  &  500 & 0.8569(6)   & 0.9941(14)   &  0.1586(6)  
\end{tabular}
\end{ruledtabular}
\end{center}
\end{minipage}
\end{tabular}
\end{table}
\end{small}

\begin{figure}[htbp]
  \begin{center}
   \includegraphics[width=8cm,clip]{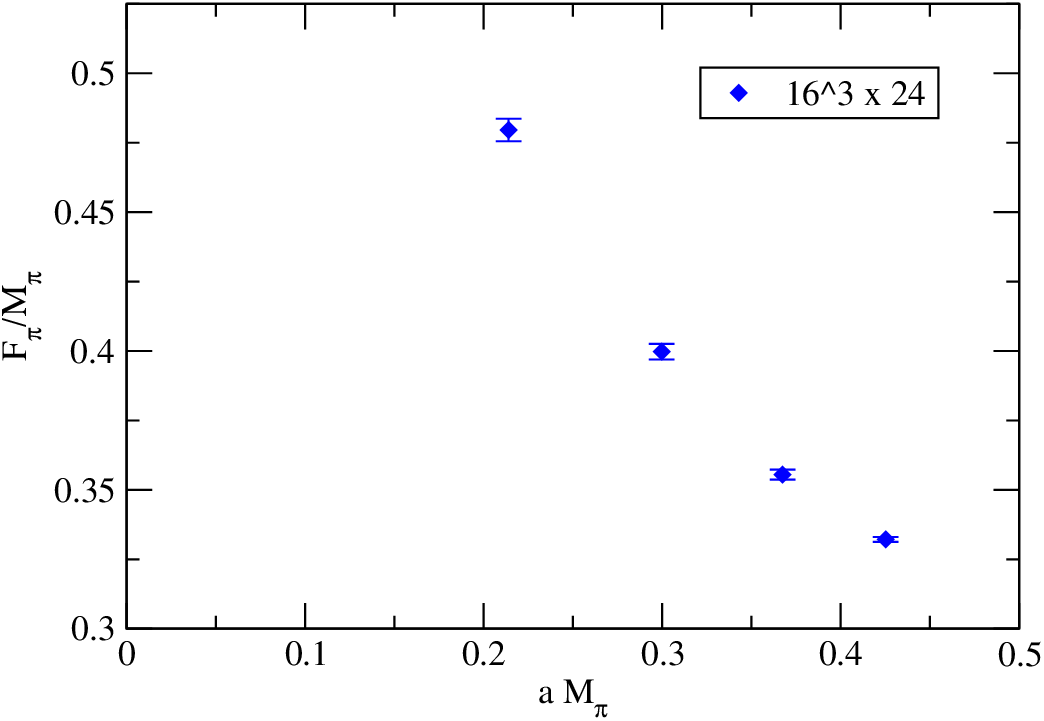} 
  \end{center}
 \caption{Dimensionless ratio $F_\pi/M_\pi$
 as a function of $aM_\pi$ for $N_f=4$ at $\beta=3.7$. Due to the 
S$\chi$SB,
the ratio diverges in the chiral limit.}
 \label{fig:ratio_mpi_nf4}
\end{figure}

\begin{figure}[htbp]
  \begin{center}
   \includegraphics[width=8cm,clip]{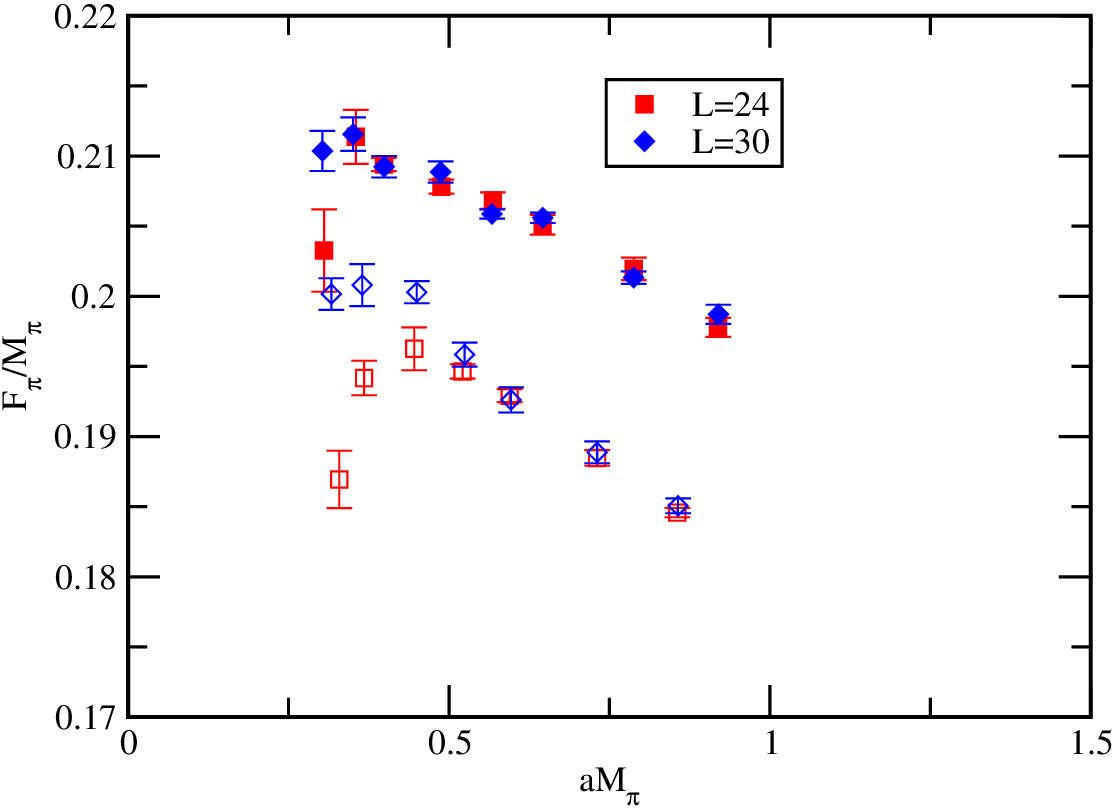}
   \includegraphics[width=8cm,clip]{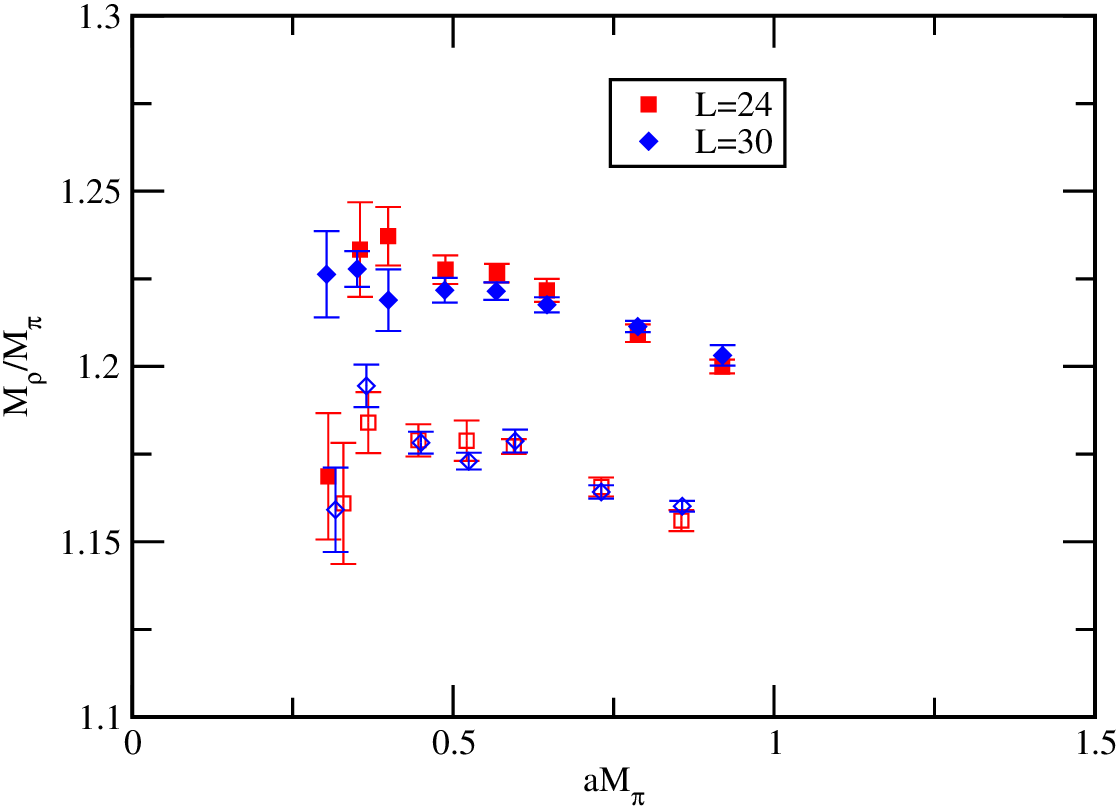} 
  \end{center}
 \caption{Dimensionless ratios $F_\pi/M_\pi$ and $M_\rho/M_\pi$
 as functions of $aM_\pi$ for $N_f=12$ at $\beta=3.7$ (filled symbol) and
 $4.0$ (open symbol) for the two largest volumes.}
 \label{fig:ratio_mpi_nf12}
\end{figure}

For a primary analysis, dimensionless ratios composed of these
measurements will be plotted against the $\pi$ mass.
In the rest of this section, 
we write lattice spacing $a$ explicitly and consider $M_\pi$ as 
dimensionful  quantity.
As a reference, we have calculated the $N_f=4$ case as plotted 
in Fig.~\ref{fig:ratio_mpi_nf4},
which resembles a familiar QCD-type behavior.
The upward tendency
of $F_\pi/M_\pi$ toward the chiral limit 
is consistent with the S$\chi$SB 
[Eq.~(\ref{eq:chpt_leading})].

The left panel of Fig.~\ref{fig:ratio_mpi_nf12} plots the same
quantity for $N_f=12$
at the largest two volumes and two bare gauge couplings $\beta=3.7$ and $4$.
If we look at the results for $\beta=3.7$ (filled symbol),
$F_\pi/M_\pi$ tends to be flat for smaller $\pi$ masses,
which shows clear contrast to $N_f=4$ case.
The behavior is consistent with the hyperscaling
Eq.~(\ref{eq:hyperscaling}). The pseudoscalar mass dependence of the ratio
at larger mass can be realized by the correction to the hyperscaling
which may be different from one quantity to another.
For $\beta=4$ (open symbol), there is no flat range without volume dependence.

Similar observation can be made for the other ratio $M_\rho/M_\pi$
shown in the right panel of Fig.~\ref{fig:ratio_mpi_nf12}.
Here, the flattening is observed for $\beta=3.7$ again, but
the range is wider than $F_\pi/M_\pi$. In this case, $\beta=4$ shows 
the flattening, also.
The difference of the constant is made possible due to a discretization
effect. 

In the following,
further detailed study using these data
is performed.  From the observation here, we note that the region
where hyperscaling is realized could be limited to the smaller masses.
Further, the scaling range of the $\pi$ decay constant may be narrower
at $\beta=3.7$, and there may be no scaling range for the decay
constant at $\beta=4$.

Existence of the scaling for $F_\pi$ at $\beta=3.7$ and absence
at $\beta=4$ at the same $aM_\pi$ can be made possible 
if the $M_\pi$ in the 
physical unit is larger (thus the correction is no longer negligible)
for $\beta=4$, {\it i.e.} the lattice  spacing decreases as $\beta$
increases. In that case, the physical volume is smaller for $\beta=4$,
which gives a reason for the volume effect observed only for $\beta=4$.

\begin{figure}[htbp]
  \begin{center}
   \includegraphics[width=8cm,clip]{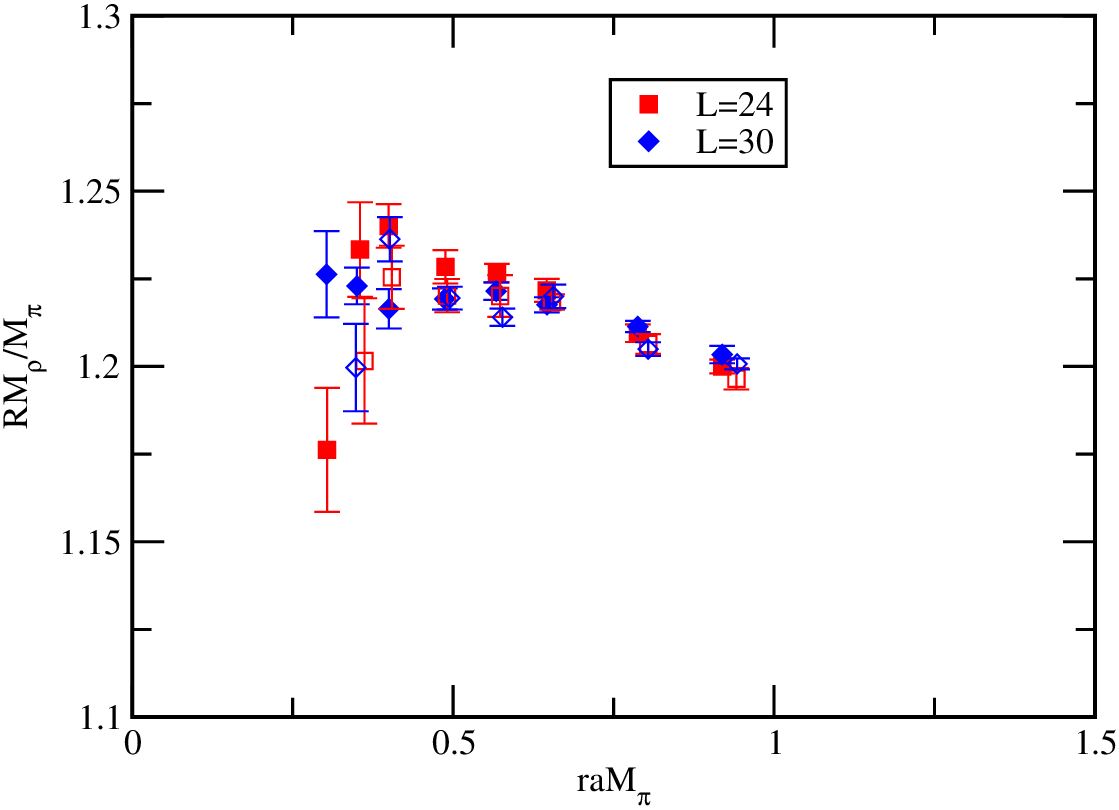} 
  \end{center}
 \caption{$R M_\rho/M_\pi$ is plotted against $r a M_\pi$ after a
 crude matching with $(R,r)=(1,1)$ for $\beta=3.7$ and $(1.035,1.1)$ for
 $\beta=4$. Legends are the same as Fig.~\ref{fig:ratio_mpi_nf12}.}
 \label{fig:ratio_mpi_nf12_match}
\end{figure}

To see if the inequality of the lattice spacing 
$a(\beta=3.7)>a(\beta=4)$ holds, a matching between the two data sets of
$M_\rho/M_\pi$ vs $aM_\pi$ is performed. 
We will not use $F_\pi/M_\pi$ for the matching due to the existence /
nonexistence of the volume effect for $\beta=3.7$ / $4$.
If an $aM_\pi$ point which describes the same physics is found
for each $\beta$ separately, it can be regarded that the
ratio of the lattice spacing as $M_\pi$ in physical units 
is the same for them.
We exploit the mass dependence at the tail for this matching.
To do that, first, the matching of the overall factor absorbing the 
discretization error in the vertical direction needs to be performed,
which can be done by introducing a factor $R$ to multiply $M_\rho/M_\pi$ 
for $\beta=4$ and to tune it to match the plateau to that of $\beta=3.7$.
Then, the remaining
difference which may appear at the tail is absorbed by the
horizontal matching factor $r$ as $r\cdot M_\pi$. Figure
\ref{fig:ratio_mpi_nf12_match} is the same as Fig.~\ref{fig:ratio_mpi_nf12},
after a crude matching is done with $(R,r)=(1.035,1.1)$.
In this particular definition of the lattice spacing and with
a crude analysis, the ratio of the lattice spacing is
$a(\beta=3.7)/(\beta=4)\sim 1.1$, and thus $a(\beta=3.7)>a(\beta=4)$
holds.
The fact that the lattice spacing decreases as $\beta=6/g^2$ increases
is consistent with being in the asymptotically free domain, even if
there is an IR fixed point in the beta function.

%% file: text_sections/fss.tex
%

\subsection{Preliminary}

In the conformal window with finite masses and volume, 
the renormalization group analysis tells us the scaling behavior 
for low-energy spectra 
which should obey the universal scaling relations~\footnote{
For reviews, see, e.g., 
Refs.~\cite{DeGrand:2009mt,DelDebbio:2010ze,DelDebbio:2010jy}.
} as
\begin{equation}
  \xi_p \equiv L M_{p} = f_p(x),  
 \label{eq:fss_mass}
\end{equation}
where $p$ distinguishes the bound state, $p=\pi$ or $\rho$ in this study, or
\begin{equation}
 \xi_F \equiv L F_{\pi} =f_F(x).
  \label{eq:fss_decay_constant}
\end{equation}
The product of bound-state mass or decay constant and linear system
size falls into a function of a single scaling variable~\footnote{
We adopt this definition, which corresponds to 
$x^{1+\gamma_*}$ of Ref.\cite{DelDebbio:2010ze}, to make its asymptotic form simpler.} 
\begin{equation}
 x=L\cdot m_f^{\frac{1}{1+\gamma_*}},
  \label{eq:scaling_variable}
\end{equation}
where $\gamma_*$ is the mass anomalous dimension at the IR
fixed point. 
We call these scaling relations ``finite-size hyperscaling'' (FSHS).
The forms of the scaling functions $f_p(x)$ are unknown in general. 
However, as the hyperscaling relation Eq.~(\ref{eq:hyperscaling}) 
must be reproduced in large volumes, 
the asymptotic form should be $f(x) \sim x$ at large $x$. 

\begin{figure}[tb]
  \begin{center}
      \includegraphics[width=5.4cm,clip]{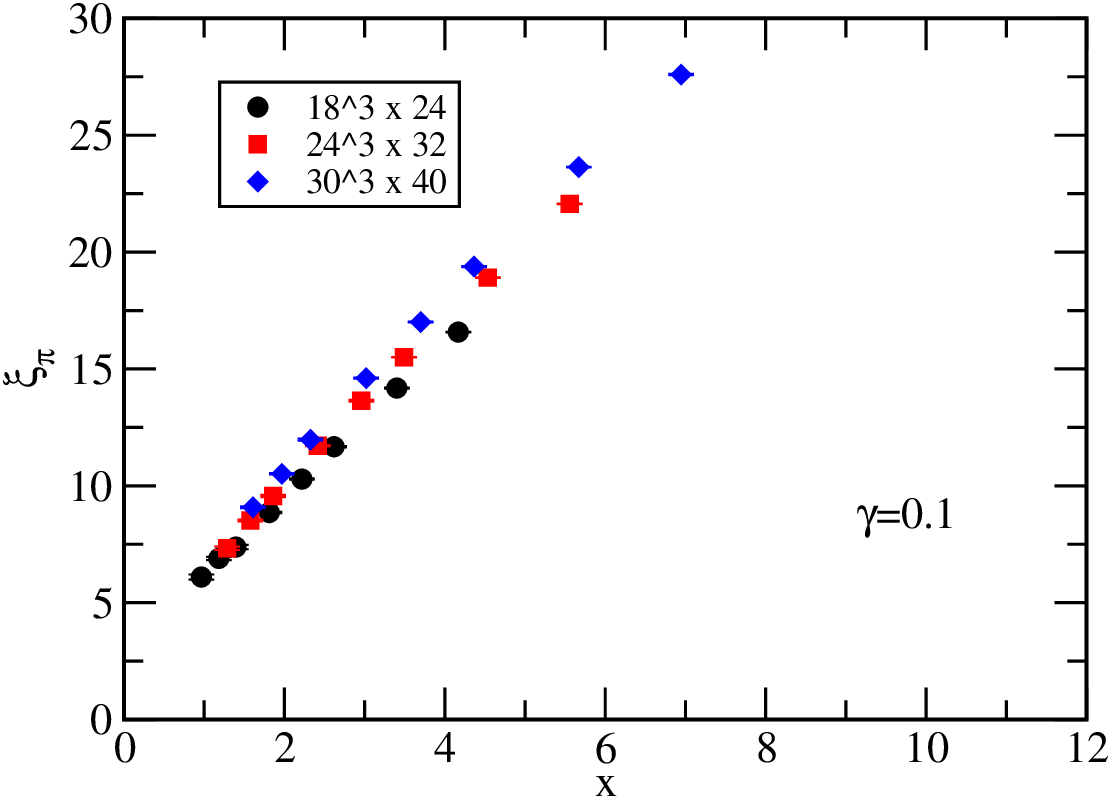}
      \includegraphics[width=5.4cm,clip]{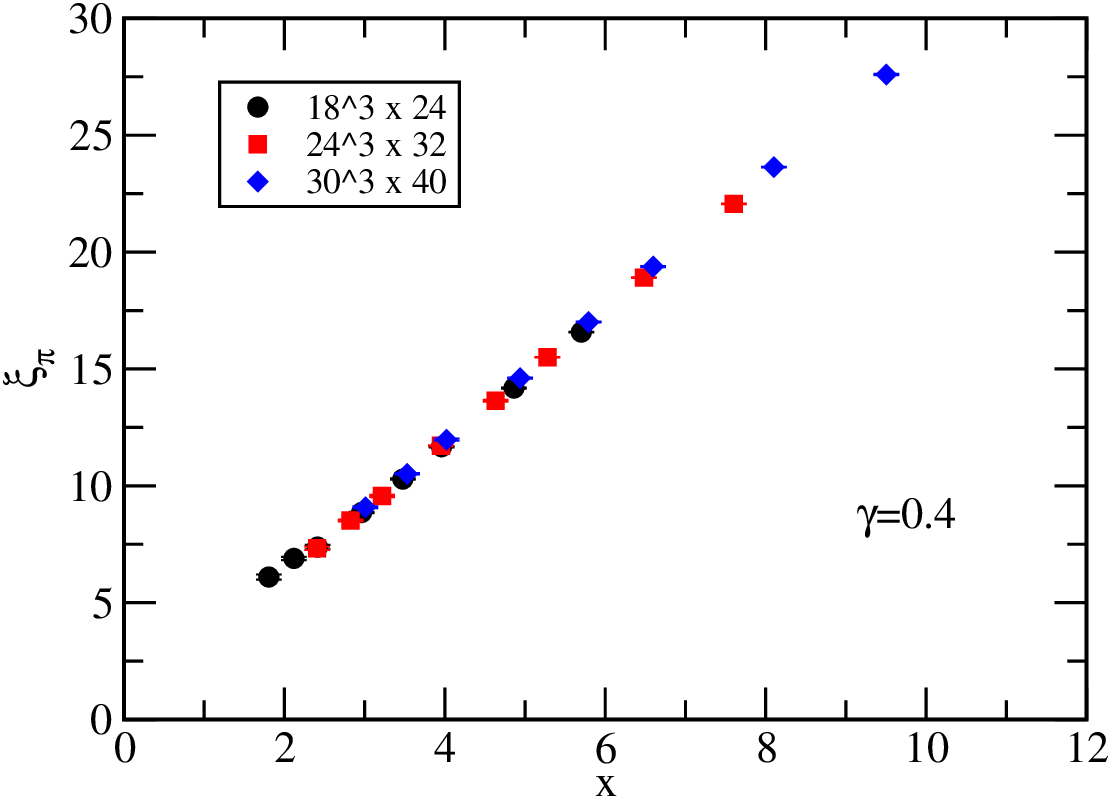}
      \includegraphics[width=5.4cm,clip]{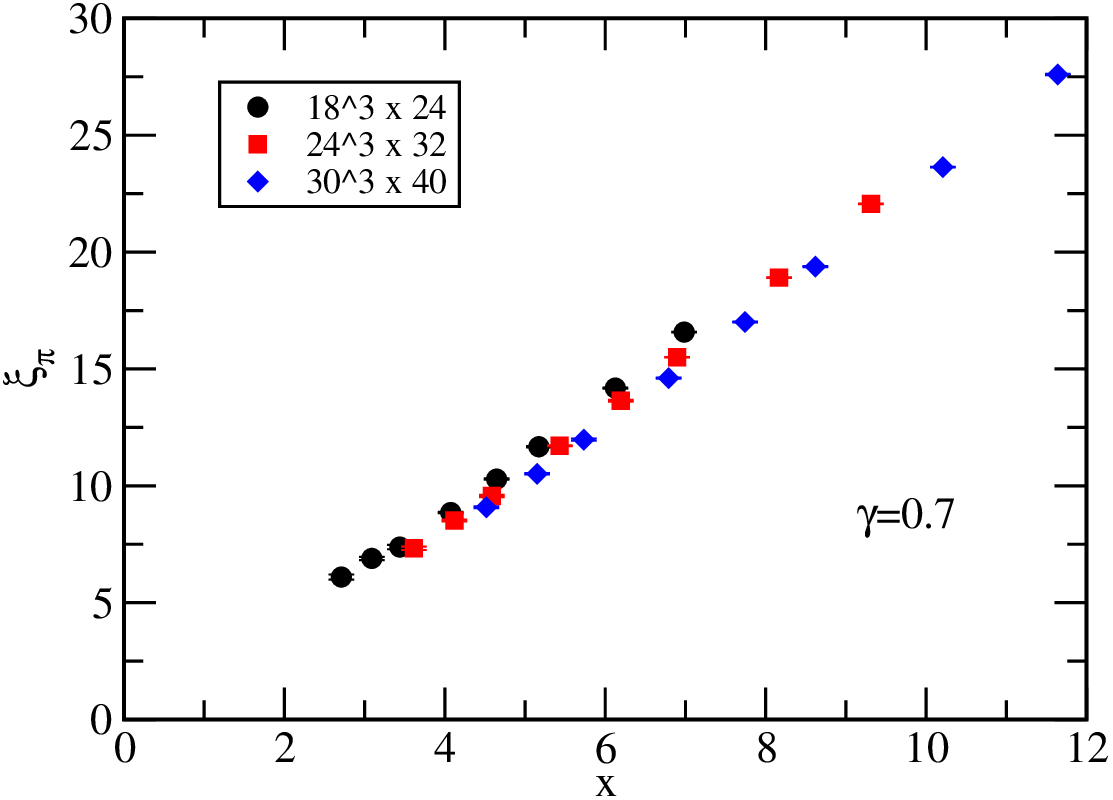}
  \end{center}
 \caption{$\xi_\pi$ is plotted as a function of the scaling variable $x$ for 
 $\gamma=0.1$, $0.4$, and $0.7$ from left to right for $N_f=12$ at
 $\beta=3.7$.  An alignment is seen for $\gamma\sim 0.4$.
  }
  \label{fig:mpi_g}
\end{figure}

\begin{figure}[tb]
  \begin{center}
      \includegraphics[width=5.4cm,clip]{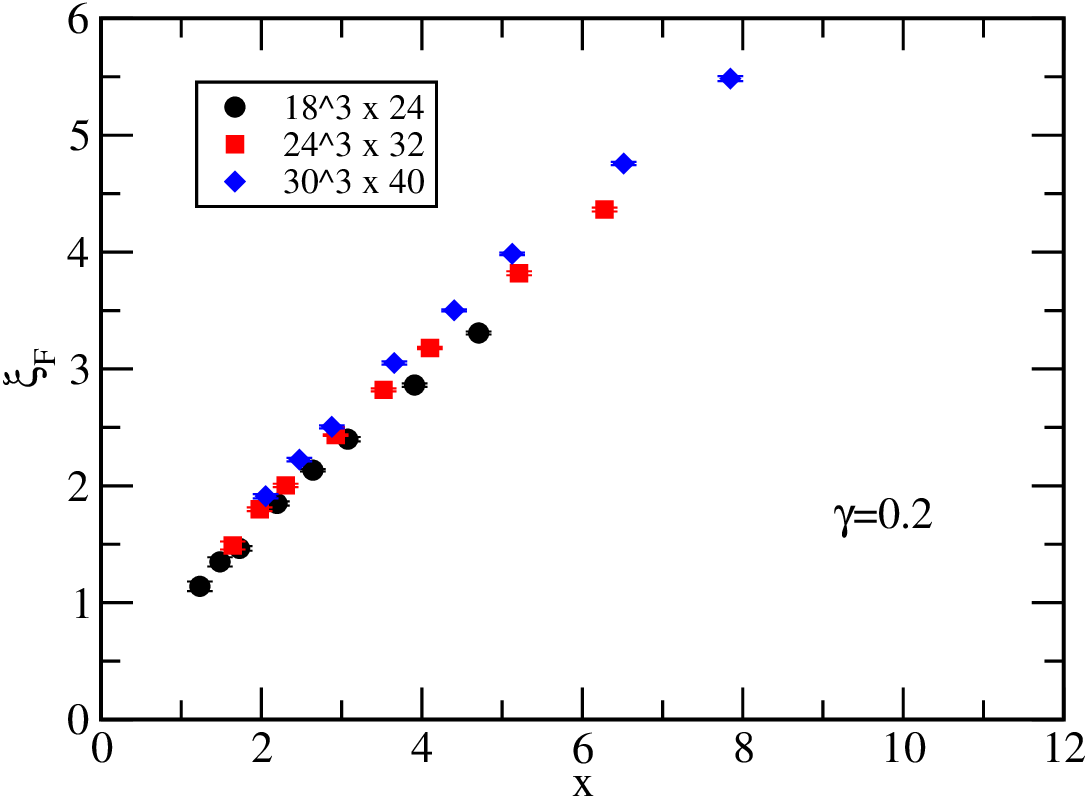}
      \includegraphics[width=5.4cm,clip]{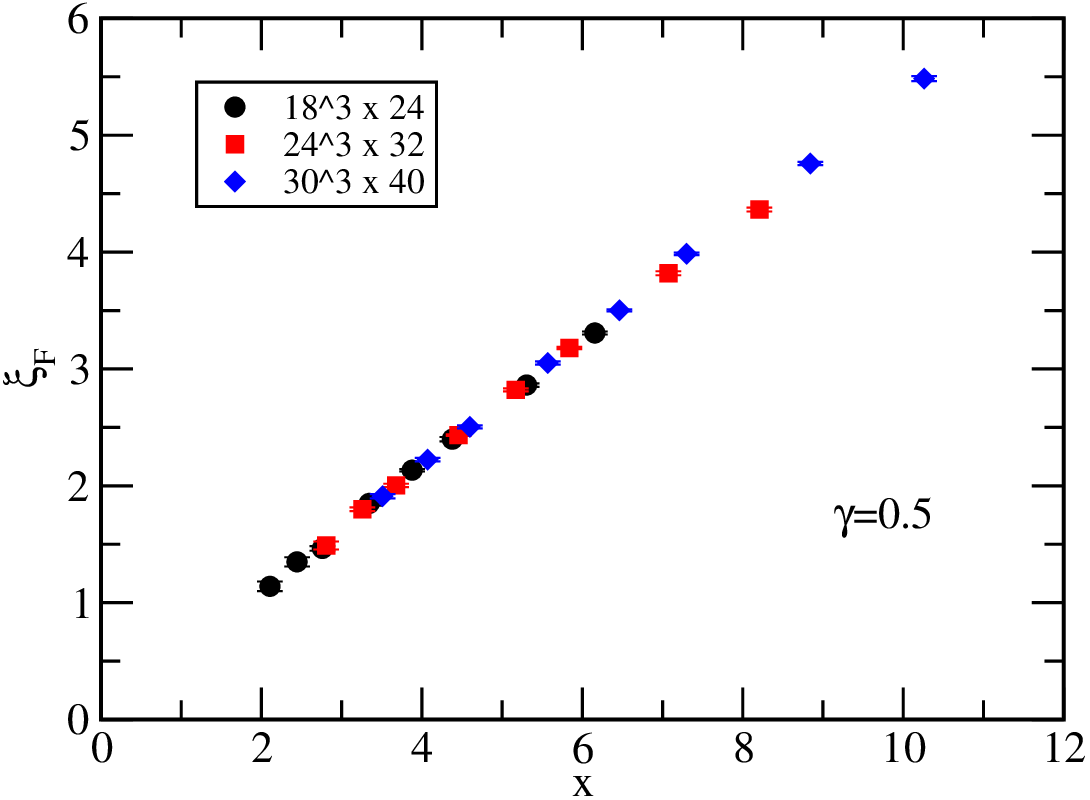}
      \includegraphics[width=5.4cm,clip]{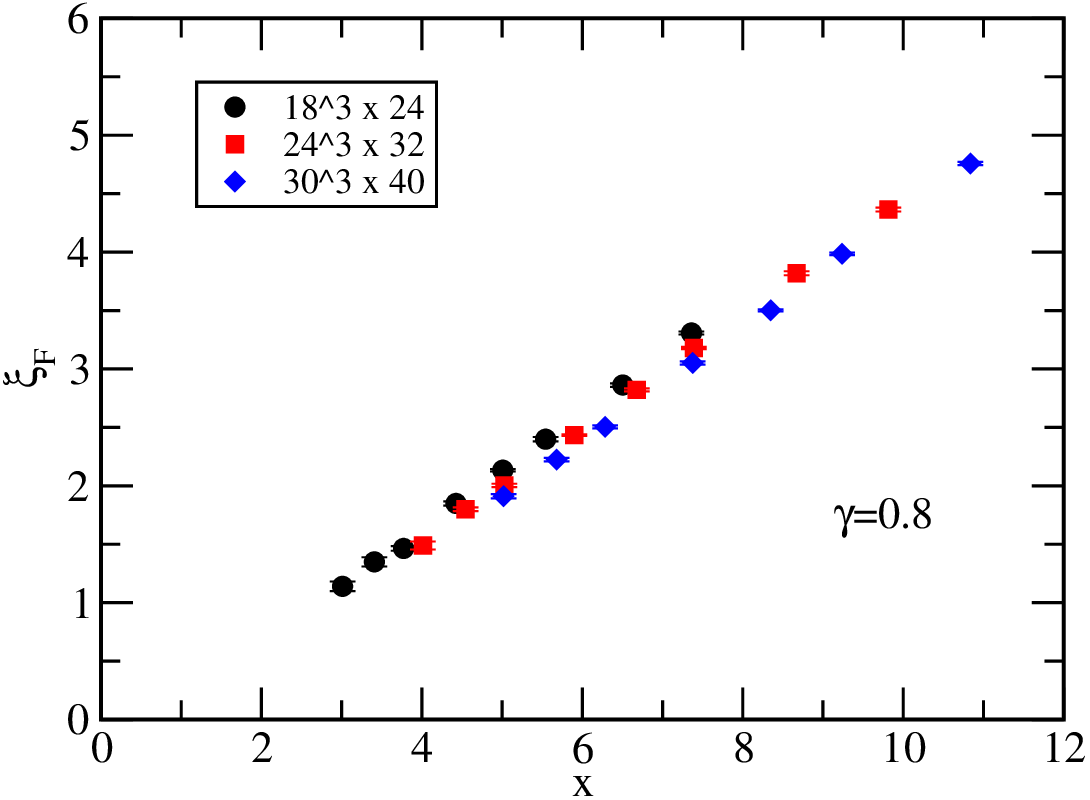}
  \end{center}
  \caption{$\xi_F$ as a function of $x$ for 
    $\gamma=0.2$, $0.5$, $0.8$ from left to right for $N_f=12$ at $\beta=3.7$. 
    An alignment is seen for $\gamma\sim 0.5$. 
}
   \label{fig:fpi_g}
\end{figure}

Now we examine whether our measurements of bound state mass and decay constant 
at different $m_f$ and $L$ obey FSHS. 
First, to visualize how the scaling works, we follow the analysis 
given, for example, in the model of the SU(3) with sextet fermions~\cite{DeGrand:2009hu}
and the SU(2) with adjoint fermions~\cite{DelDebbio:2010hu}.
Panels in Fig.~\ref{fig:mpi_g} show 
$\xi_\pi$ as functions of $x = L \cdot m_f^{1/(1+\gamma)}$ 
for several values of $\gamma$. 
It is observed that the data points align 
at 
around $\gamma=0.4$, while they become scattered for $\gamma$ 
away from that value. This indicates a possible existence of FSHS with 
$\gamma\sim 0.4$.
A similar alignment is observed for $\xi_F$ in Fig.~\ref{fig:fpi_g} as well.
In this case, one finds the optimal scaling at around $\gamma\sim 0.5$.

\begin{figure}[tb]
  \begin{center}
      \includegraphics[width=5.4cm,clip]{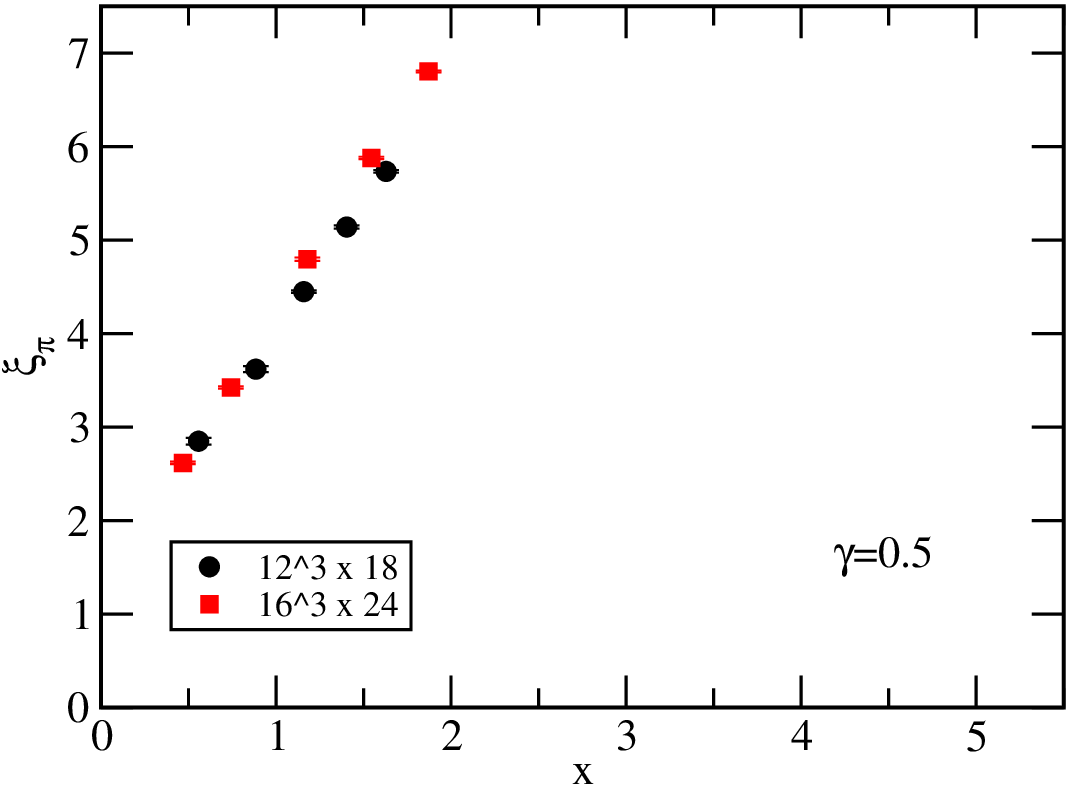}
      \includegraphics[width=5.4cm,clip]{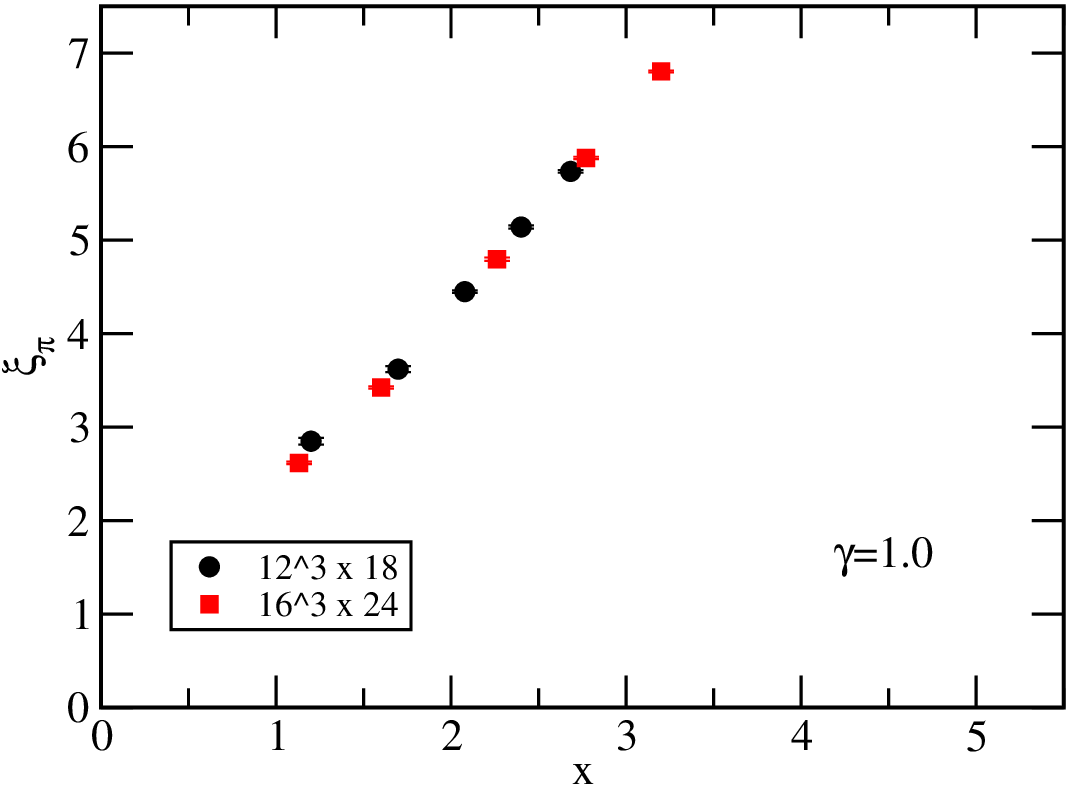}
      \includegraphics[width=5.4cm,clip]{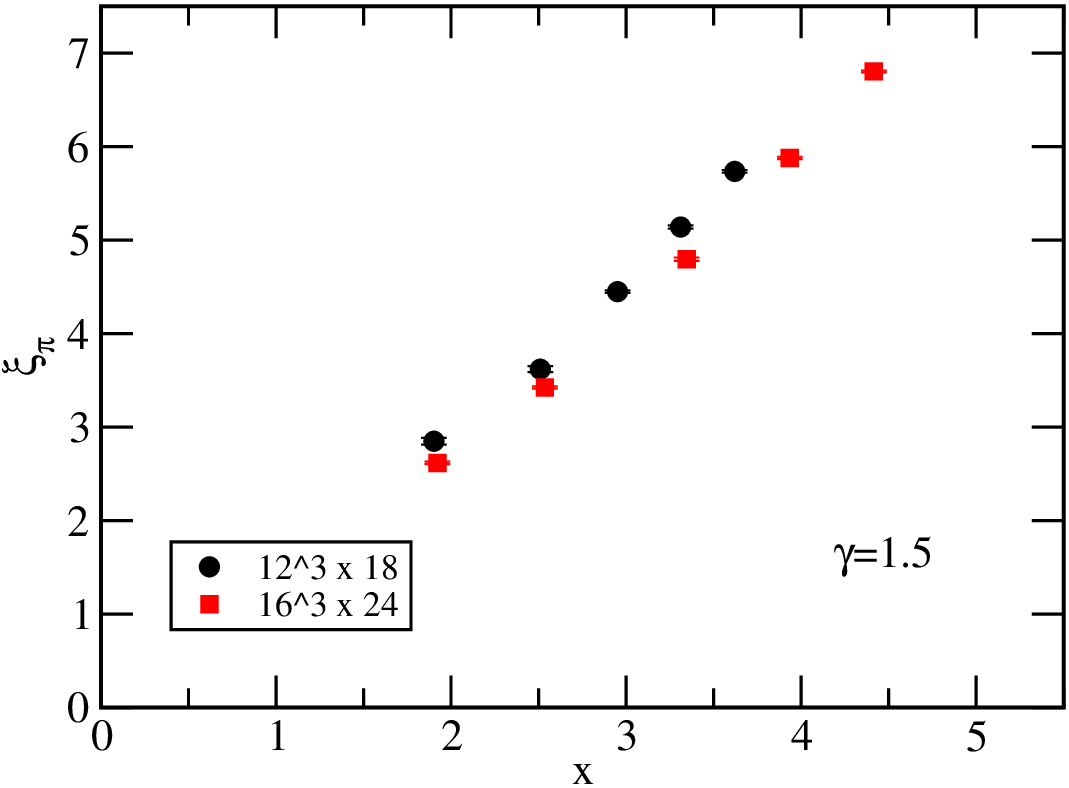}
  \end{center}
  \caption{$\xi_\pi$ is plotted as a function of the scaling variable $x$ 
    for $\gamma=0.5$, $1.0$, and $1.5$ from left to right for $N_f=4$ 
    at $\beta=3.7$, where the S$\chi$SB occurs.
    An alignment found at $\gamma=1$ is consistent 
    with Eq.~(\ref{eq:chpt_leading}). 
}
   \label{fig:nf4_mpi_g}
\end{figure}

\begin{figure}[tb]
  \begin{center}
      \includegraphics[width=5.4cm,clip]{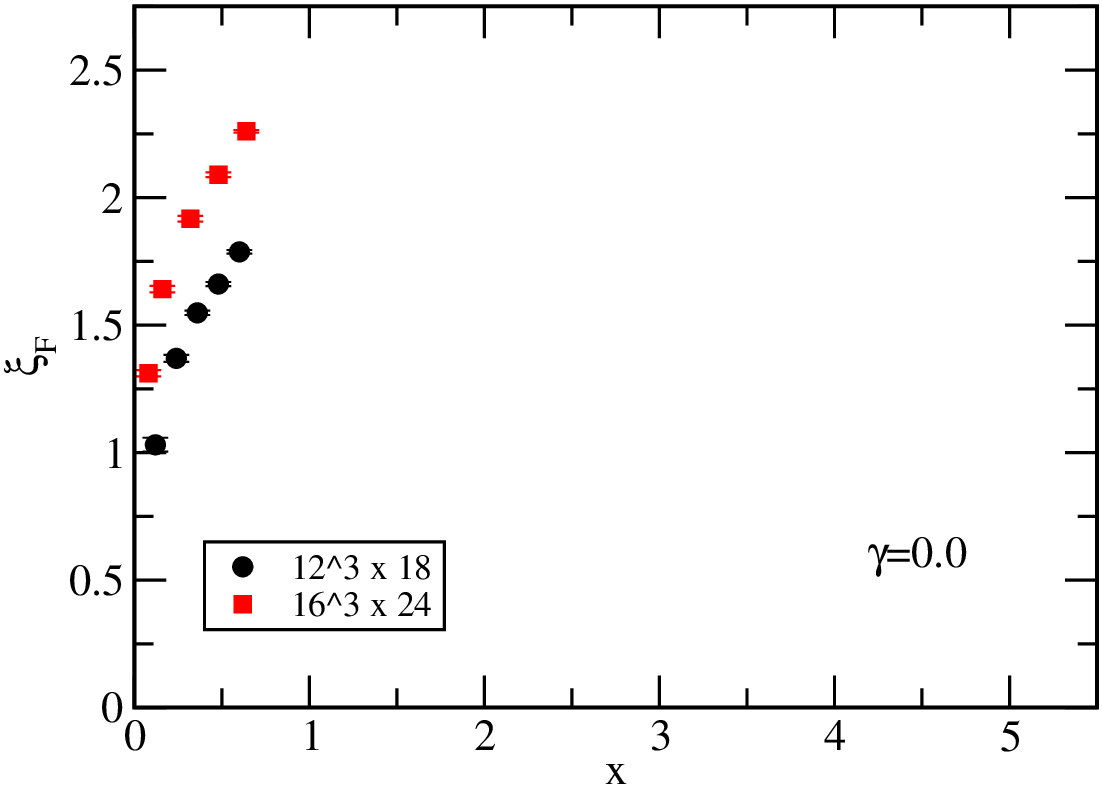}
      \includegraphics[width=5.4cm,clip]{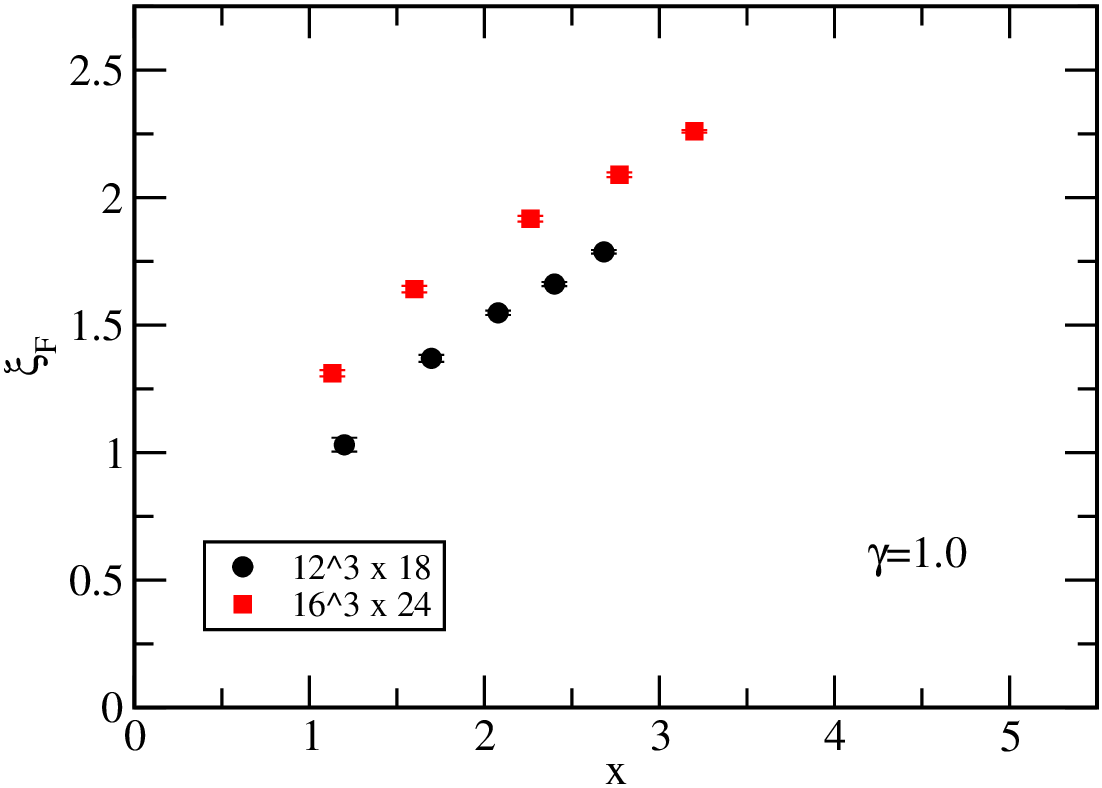}
      \includegraphics[width=5.4cm,clip]{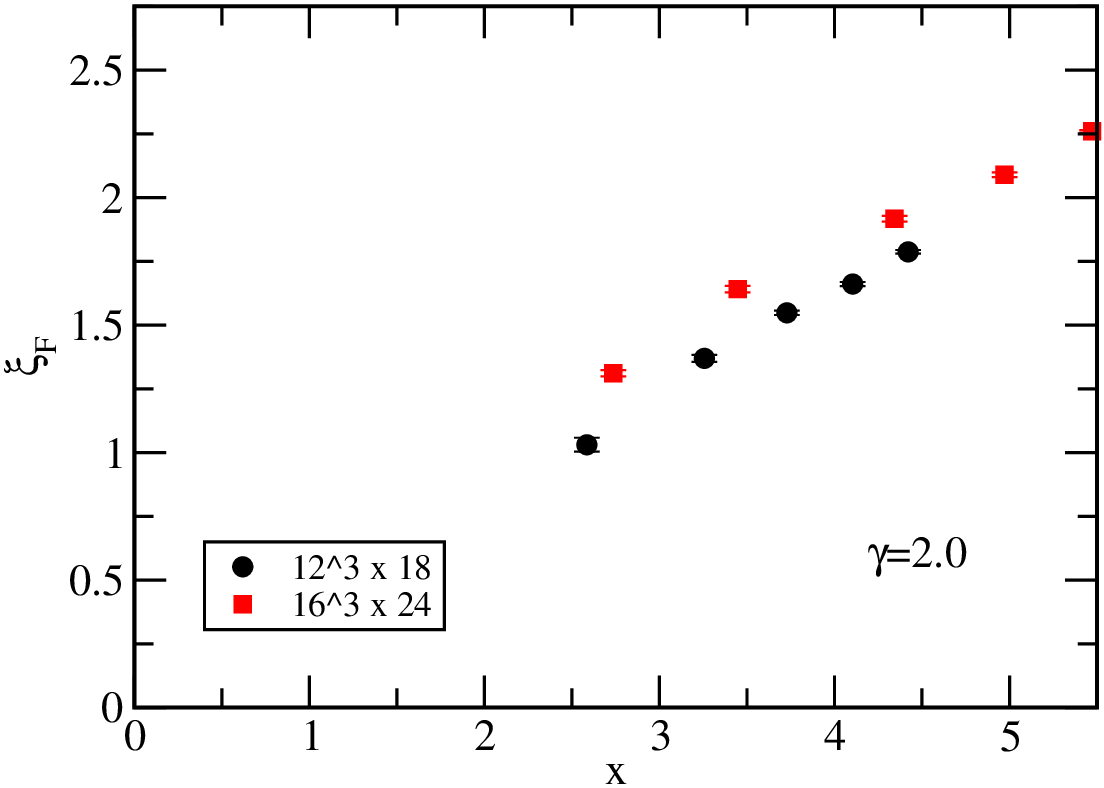}
  \end{center}
  \caption{$\xi_F$ as a function of $x$ for $\gamma=0$, $1$, and $2$ 
    from left to right for $N_f=4$ at $\beta=3.7$. 
    No alignment is seen in this case. 
}
   \label{fig:nf4_fpi_g}
\end{figure}

These results are in contrast to our results for the $N_f=4$ system 
with the same lattice action with $\beta=3.7$, 
in which the chiral symmetry is spontaneously broken. 
It is found in Fig.~\ref{fig:nf4_mpi_g} that the alignment is 
observed at $\gamma=1$, which is interpreted 
as a realization of Eq.~(\ref{eq:chpt_leading}).
The $\pi$ decay constant does not exhibit alignment at any value of 
$\gamma$ allowed for the unitarity requirement $0\le\gamma\le 2$
\cite{Mack:1975je,Flato:1983te,Dobrev:1985qv}
(Fig.~\ref{fig:nf4_fpi_g}). 

\subsection{Quantitative analysis}\label{sec:Q}

To quantify the ``alignment'' we introduce an evaluation function 
$P(\gamma)$ for an observable $p$ as follows. 
Suppose $\xi^j$ is a data point of the measured observable $p$ at 
$x_j = L_j \cdot m_j^{1/(1+\gamma)}$ and $\delta\xi^j$ is the error of $\xi^j$. 
$j$ labels distinction of parameters $L$ and $m_f$. 
Let $K$ be a subset of data points $\{(x_k, \xi_k)\}$ from which 
we construct a function $f^{(K)}(x)$ which represents the subset of data. 
Then, the evaluation function is defined as 
\begin{equation}
  \label{eq:p}
  P(\gamma) =
  \frac{1}{\mathcal{N}}
  \sum_{L}
  \sum_{j \not\in{K_L}}
  \frac{\left|\xi^{j} - f^{(K_L)}(x_j)\right|^2}
       {\left|\delta \xi^{j}\right|^2}, 
\end{equation}
where $L$ runs through all the lattice sizes we have,
the sum over $j$ is taken for a set of data points which do not 
belong to $K_L$ which includes all the data obtained on the lattice 
with size $L$.
$\mathcal{N}$ denotes the total number of summation. 
Here, we choose for the function $f^{(K_L)}$ a linear interpolation 
of the data points of the fixed lattice size $L$ for simplicity, 
which should be a good approximation of $\xi$ for large $x$. 

This evaluation function takes a smaller value when the data points are 
more closely collapsed to the line $f^{(K_L)}$ and thus provides 
a measure of the alignment. 
$P(\gamma)$ varies as the choice of parameter $\gamma$ and should show 
a minimum at a certain value of $\gamma$ when the optimal alignment of 
data is achieved. 
We take it as the optimal value of $\gamma$. 

We then estimate the uncertainty of the optimal $\gamma$ 
by properly taking account of the statistical fluctuation of 
$\xi_i$ as well as its effect to the line $f^{(K_L)}$. 
For this purpose, we employ the parametric bootstrap method,
in which the data point is simulated by a random sample 
generated by Gaussian distribution 
with the mean $\xi^j$ and the standard deviation $\delta\xi^j$. 
The distribution of $\gamma$ is thus obtained for a large number 
of these samples, from which the variance of $\gamma$ is estimated. 
The systematic error associated with the interpolation will be 
estimated by choosing different functional form with linear or 
quadratic splines as will be discussed subsequently. 

Our method for lattice study is similar to those used 
in the literature~\cite{DeGrand:2009hu,DeGrand:2011cu,Giedt:2012rj}, 
based on the study of the critical phenomena 
in a finite-volume system~\cite{0305-4470-34-33-302}~\footnote{
After submitting this paper, we were informed that a similar method
had been considered in the literature~\cite{JPSJ.62.435, PhysRevB.70.014418}.
We thank David Schaich for the information.}.
We incorporated the uncertainty of data points as the weights 
in the evaluation function so that it is normalized to one when the 
distance between the data points and the interpolated line is 
equal to the standard deviation of the data point. 
The unweighted version of the evaluation function has also been examined 
in our analysis, which resulted in values consistent with the optimal $\gamma$.

In the evaluation function Eq.~(\ref{eq:p}), the data points need to be 
taken for a range of $x = L \cdot m_f^{1/(1+\gamma)}$ 
in which there is an overlap of available data 
for all volumes, $L=18$, $24$, and $30$. 
The maximum value of $m_f$ is chosen so that $M_\pi\simle 1$ is satisfied, 
and the minimum is chosen so that the finite-volume effect on the mass of 
the bound state is not too large. 
Therefore, for all the values of $\gamma$ to be examined, 
we consider the range of $m_f$ as follows: 
The minimum $x_\text{min}$ should be such that 
$m_f=0.04$ for the largest volume $L=30$ at $\beta=3.7$, 
or $m_f=0.05$ at $\beta=4$, 
and the maximum $x_\text{max}$ should be such that 
$m_f=0.2$ for the smallest volume $L=18$ at $\beta=3.7$, 
or $m_f=0.24$ at $\beta=4$. 
Around the optimal value of $\gamma$, we have 12 data points for $\beta=3.7$ 
and 11 for $\beta=4$ within the range $[x_\text{min}, x_\text{max}]$. 
Note, however, we may need to incorporate some neighboring data 
outside this range to obtain the interpolated value $f^{(K)}(x)$ 
by the spline functions. 
\begin{figure}[tb]
\rotatebox{0}{
      \includegraphics[width=8cm,clip]{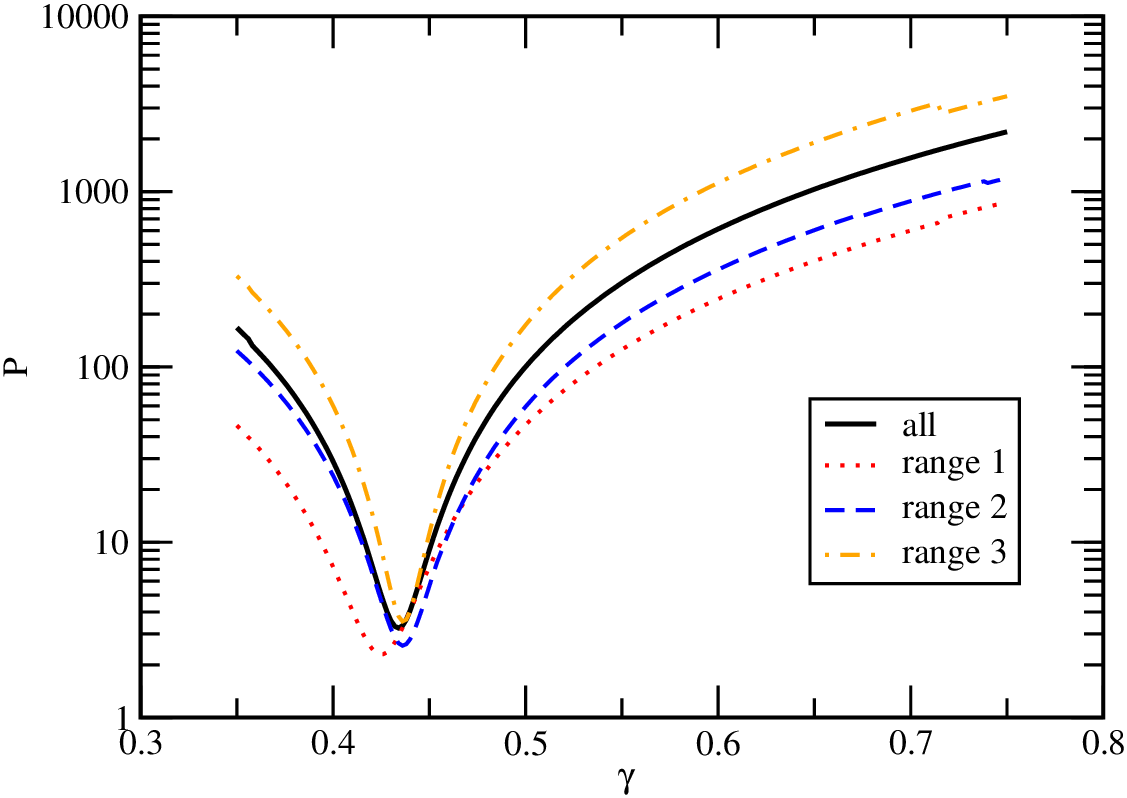} 
      \includegraphics[width=8cm,clip]{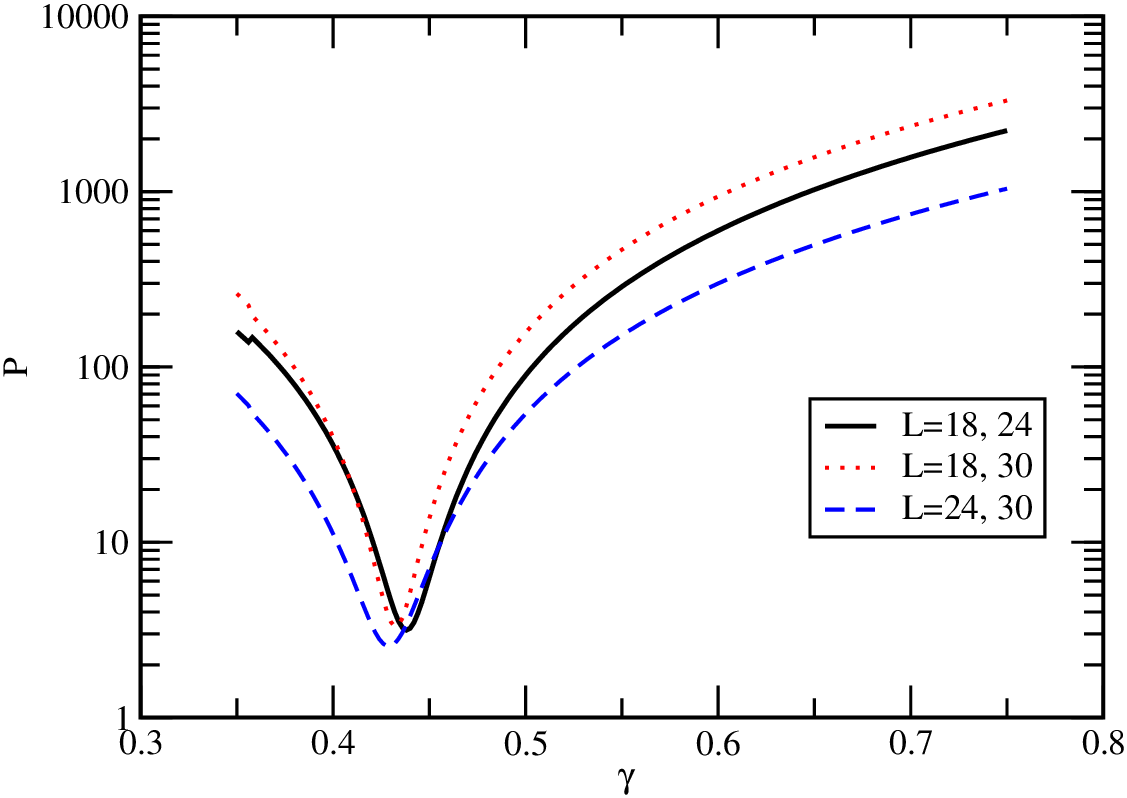} 
}
\caption{
(Left): The $\gamma$ dependence of the evaluation function $P$ 
for $M_\pi$ at $\beta=3.7$ is shown. 
The vertical axis shows the central values of $P$ as a function of $\gamma$. 
Each curve shows the results of $P(\gamma)$ with the corresponding range. 
\ 
(Right): The results of $P(\gamma)$ using data sets with two different 
volumes as $L=18, 24$, $L=18, 30$, and $L=24, 30$, respectively, 
in which the whole $x$ range is considered. 
}
\label{fig:P_mpi}
\end{figure}
 
\begin{figure}[tb]
      \includegraphics[height=3.9cm,clip]{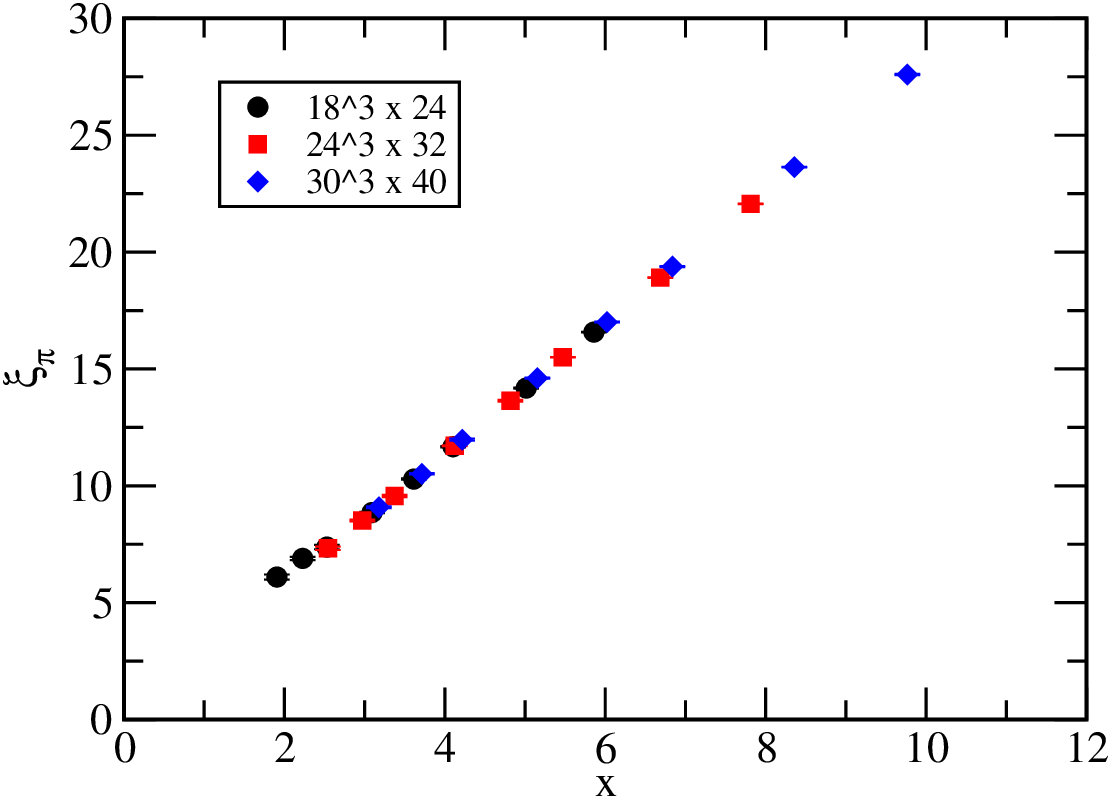}
      \includegraphics[height=3.9cm,clip]{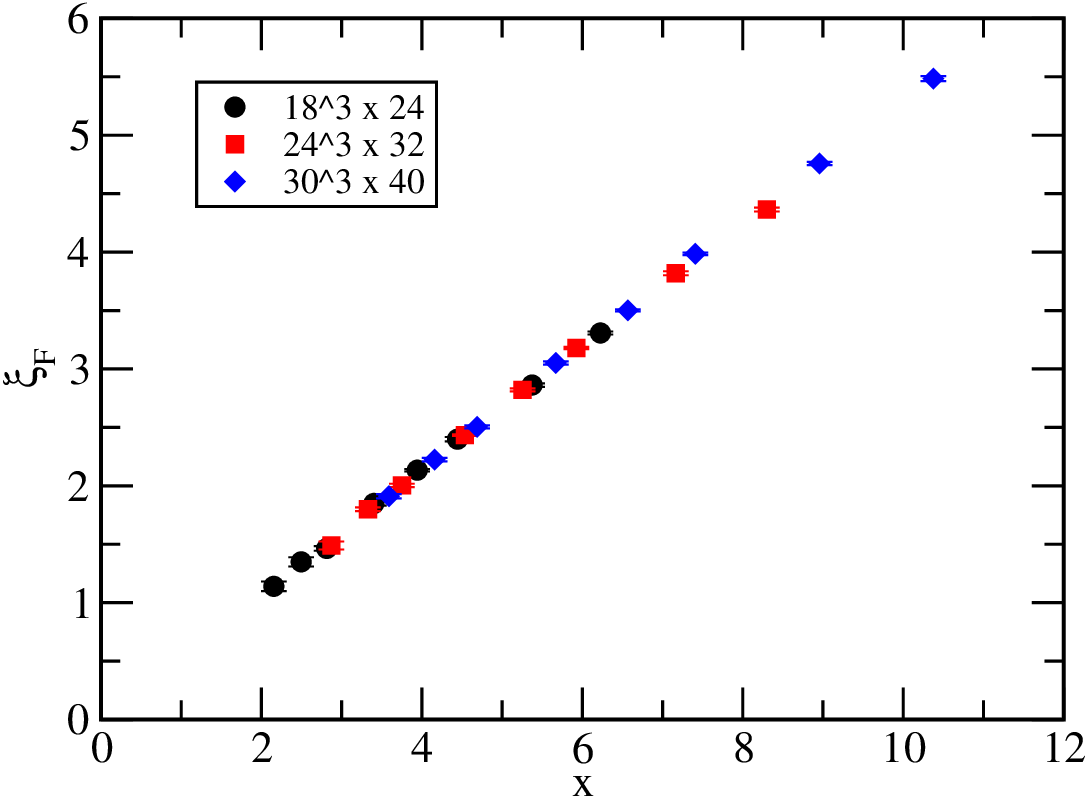} 
      \includegraphics[height=3.9cm,clip]{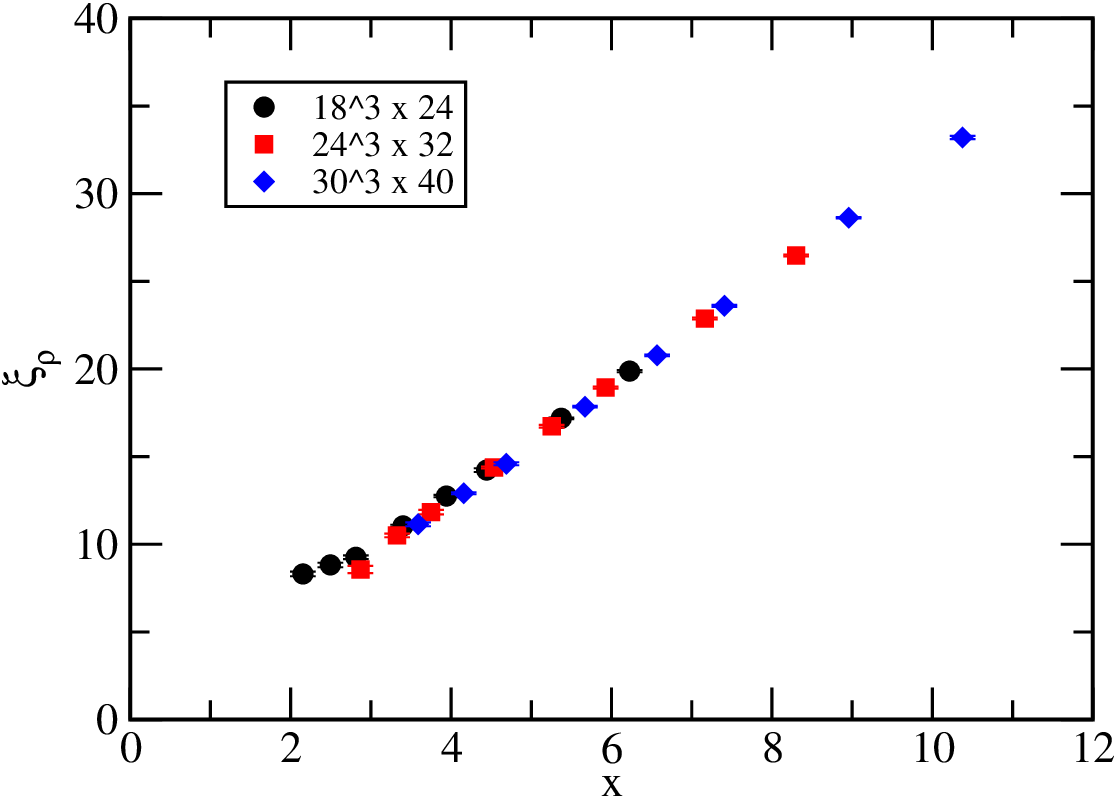} 
  \caption{
The data for $M_\pi$ (left), $F_\pi$ (center), and $M_\rho$ (right) 
are plotted as a function of $x$ at $\beta=3.7$. 
The vertical axis shows the values for $\xi$. 
The horizontal axis shows the $x=L \cdot m_f^{1/(1+\gamma)}$
where the values of $\gamma$ are $0.436$, $0.513$, and $0.476$ 
for $M_\pi$, $F_\pi$, and $M_\rho$, respectively. 
}
\label{fig:xi}
\end{figure}

\begin{figure}[tb]
      \includegraphics[height=3.9cm,clip]{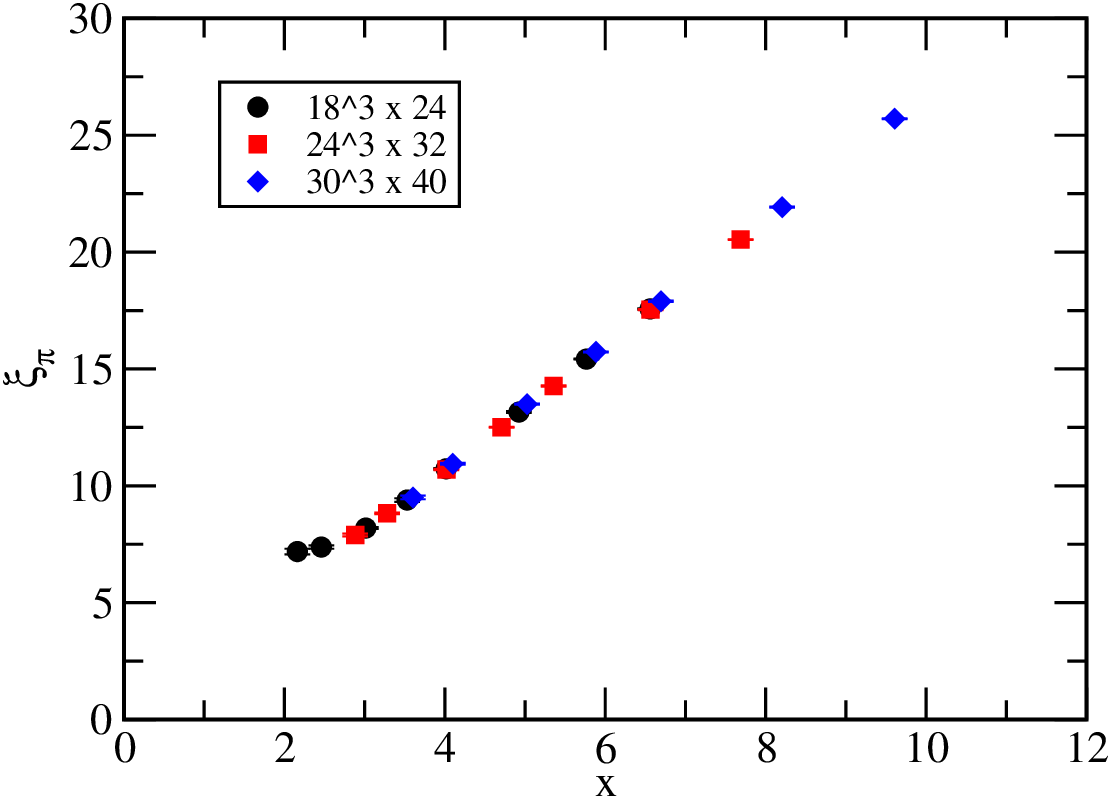}
      \includegraphics[height=3.9cm,clip]{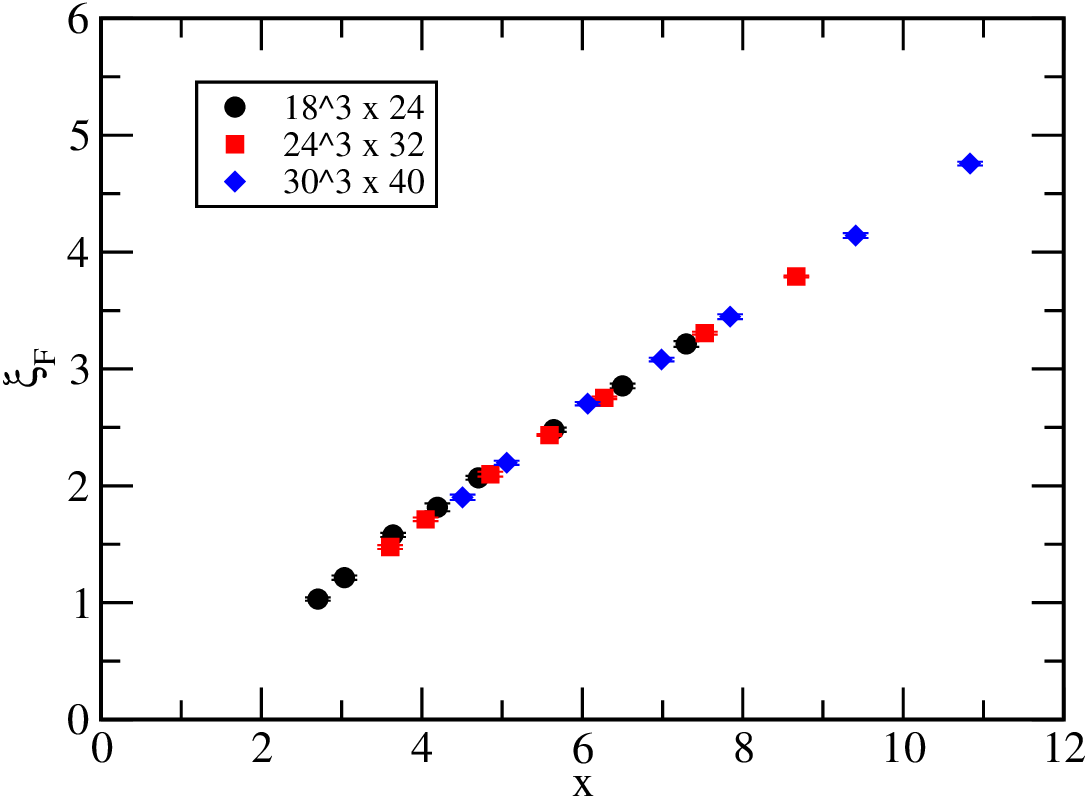} 
      \includegraphics[height=3.9cm,clip]{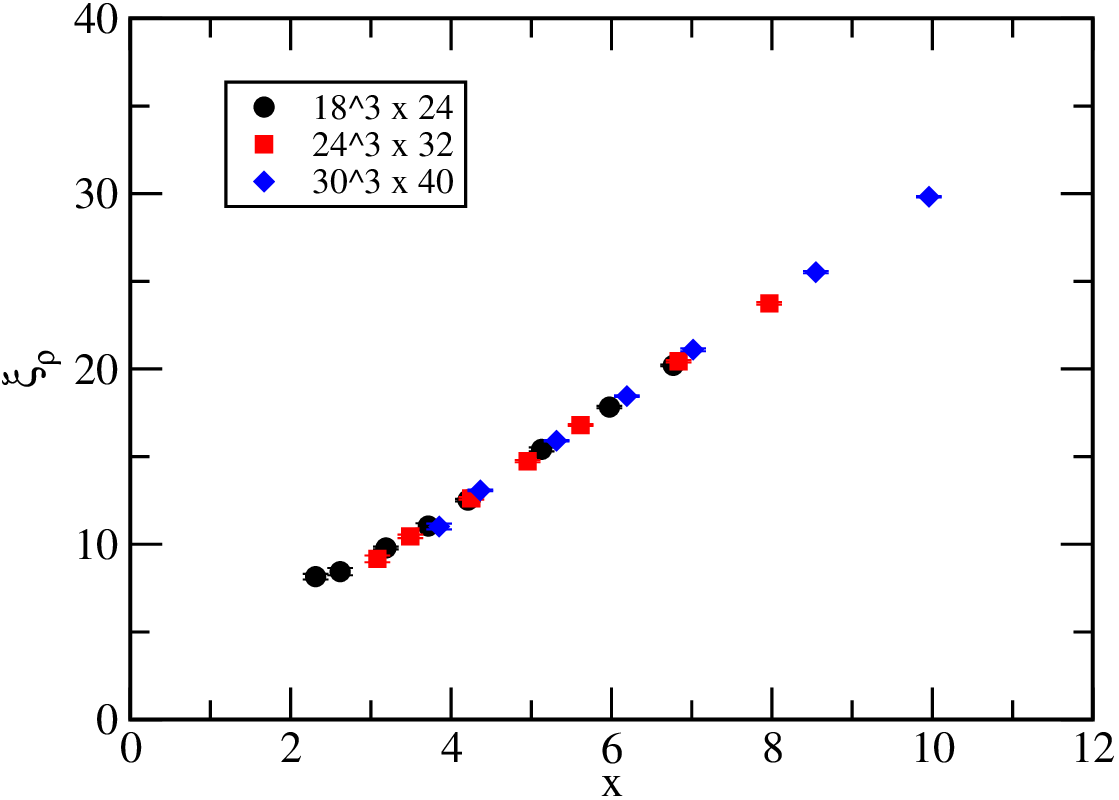} 
  \caption{
The data for $M_\pi$ (left), $F_\pi$ (center), and $M_\rho$ (right) 
are plotted as a function of $x$ at $\beta=4$. 
The vertical axis shows the value for $\xi$. 
The horizontal axis shows the $x=L \cdot m_f^{1/(1+\gamma)}$ 
where the values of $\gamma$ are $0.408$, $0.560$, and $0.479$ 
for $M_\pi$, $F_\pi$, and $M_\rho$, respectively. 
}
\label{fig:xi_b4}
\end{figure}

The line marked by ``all'' in Fig.~\ref{fig:P_mpi} plots the values of 
the evaluation function using all data in $x_{min} \le x \le x_{max}$ 
for $\beta=3.7$. 
A clear minimum is observed, at which the optimal alignment of the data 
is achieved. 
We repeat this analysis for each observable and at each value of $\beta$. 
The results are tabulated in the column labeled by ``all'' 
of Table~\ref{tab:gamma_summary}. 
Figures \ref{fig:xi} and~\ref{fig:xi_b4} plot the values of $\xi$ for 
each case as a function of $x$ with the optimal $\gamma$. 

Let us now recall that from the naive analysis of the ratio 
in Sec.~\ref{sec:primary_results}, the scaling region may be 
restricted to a range of smaller masses. 
To capture such an effect, we need to study systematically 
how the range of fermion mass affects the evaluation function 
and the resulting optimal $\gamma$. 
In this regard, we define a window of the parameters $x$ and $L$ 
to which the data are restricted in evaluating the evaluation function 
and see if the results change against the choice of the window. 

We consider a window in the $x$ direction which has a span $\Delta x/2$, 
where $\Delta x = x_\text{max} - x_\text{min}$ and slide it with an 
interval ${\Delta x}/4$. 
There are three choices of windows denoted by range 1, 2, and 3, which are 
$[x_\text{min}, x_\text{min}+{\Delta x}/2]$, 
$[x_\text{min}+{\Delta x}/4, x_\text{max}-{\Delta x}/4]$,
and $[x_\text{max}-{\Delta x}/2, x_\text{max}]$, respectively. 
In these windows, the data are taken for all three volumes. 
For the $L$ direction we have three windows: The first window consists of 
data for $L=18$ and $24$, the second window for $L=18$ and $30$, 
and the third window for $L=24$ and $30$. 
In these windows in $L$ directions, the whole $x$ range 
$[x_\text{min}, x_\text{max}]$ is considered. 

\begin{table}[tb]
\caption{Summary of the optimal values of $\gamma$. See the text for details.}
\label{tab:gamma_summary}
\begin{ruledtabular}
\begin{tabular}{c l l lll lll}
         &
         &
         &
\multicolumn{3}{c}{$x$} &  
\multicolumn{3}{c}{$L$} \\
\cline{4-6}\cline{7-9}
\multicolumn{1}{c}{quantity}  &
\multicolumn{1}{c}{$\beta$}   &
\multicolumn{1}{c}{all}       & 
\multicolumn{1}{c}{range 1}  &
\multicolumn{1}{c}{range 2}  &
\multicolumn{1}{c}{range 3}  &
\multicolumn{1}{c}{(18,24)}   &
\multicolumn{1}{c}{(18,30)}   &
\multicolumn{1}{c}{(24,30)}   \\  
\hline
$M_\pi$  & 3.7 & 0.434(4)  & 0.425(9)  & 0.436(6)   & 0.437(4) &
                        0.438(6)  & 0.433(4)   & 0.429(8) \\
$M_\pi$  & 4   & 0.414(5)  & 0.420(7)  & 0.418(6)   & 0.411(5) &
                        0.397(7)  & 0.414(4)   & 0.447(9) \\
$F_\pi$  & 3.7 & 0.516(12) & 0.481(19) & 0.512(19)  & 0.544(14) &
                        0.526(18)  & 0.514(11)  & 0.505(24) \\
$F_\pi$  & 4   & 0.580(15) & 0.552(21) & 0.602(20)  & 0.605(19) &
                        0.544(27) & 0.577(14)  & 0.645(32) \\
$M_\rho$ & 3.7 & 0.459(8)  & 0.411(17) & 0.461(10)  & 0.473(8) &
                        0.491(15) & 0.457(8)   & 0.414(18) \\
$M_\rho$ & 4   & 0.460(9)  & 0.458(13) & 0.455(14)   & 0.460(8) &
                        0.457(16) & 0.459(8)   & 0.463(15)  \\
\end{tabular}
\end{ruledtabular}
\end{table}

\begin{figure}[tb]
 \includegraphics[width=8.cm,clip]{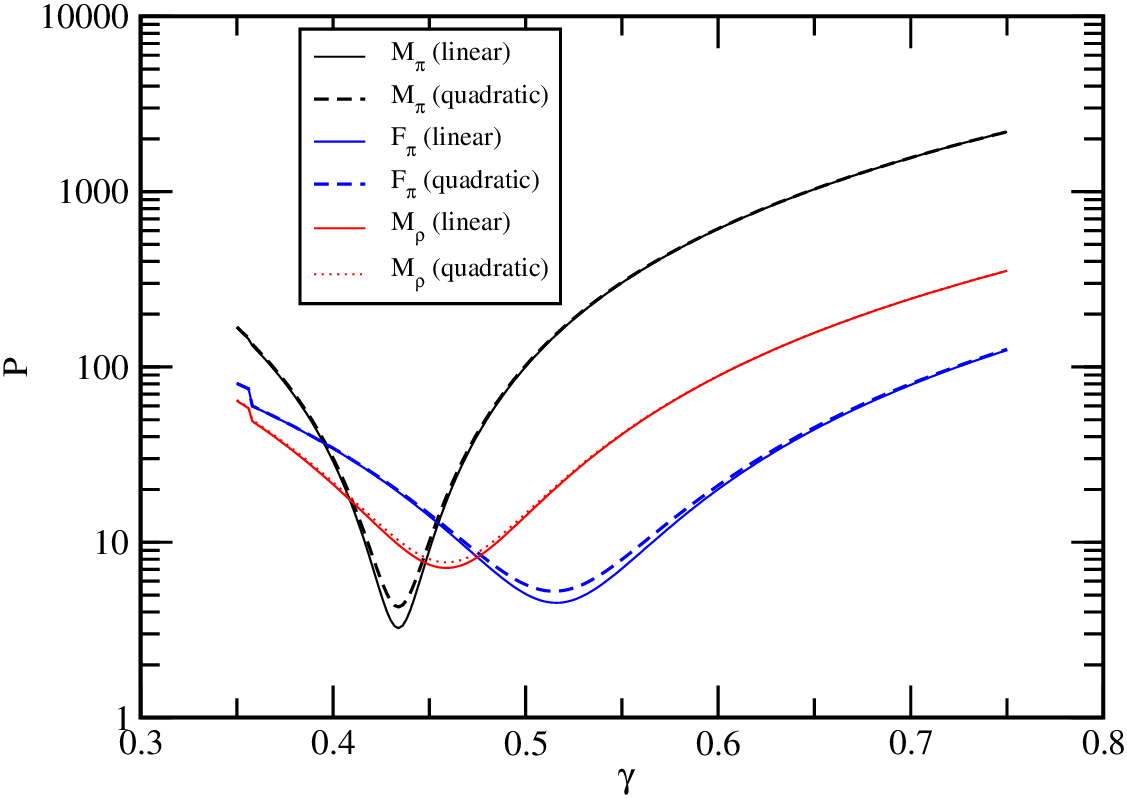} 
\caption{
The $\gamma$ dependence of the evaluation function $P$ 
for $M_\pi$, $F_\pi$, and $M_\rho$ at $\beta=3.7$ is plotted. 
The vertical axis shows the values of $P$ at each of $\gamma$ where 
the three volumes and full range of $x$ for the data are considered. 
The solid and dashed curves show the results of $P(\gamma)$ 
with the interpolation functions $f(x)$ 
by the linear and quadratic functions, respectively. 
}
\label{fig:P_io}
\end{figure}

The evaluation function $P(\gamma)$ of the $\pi$ mass for each window 
is also plotted in Fig.~\ref{fig:P_mpi}. 
It is noted that the value of $P(\gamma)$ is $\mathcal{O}(1)$ at the 
minimum. 
The optimal values of $\gamma$ which minimize $P(\gamma)$ are summarized 
in Table~\ref{tab:gamma_summary}, where the numerals in the parentheses 
include the statistical and systematic errors. 
The systematic error due to the ambiguity of the interpolation is 
estimated by the difference of the optimal $\gamma$'s obtained 
with linear and quadratic spline interpolations. 
The comparison of these $P(\gamma)$ is shown in Fig.~\ref{fig:P_io}. 
The minima for the quadratic spline interpolation appear approximately 
at the same place as those for the linear one. 
The systematic error is thus small, and it is found to be always 
smaller than the statistical error in the analysis with the whole data points. 

Let us first look at the case of $M_\pi$ at $\beta=3.7$ in 
Table~\ref{tab:gamma_summary}. 
We do not observe the window dependence beyond the error bars, and 
thus $\xi_\pi$ is well-described by a function of a single variable $x$. 
If there is IR conformality, 
the nonuniversal correction to the hyperscaling is negligible 
at this precision. 
Then, from the fact that in Sec.~\ref{sec:primary_results}, 
the scaling is observed in the small mass range 
for the ratio $M_\rho/M_\pi$ at $\beta=3.7$, 
it is suggested that there should be certain window dependence 
of $\gamma$ from $M_\rho$. 
As 
$\gamma(M_\pi)=0.434(4)$ and 
$\gamma(F_\pi)=0.459(8)$ at $\beta=3.7$, 
if one restricts the mass range for $M_\rho$ to the smaller side, 
then the value of $\gamma(M_\rho)$ should get closer to that of
$\gamma(M_\pi)$.  
This is actually the case, as is observed from Table~\ref{tab:gamma_summary} 
in which $\gamma(M_\rho)$ reduces for smaller mass range (toward range 1) 
and larger volume (toward $L=24$, $30$). 

However, such trends are not observed for the $M_\rho$ at $\beta=4$, where one
expects the similar $x$ and $L$ range dependence. As the number of samples
have gotten reduced for the fixed range analysis, a statistical instability
might have spoiled the result.

The similar trend for the $x$-range dependence as for $M_\rho$ at
$\beta=3.7$ is observed for $F_\pi$ at $\beta=3.7$, too.
The direction of the movement is correct, but it does not get close
enough to the value of $\gamma(M_\pi)$.
Moreover, the $L$ range dependence is too weak to conclude that it will
get close to $\gamma(M_\pi)$.
These results may be understood from the fact that in
Sec.~\ref{sec:primary_results},  
the scaling is observed only in the very small mass range. 
For $F_\pi$ at $\beta=4$, the $L$ dependence appears to be opposite to the
expectation, which can be understood as the result of unobserved scaling
in the analysis in Sec.~\ref{sec:primary_results}.

As we cannot completely resolve these trends in the mass dependence, 
we regard these variations of $\gamma$ with respect to the change of 
the window as the systematic error on the central value of $\gamma$ 
obtained with ``all'' data. 
We put the asymmetric error for both $x$ and $L$ directions separately 
estimated by the maximum variations from the central value.
The results read 
\begin{equation} \label{eq:g}
\gamma = \left\{
\begin{array}{l}
 \left.
 \begin{array}{ll}
 0.434(4)(^{+5}_{-5})(^{+3}_{-10})   \qquad\quad & \text{for $M_\pi$}  \\
 0.516(12)(^{+10}_{-11})(^{+28}_{-35})  & \text{for $F_\pi$}  \\
 0.459(8)(^{+33}_{-45})(^{+15}_{-48})   & \text{for $M_\rho$}  \\
 \end{array}
 \right\} \text{for $\beta=3.7$}, \\ 
 \left.
 \begin{array}{ll}
 0.414(6)(^{+34}_{-17})(^{+7}_{-3})  \qquad\quad & \text{for $M_\pi$}  \\
 0.580(15)(^{+65}_{-37})(^{+26}_{-29})  & \text{for $F_\pi$}  \\
 0.460(8)(^{+4}_{-3})(^{+0}_{-5})       & \text{for $M_\rho$} \\
 \end{array}
 \right\} \text{for $\beta=4$},
\end{array}
\right.
\end{equation}
where the numerals in parentheses are 
the combined errors of the statistics and the interpolation, 
the systematic uncertainties due to the dependence on the volume, 
and the systematic errors due to the dependence on the $x$ range, respectively.  
\begin{figure}[tb]
  \begin{center}
   \includegraphics[width=8cm,clip]{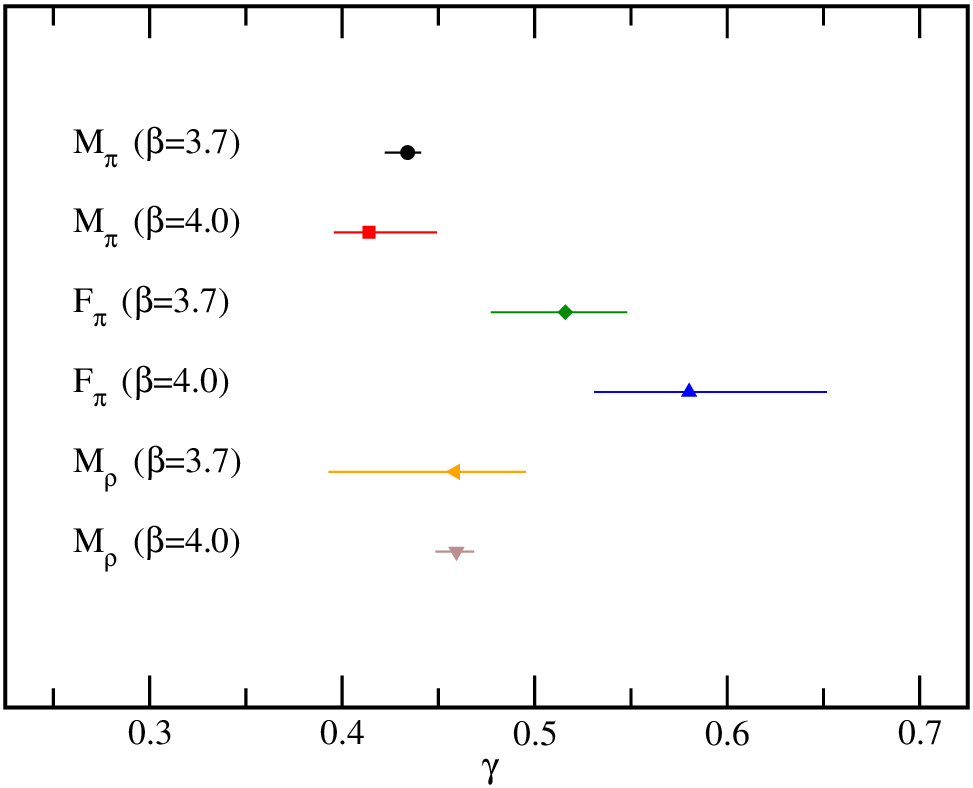}
  \end{center}
  \caption{
The results of the values of $\gamma$ for three observables at 
two $\beta$ are summarized, 
where the statistical and systematic errors are added in quadrature. 
All the results are consistent with each other within $2\sigma$ level, 
except for the $\gamma$ from $F_\pi$ at $\beta=4$, where 
the scaling region is suspected to be outside of the parameter range we have. 
See the text.
}
   \label{fig:gamma}
\end{figure}
The results with all the errors added in quadrature are summarized
in Fig.~\ref{fig:gamma}. 
All the results are consistent with each other within $2 \sigma$ level, 
except for $\gamma$ from $F_\pi$ at $\beta=4$ 
for which the scaling region was suspected to be outside of the 
parameter range we have examined. (See Sec.~\ref{sec:primary_results}.) 

From these analyses, we conclude that our data for the $N_f=12$ theory 
by the lattice simulations are reasonably consistent with the FSHS 
and, thus, with the IR conformality, 
if we exclude $F_\pi$ at $\beta=4$ from the analysis. 
The resulting mass anomalous dimensions from different quantities 
and two lattice spacings 
are also reasonably consistent. 
We quote $0.4\simle \gamma_* \simle 0.5$ for the value of 
the mass anomalous dimension at the IR fixed point.

%% file: text_sections/fshp.tex
%
We found in the previous sections that 
our data is consistent with the FSHS once
the systematic effect due to the limited parameter range is taken into
account. The analyses also suggested that 
the nonuniversal correction to the hyperscaling could be important.
In this section, we try to test two plausible models for the correction.  
To do this test, we need to fix the term for the universal scaling in the
following. Therefore, the approach loses generality that the analysis
in the previous section had. Thus, the result here is not going to be
the main result in this paper, but it still provides useful information.

For this purpose, we use the following formulae for the fit:
\begin{eqnarray}  \label{eq:f} 
\xi &=& c_0 + c_1 L m_f^{1/(1+\gamma)} \ \cdots \textrm{fit A}, \\
\xi &=& c_0 + c_1 L m_f^{1/(1+\gamma)} 
+ c_2 L m_f^{\alpha} \ \cdots \textrm{fit B}   .   
\end{eqnarray}
The fit A uses a naive fit form based on the hyperscaling
relation which is described by the function
form of $f(x)=c_0 +c_1x$ with $x=L m_f^{1/(1+\gamma)}$. 
This formula is motivated from the results obtained in Fig.~\ref{fig:xi}, 
since the clear linearity of the data for large $x$ can be found near the
optimal value of $\gamma$.
The second formula for fit B includes a mass correction term.

As discussed in the previous sections, 
there may exist some corrections beyond the hyperscaling relations 
in the region we simulated, so we try to include such contributions.
In general, finite-volume corrections exist, and the form
is given {\it \`a la} Fisher \cite{Fisher:1972zza,Fodor:2012uu}.
We do not take into account these effects, however, as the analysis
in the previous sections indicates the mass correction is more important
for the parameter space we simulated.

Among various possible choices, we take
$\alpha=(3-2\gamma)/(1+\gamma)$, which is inspired by the analytic expression 
of the solution of the SD equation given in Ref.~\cite{Aoki:2012ve}.
We also consider $\alpha=2$ case,
which is regarded as the small mass correction caused by
the explicit chiral-symmetry-breaking effects or due to the
lattice discretization artifact.  
It is noted that, in both cases, the fit function cannot be described
by a single scaling variable $x=Lm_f^{1/(1+\gamma)}$. 
We denote these fit functions with $\alpha=(3-2\gamma)/(1+\gamma)$ and 
$\alpha=2$ as fit B1 and fit B2, respectively.
All the details of the fit results are shown in
Appendix \ref{sec:conformal_fit}, which also includes other fits,
for example, the case of $\alpha=1$ \cite{Appelquist:2011dp}.
We only give a digest of the results.

\begin{figure}[htbp]
\rotatebox{0}{
            \includegraphics[height=4.0cm,clip]{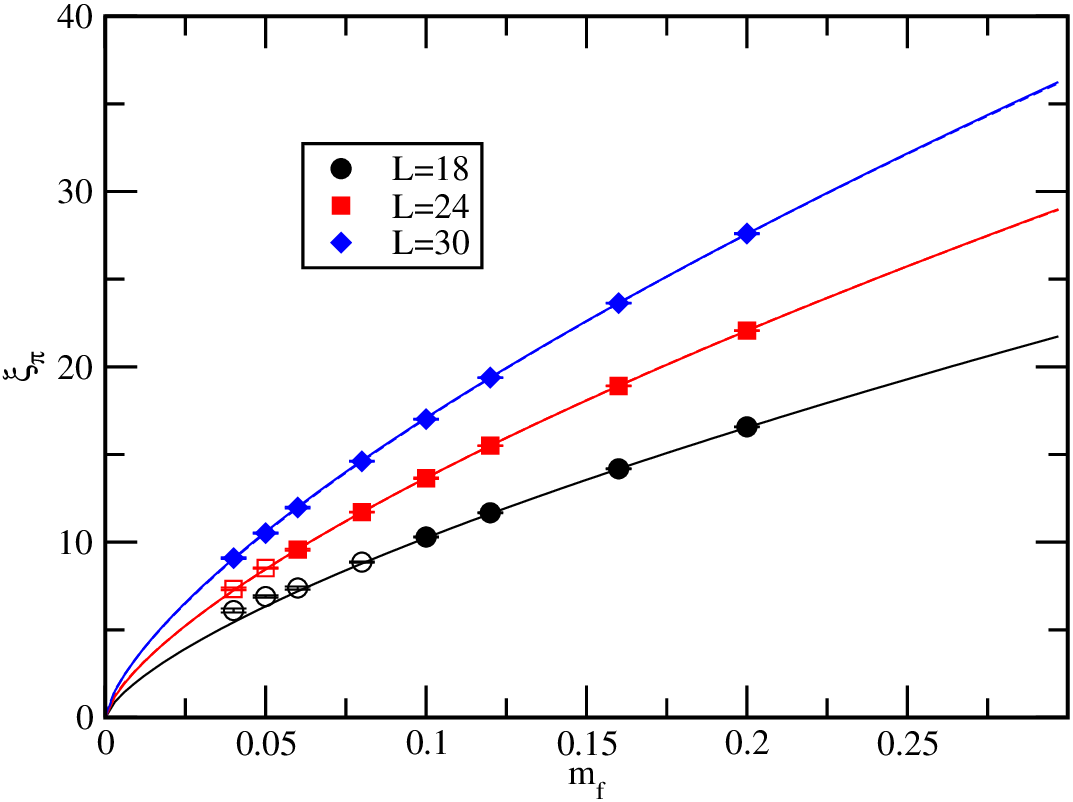} 
            \includegraphics[height=4.0cm,clip]{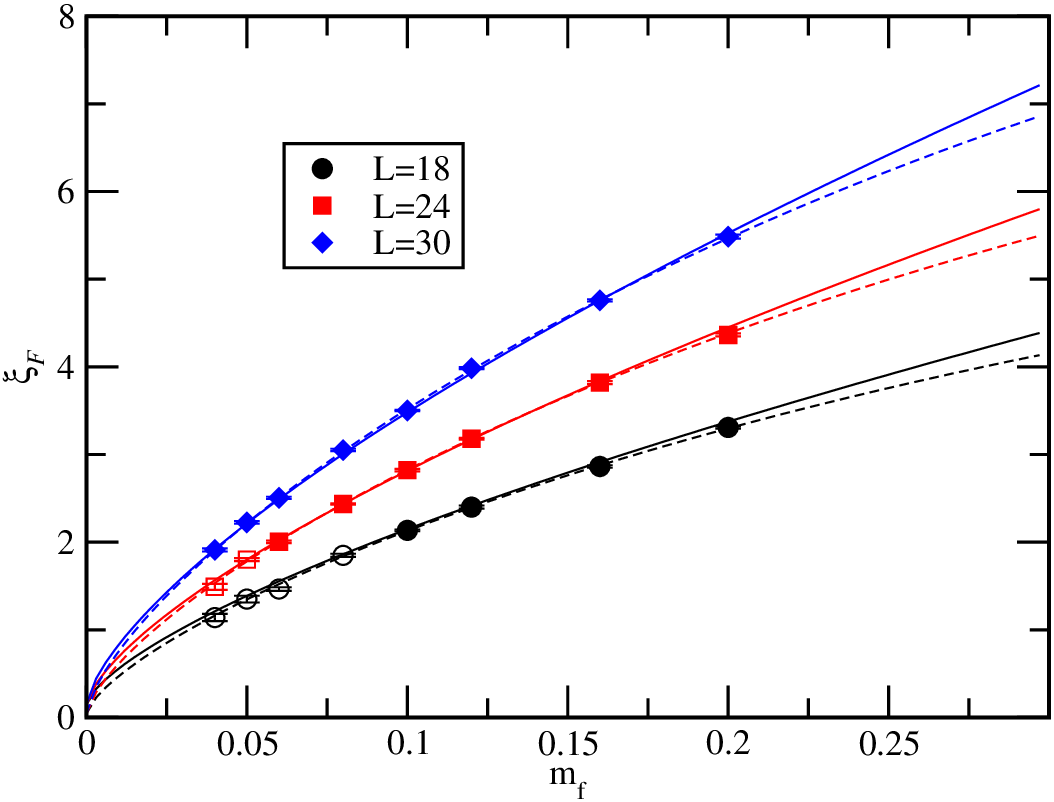} 
            \includegraphics[height=4.0cm,clip]{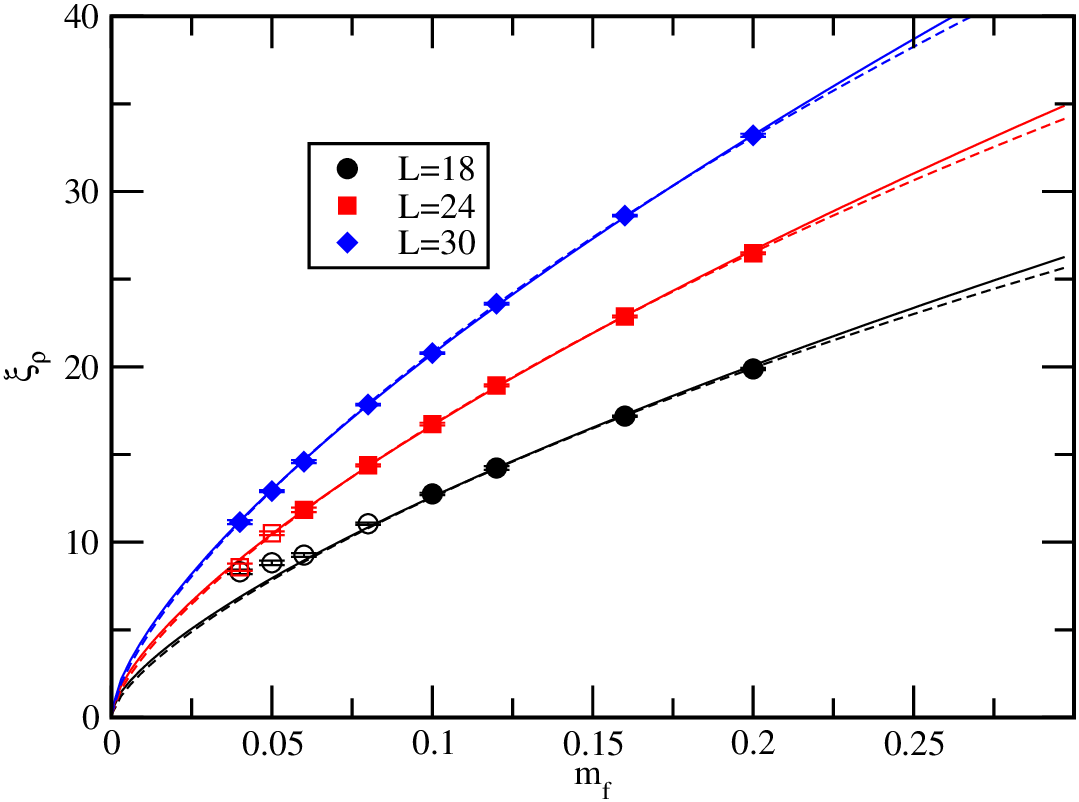} 
}
\caption{
Spectra $\xi_\pi$(left), $\xi_F$(center), and 
$\xi_\rho$(right) as a function of $m_f$ at $\beta=3.7$.
For simplicity, we only show two fit results of fit A and fit B2, 
by the solid and dotted curves, respectively.
The data with empty symbols are not used in the fit.
}
\label{fig:mpiL}
\end{figure}

\begin{figure}[htbp]
\rotatebox{0}{
            \includegraphics[height=4.0cm,clip]{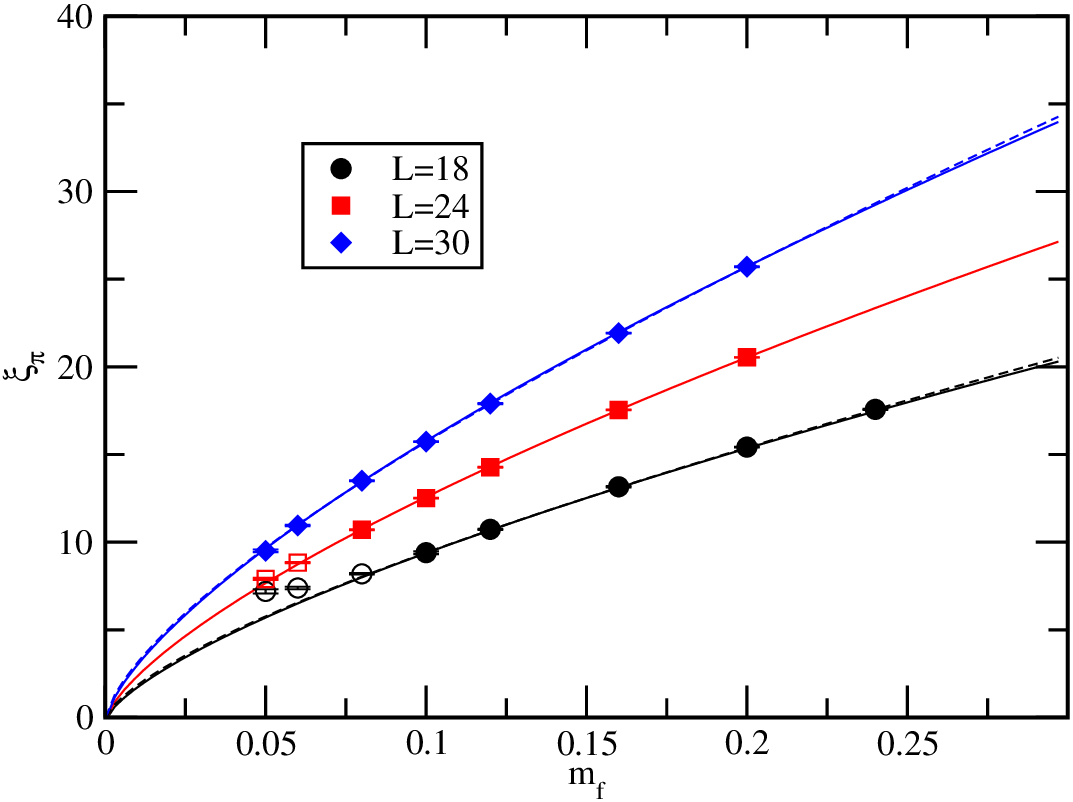} 
            \includegraphics[height=4.0cm,clip]{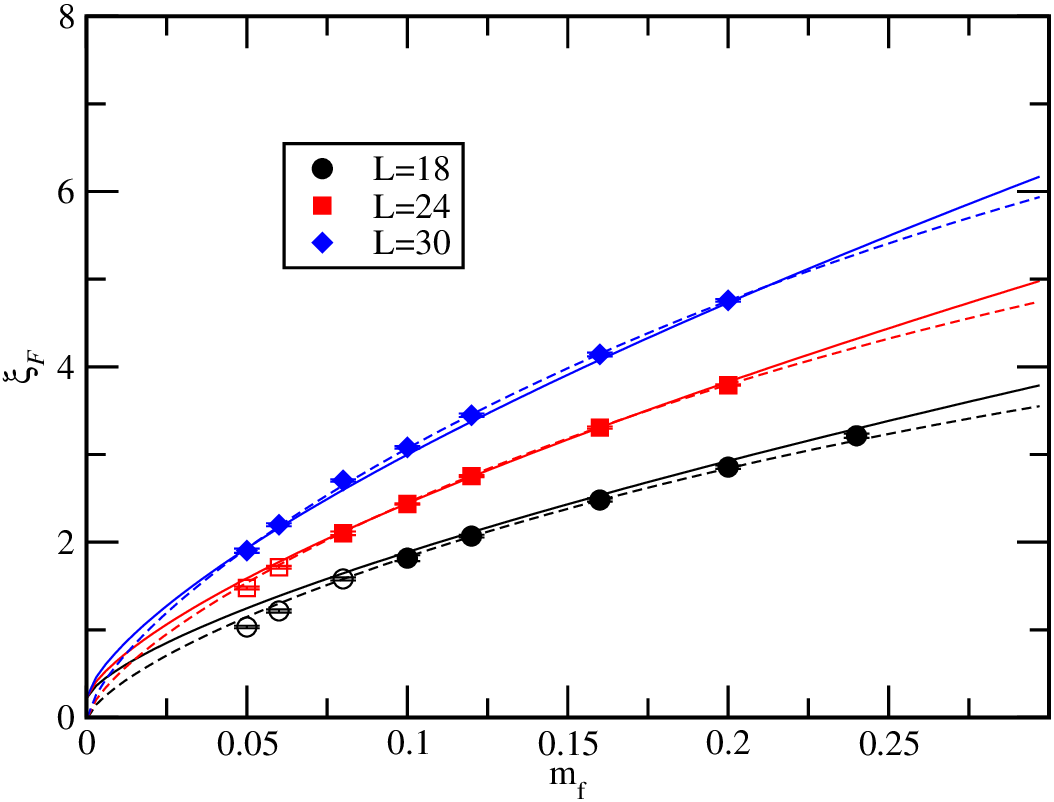} 
            \includegraphics[height=4.0cm,clip]{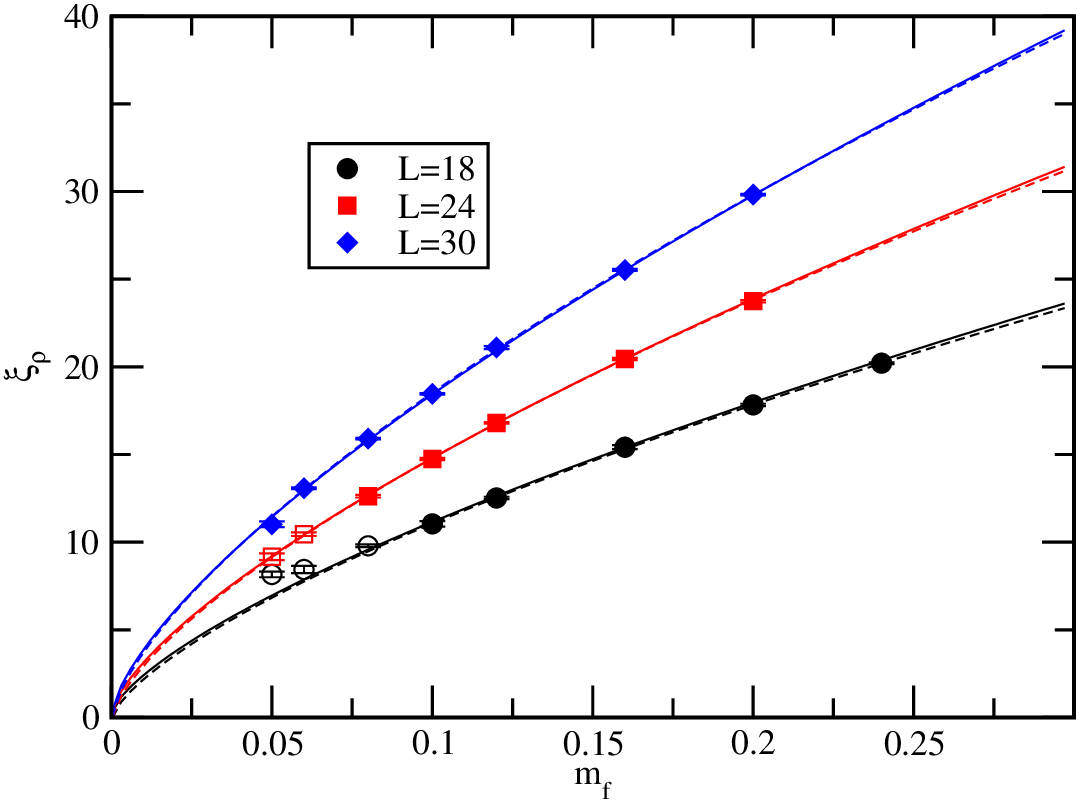} 
}
\caption{
Spectra $\xi_\pi$(left), $\xi_F$(center), and 
$\xi_\rho$(right) as a function of $m_f$ at $\beta=4$.
For simplicity, we only show two fit results of fit A and fit B2, 
by the solid and dotted curves, respectively.
The data with empty symbols are not used in the fit.
}
   \label{fig:mpiL_b4}
\end{figure}

We perform simultaneous fits using $M_\pi$, $M_\rho$ and $F_\pi$,
with common $\gamma$ for each fit ansatz: fit A, B1, B2, and for each
$\beta$ separately. 
In these fit analyses, we assume for simplicity that 
possible correlation between the data of $\xi_\pi$, $\xi_F$ and $\xi_\rho$ 
is absent.
Note that, here, we include the $F_\pi$ data at $\beta=4$, which were excluded
in the analysis in the previous section due to the absence of manifest
hyperscaling (Fig.~\ref{fig:ratio_mpi_nf12}). This is because the mass
correction in the hyperscaling analysis, which is included here but not
there, could amend the possible disturbance to the hyperscaling from
the mass correction.
To avoid the large finite-size effects, the data used in the fits
are restricted to those which satisfy $\xi_\pi=LM_\pi\ge 9$
determined by trial and error.
Figures \ref{fig:mpiL} and \ref{fig:mpiL_b4} show the fit B2
as examples. Table \ref{tab:fsc} shows the resulting fitting parameters.

The $\chi^2/{\rm dof}$ is large $\simge 5$ for the fit A, while
it is reasonable ($\chi^2/{\rm dof}\le 2$) for the fit B1 and B2.
This implies that the correction is necessary, and both types of mass
correction work.
Let us look into the fit B2 at $\beta=3.7$, for example.
The ratio of the terms proportional to $c_1$ (leading scaling) and $c_2$ 
(mass correction) at the heaviest point ($m_f=0.2$) is $-0.008(4)$ ($M_\pi$),
$-0.034(6)$ ($M_\rho$) or $-0.067(7)$ ($F_\pi$). 
The small correction for $M_\pi$ is consistent with the fact that
the value of $\gamma(M_\pi)$ was stable against the $x$ and $L$ range
used in the analysis in Sec.~\ref{sec:fss}.
Furthermore, the order of the size of the correction for $M_\pi$,
$M_\rho$, and $F_\pi$ is consistent with what we observed in 
Secs.~\ref{sec:primary_results} and \ref{sec:fss}.

We obtain consistent $\gamma$ for all the fit B1 and B2.
The resulting values are in the range $\gamma\sim 0.41-0.46$,
which is consistent with the main result 
presented in Sec.~\ref{sec:fss}: $0.4\simle \gamma_*\simle 0.5$.

\begin{table}[tbp]
\caption{The fit results of finite-size conformal hypothesis at $\beta=3.7$(left) and
$\beta=4$(right).
The values in the brackets mean the input in the fit.
} 
\label{tab:fsc}
\begin{tabular}{cc}
\begin{minipage}{0.5\hsize}
\begin{ruledtabular}
\begin{tabular}{ccccc}
& $\gamma$ & $\alpha$ &  $\chi^2/$dof & dof \\ \hline 
fit A    & 0.449(3) & -  &  4.52 & 47 \\
fit B1 &  0.411(9) & $\frac{(3-2\gamma)}{(1+\gamma)}$ & 1.23 & 44\\ 
fit B2 & 0.423(7) &  [2] &  1.15 & 44
\end{tabular} 
\end{ruledtabular}
\end{minipage}
\ 
\begin{minipage}{0.5\hsize}
\begin{ruledtabular}
\begin{tabular}{ccccc}
& $\gamma$ & $\alpha$ &  $\chi^2/$dof & dof \\ \hline 
fit A    & 0.430(3) & -  &  6.78 & 44 \\
fit B1 &  0.461(18) & $\frac{(3-2\gamma)}{(1+\gamma)}$ & 1.86 & 41\\ 
fit B2 & 0.453(11) &  [2] &  2.00 & 41\\ 
\end{tabular}
\end{ruledtabular}
\end{minipage}
\end{tabular}
\end{table}

%% file: text_sections/chpt_summary.tex
%

A counter scenario to the conformal hyperscaling is the S$\chi$SB.
Here, we give a digest of the
analysis, the detail of which is described in Appendix \ref{sec:chpt}.

As the current data set of $\beta=3.7$ simulation has physically lighter
quark mass, the property of the chiral limit is captured better than 
those at $\beta=4$. Therefore, we focus on the former.
First, we take the infinite volume limit of $M_\pi$ and $F_\pi$ using a
formula inspired 
by ChPT or finite-size effect of the multiparticle state 
{\it \`a la}
L\"uscher. Then, the $\pi$ decay constant and mass squared are fit with
polynomial functions in $m_f$.
Logarithmic mass dependence of these quantities would emerge
at the chiral regime and play an important role there. As it turns out, 
however, our mass range is far from it. 

Several fitting ranges are examined for $M_\pi^2$ with second order
polynomial in $m_f$. With fixed minimum $m_f^{\rm min}=0.04$, 
the maximum is changed as $m_f^{\rm max}=0.12$, 0.1, and 0.08.
Only the $m_f^{\rm max}=0.08$ fit is consistent with the vanishing
intercept, while others end up with having negative intercept.
A fit with the intercept constrained to be zero gives good
$\chi^2/{\rm dof}=0.88$ for $m_f^{\rm max}=0.08$, though including
heavier mass gives unacceptably large $\chi^2/{\rm dof}>8$.

Similar exercises are done for the $F_\pi$, where the intercept is 
always positive, and for $m_f^{\rm max}=0.1$ and 0.08, reasonable
$\chi^2/{\rm dof}$ yields $\chi^2/{\rm dof}=0.37$ and $0.29$,
respectively.

The $\pi$ mass squared and decay constant behave in the way reasonably 
consistent with the second-order polynomial, when the mass range is 
restricted to smaller side  $0.04\le m_f \le 0.08$ where we have four
mass points. The $\pi$ mass vanishes in the chiral limit, and the decay
constant is nonzero.  However, we are not able to conclude these
results are consistent with S$\chi$SB.  The problem can be appreciated if
the expansion parameter in ChPT is
evaluated. Using the value of $F_\pi$ in the chiral limit $F_\pi=0.0190(52)$,
the natural expansion parameter
\cite{Soldate:1989fh,Chivukula:1992gi,Harada:2003jx}
of ChPT in the large $N_f$ theory as given in Eq.~(\ref{eq:X}) is 
\begin{equation}\label{eq:X2}
 \mathcal{X} = N_f\left(\frac{M_\pi}{4\pi F_\pi/\sqrt{2} }\right)^2 \simeq 39,
\end{equation}
which has been evaluated at the lightest point $m_f=0.04$ on $L=30$
lattice, where $M_\pi=0.3028(14)$.
For the expansion to be consistent, it should satisfy $\mathcal{X}<1$.
With this large expansion parameter at the smallest mass in the deposit,
we are not in the position to judge if the decay constant is really
nonzero at the chiral limit from the extrapolation performed here.

%% file: text_sections/summary.tex
We have studied the SU(3) gauge theory with 12-flavor fermions in the
fundamental representation of the gauge group. 
We tried to determine if the massless theory undergoes
S$\chi$SB or exhibits IR conformality due to the existence 
of an IR fixed point of the beta function, through the fermion mass $m_f$ 
dependence of the spectrum.
A type of HISQ action was adopted to 
maximally suppress the staggered flavor-symmetry violation as well as
other discretization errors. Three volumes with fixed aspect ratio
and various $m_f$ were examined. Simulations were repeated 
for two values of the bare gauge coupling, corresponding to two
different lattice spacings.
We observed that the lattice spacing became finer when the bare gauge
coupling was made small, which is expected when the simulation
points are in the asymptotically free domain.

A primary analysis of the masses of pseudoscalar and
vector channel and the pseudoscalar decay constant revealed
a range of small $m_f$ where the ratios of composite masses and
decay constant, $F_\pi/M_\pi$ and $M_\rho/M_\pi$, 
are independent of $m_f$. This is consistent with the 
hyperscaling characteristic to the IR conformality and was observed for
all the quantities except for the one which involves the decay constant at a
finer lattice.

Further detailed study adopting a finite-size scaling have shown that
our data are reasonably consistent with the FSHS, 
where the product of linear system size and the composite masses
(or decay constant) fall into a function of a universal scaling variable
composed of the $m_f$, linear system size $L$, and mass anomalous dimension
$\gamma_*$ at the IR fixed point at low energy. The resulting $\gamma_*$'s obtained 
from the introduced evaluation function, $P(\gamma)$, 
of the scaling were reasonably
consistent with each other for three observables and two lattice
spacings when only the decay constant at finer lattice was
excluded as before. We conclude the mass anomalous
dimension is in the range $0.4\simle \gamma_* \simle 0.5$ 
at IR fixed point if it exists.

Existence of the nonuniversal correction indicated by the FSHS 
motivated a study of global fit with models assuming some corrections.
By adding a correction term to the lowest-order universal term,
the global fit works well with a reasonable $\chi^2$, though possible
correlation was assumed to be absent. Among the tested, the one
with the power of the fermion mass fixed {\it \`a la}
ladder SD study \cite{Aoki:2012ve}
and another with $O((ma)^2)$ discretization error resulted in 
$\gamma_*$ consistent with FSHS.

In the test of our data against the S$\chi$SB scenario, the $m_f$ dependence
of the $\pi$ mass squared  
and decay constant appeared to be consistent with the tree-level mass
dependence to second order from chiral perturbation theory,
if the data are restricted to sufficiently small $m_f$.
The decay constant remains nonzero at the chiral limit with this
analysis.
It turns out, however, the natural expansion parameter of ChPT
estimated using the value of the decay constant at the chiral limit
extracted in this analysis is very large $\mathcal{X}$ 
even at the smallest $\pi$ mass simulated.
Therefore, we cannot judge if the decay constant is really
nonzero at the chiral limit from the extrapolation performed in this study.

A possibility of S$\chi$SB in $N_f=12$ is not excluded yet.
Further efforts would be required to arrive at a decisive conclusion.
One possible direction is to use the brute force (larger volume and
lighter mass) calculations, which seems very hard.
Another may be to gather  more information 
about the spectrum, for example, the mass of glueball and flavor singlet
composite state, which could have distinctive signature across the
phase boundary.

However, we have confirmed that the general property of the spectrum in
$N_f=12$ theory as functions of $m_f$ is completely different 
from the $N_f=4$ theory~\cite{Aoki:2012ep}, where we found that
the ratio $F_\pi/M_\pi$ gets divergent towards the chiral limit 
in accord with the ordinary QCD but in sharp contrast to the conformal 
behavior we have observed for $N_f=12$.
Although the analyses support the conformal scenario,
even if the $N_f=12$ theory breaks chiral symmetry,
the breaking scale is very small, so it must be very close to 
the boundary of the conformal phase transition point.
In this case, our result of $\gamma_m$ obtained with assuming 
the IR fixed point can be regarded as an approximate mass anomalous dimension
at the walking regime.

There have already been several studies focusing on the conformality
of the $N_f=12$ SU(3) gauge theory
by investigating the masses of composite states of the mass-deformed theory:

Ref.~\cite{Fodor:2011tu} 
performed a large-scale simulation using the stout-smeared
staggered fermions at a single lattice spacing.
They performed the global fits of ChPT and conformal scenario.
ChPT fit (without logarithmic terms as in ours), 
which resulted in a nonzero pion decay constant and chiral  
condensate, was favored over conformal scenario by examining the 
$\chi^2/{\rm dof}$ of the fits.
However, the expansion parameter 
at their lightest quark mass reads $\mathcal{X}\simeq 34$, 
which is as large as ours [Eq.~(\ref{eq:X2})].
We note that we have concluded that for such a large $\mathcal{X}$,
ChPT is not self-consistent, and hence
it is difficult to conclude what is really happening at the chiral limit.

As to their conformal fit, it is equivalent to our fit A in Eq.~(\ref{eq:f}),
by which we obtained a $\chi^2/{\rm dof}$ similar to theirs. 
(Their estimate of the effective mass anomalous dimension reads
$\gamma_\ast=0.395(52)$, which is consistent with ours.)
Although it is not our main analysis, we introduced a mass correction
to the conformal type fit (fit B1 or B2), which was required by our
data, and obtained a reasonable $\chi^2/{\rm dof}$.  
Looking into more details, we observe a difference in the quark-mass
dependence of pion decay constant.  
Their decay constant is perfectly fitted with the linear function in
$m_f$ (Fig.~1 middle in Ref.~\cite{Fodor:2011tu}), while ours exhibits
a curvature (Fig.~\ref{fig:mpiL} middle or Fig.~\ref{fig:chpt_fpi}).
The origin of different behavior is so far unknown.

There have also been studies analyzing the data of Ref.~\cite{Fodor:2011tu}
for tests of the conformal hypothesis.
Ref.~\cite{Appelquist:2011dp} performed fits with the single
power corresponding to 
the fit A in Eq.~(\ref{eq:f}) without the constant term and examined the 
dependence of $\chi^2$ as function of input $\gamma$. The optimal vale
of $\gamma$ which minimizes $\chi^2$ reads $\gamma\sim 0.4$ for 
$\pi$ and $\rho$ masses, which is consistent with our result.
However, they obtained $\gamma\sim 0.25$ for pion decay constant, while
our $\gamma(F_\pi)$ obtained without fixing the function form in
Sec.~\ref{sec:fss} tends to be even larger than $\gamma(M_\pi)$.
This inconsistency likely originates from the aforementioned difference
in mass dependence. They tried the global fit to various observables
with a common $\gamma$ introducing a volume correction term to every
observable and a mass correction term to the pion decay constant.
By doing that, a model dependence is introduced, 
but our result is consistent with their result:
$\gamma=0.403(13)$ with a reasonable $\chi^2/{\rm dof}=42/44$.

Ref.~\cite{DeGrand:2011cu}
tried a similar analysis to what is presented in Sec.~\ref{sec:Q}, 
using the data of Ref.~\cite{Fodor:2011tu},
and obtained $\gamma\sim 0.35$. The error is large ($\pm 0.23$ for pion
mass), thus consistent with our result.

Ref.~\cite{Deuzeman:2009mh} performed a pure power-law fit 
to $\pi$ and $\rho$ masses on a single volume but over two lattice
spacings. The results are $\gamma\sim 0.52-0.64$, which is only a bit
higher than what we obtained.

To conclude, the (effective) $\gamma_*$ did not appear as large as the walking
theory should exhibit. By this, the hunt for the realistic walking theory 
should anyway alter the direction of the number of flavors towards
smaller.  The $N_f=8$ theory, 
which could have large anomalous dimension $\gamma_* \sim 1$ 
as a candidate of the walking technicolor model, 
is under investigation with 
exactly the same setting with HISQ action~\cite{Aoki:2012ep}.

%% file: text_sections/other_spectra.tex
\begin{figure}[htbp]
  \begin{center}
\rotatebox{0}{
      \includegraphics[width=7.5cm,clip]{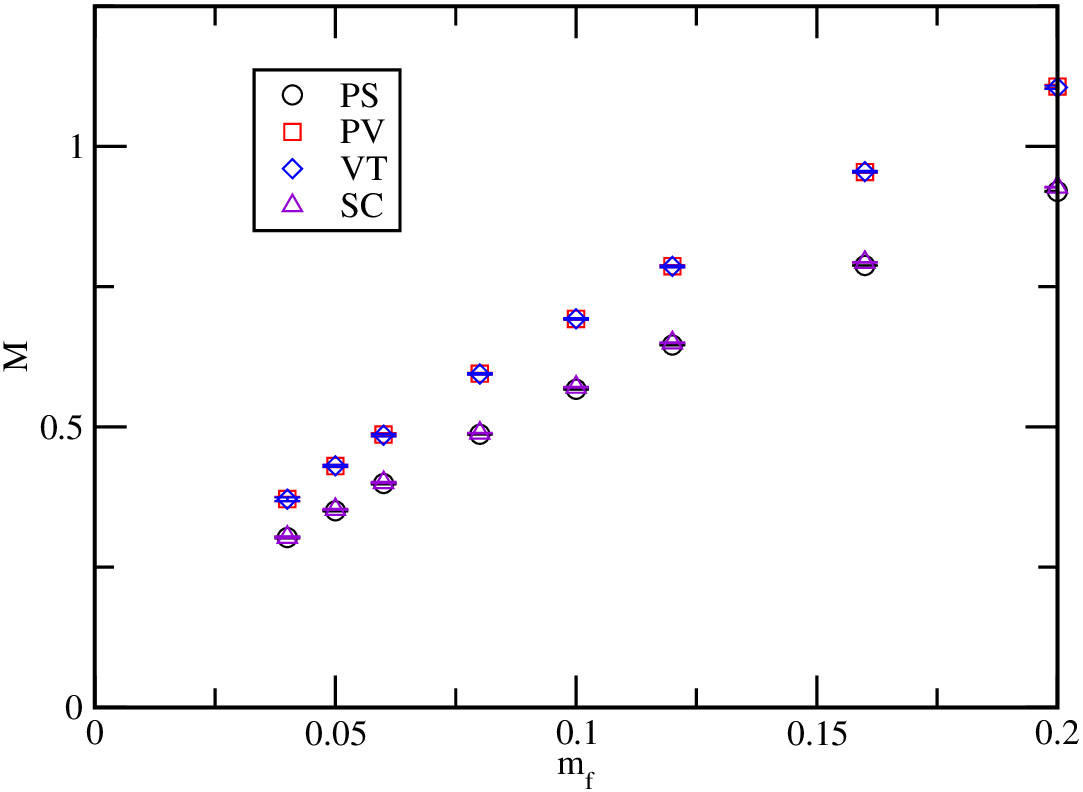}
\ \   \includegraphics[width=7.5cm,clip]{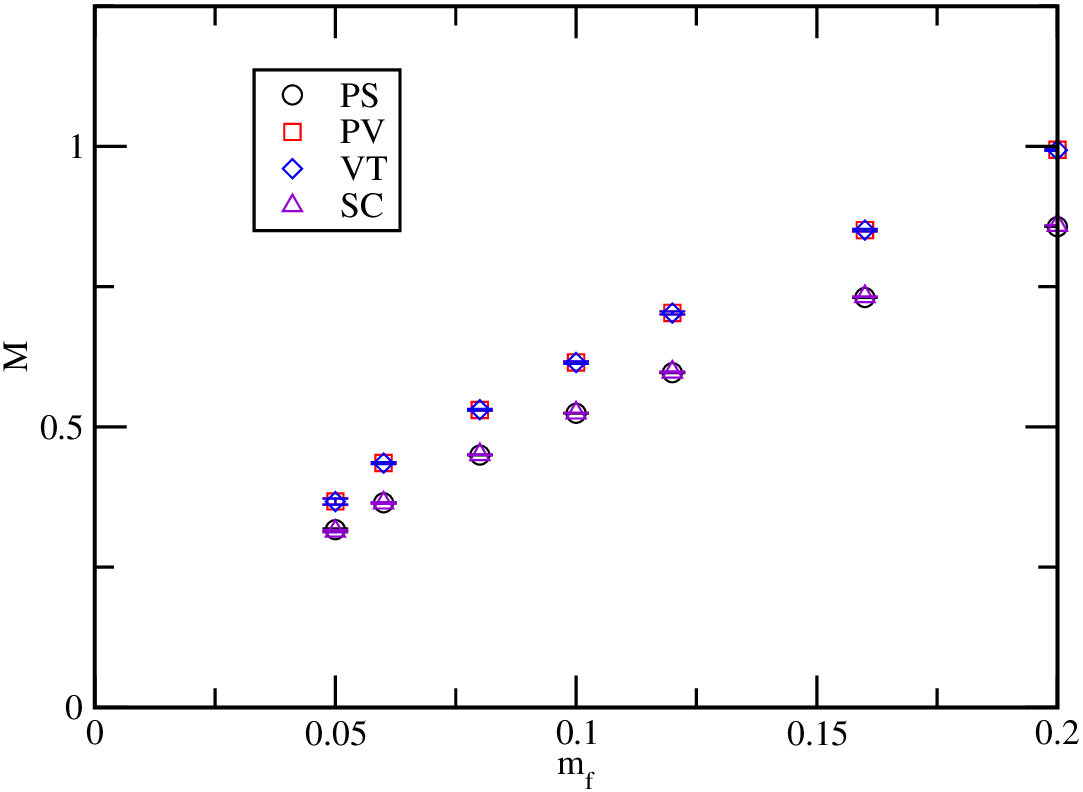} 
}
  \end{center}
  \caption{
Result of mass spectra of four operators at $\beta=3.7$ (left panel),
$4$ (right panel) on $(L,T)=(30,40)$. 
}
   \label{fig:spectrum}
\end{figure}

%
We show some other staggered meson spectra e.g. 
non-NG pseudoscalar $(\gamma_5\gamma_4\otimes\xi_5\xi_4)$ denoted as SC
and a vector channel $(\gamma_i\otimes\xi_i)$ denoted as VT in comparison with
the corresponding (NG) $\pi$, PS: $(\gamma_5\otimes\xi_5)$ and the
vector, PV: $(\gamma_i\gamma_4\otimes\xi_i\xi_4)$, which are used in the analysis
given in the main text.
This is to see a staggered flavor-symmetry-breaking effect 
in our HISQ simulation.
The results are shown in Fig.~\ref{fig:spectrum}.
As we expected, the two masses for each pseudoscalar and vector meson
channel are almost degenerate.

%% file: text_sections/conformal_fit.tex
%
Numerical detail of fit results on the conformal hypothesis
are given here.
In the conformal hypothesis with a finite volume, 
we make an attempt to use the fit functions given in Eq.~\ref{eq:f}. 
We fix the value of the exponent $\alpha$ 
to a certain value or to relate with $\gamma$
in the fit since having two exponents free makes the fit unstable.
We consider three possible cases:
$\alpha=(3-2\gamma)/(1+\gamma)$, $2$, and $1$;
we denote these fit functions as fit B1, fit B2, and fit B3, respectively.
We carry out simultaneous fits using all the data for 
$M_\pi$, $F_\pi$, and $M_\rho$ with 
common anomalous dimension $\gamma$ and/or the parameter $\alpha$
where possible correlation is assumed to be absent.
We use same data set for the fit as given in Sec.~\ref{sec:fshp}. 
The fit results are shown in Table \ref{tab:fsc_app} 
and \ref{tab:fsc_b4_app} for $\beta=3.7$ and $\beta=4$.
As already discussed in Sec.~\ref{sec:fshp}, 
additional correction terms improve $\chi^2/{\rm dof}$ for both cases  
of $\beta=3.7$ and $\beta=4$.
In both cases of fit B1 and fit B2, the values of $\gamma$ are
consistent with the value  
from the analysis of the FSHS test given in Sec.~\ref{sec:fss}.
On the other hand, the result of fit B3 gives a slightly smaller
value of the $\gamma$, but with larger errors.

\begin{table}[tbp]
\caption{The fit results of finite-size conformal hypothesis at $\beta=3.7$.} 
\label{tab:fsc_app}
\small
\begin{tabular}{cc}
\begin{minipage}{0.5\hsize}
\begin{ruledtabular}
\begin{tabular}{ccc}
fit A &  $c_0$  & $c_1$ \\ \hline
$\xi_\pi$  & -0.029(28) &  2.793(5)    \\
$\xi_F$ & 0.147(12) &  0.544(3) \\ 
$\xi_\rho$ & 0.363(58) & 3.325(10)  \\ \hline
\multicolumn{3}{c}{ $\gamma=0.449(3)$,  $\chi^2/$dof=4.52, dof=47}
\end{tabular}
\end{ruledtabular}
\end{minipage}
\begin{minipage}{0.5 \hsize}
\begin{ruledtabular}
\begin{tabular}{cccc}
fit B1 &   $c_0$  & $c_1$ & $c_2$ 
\\ \hline 
$\xi_\pi$  & 0.095(30) & 2.924(39)  & -0.227(83) \\
$\xi_F$ &  
0.627(10) & -0.230(27) & 0.037(17)   \\ 
$\xi_\rho$ & 0.146(77) & 3.634(52)  & -0.74(14) \\ \hline
\multicolumn{4}{c}{$\gamma=0.411(9)$,  $\chi^2/$dof=1.23, dof=44}
\end{tabular}
\end{ruledtabular}
\end{minipage}  
\\ \\
\begin{minipage}{0.5\hsize}
\begin{ruledtabular}
\begin{tabular}{cccc}
fit B2 &  $c_0$  & $c_1$ & $c_2$ \\ \hline
$\xi_\pi$  & 0.089(30) & 2.860(26)  & -0.181(96) \\
$\xi_F$ & 0.045(16) & 0.600(7)  & -0.325(33) \\ 
$\xi_\rho$ & 0.147(75) & 3.524(37)  & -0.97(16) \\ \hline
\multicolumn{4}{c}{$\gamma=0.423(7)$,  $\chi^2/$dof=1.15, dof=44}
\end{tabular}
\end{ruledtabular}
\end{minipage}  
\begin{minipage}{0.5 \hsize}
\begin{ruledtabular}
\begin{tabular}{cccc}
fit B3 &   $c_0$  & $c_1$ & $c_2$  \\ \hline
$\xi_\pi$  & 0.095(30) &  3.53(25)  & -0.81(27) \\
$\xi_F$  & 0.030(17) &  0.829(59)  &  -0.356(67) \\ 
$\xi_\rho$ & 0.149(79) & 4.56(32)  & -1.45(36)   \\ \hline 
\multicolumn{4}{c}{$\gamma=0.356(22)$,  $\chi^2/$dof=1.45, dof=44}
\end{tabular}
\end{ruledtabular}
\end{minipage}  
\end{tabular}
\end{table}

\begin{table}[tbp]
\caption{The fit results of finite-size conformal hypothesis at $\beta=4$.} 
\label{tab:fsc_b4_app}
\begin{tabular}{cc}
\small
\begin{minipage}{0.5\hsize}
\begin{ruledtabular}
\begin{tabular}{ccc}
fit A &  $c_0$  & $c_1$ \\ \hline
$\xi_\pi$   & -0.202(31) & 2.662(5)    \\
$\xi_F$ & 0.212(15) & 0.464(3)  \\ 
$\xi_\rho$ & 0.228(60) & 3.036(9)  \\ \hline    
\multicolumn{3}{c}{$\gamma=0.430(3)$,  $\chi^2/$dof=6.78, dof=44}
\end{tabular}
\end{ruledtabular}
\end{minipage}
\begin{minipage}{0.5 \hsize}
\begin{ruledtabular}
\begin{tabular}{cccc}
fit B1 & $c_0$  & $c_1$ & $c_2$ \\ \hline 
$\xi_\pi$  & -0.083(33) &  2.486(74)  & 0.33(12) \\
$\xi_F$  & -0.040(25) &  0.531(15)  & 0.166(34) \\ 
$\xi_\rho$ & -0.146(84) & 3.013(85)  & -0.03(16) \\ \hline
\multicolumn{4}{c}{$\gamma=0.461(18)$,  $\chi^2/$dof=1.86, dof=41}
\end{tabular}
\end{ruledtabular}
\end{minipage}
\\ \\
\begin{minipage}{0.5\hsize}
\begin{ruledtabular}
\begin{tabular}{cccc}
fit B2 &  $c_0$  & $c_1$ & $c_2$  
\\ \hline
$\xi_\pi$  & -0.096(33) & 2.551(34)  & 0.45(12) \\
$\xi_F$  & -0.026(25)  & 0.518(9) & -0.291(36) \\ 
$\xi_\rho$ & -0.123(81) & 3.039(42) & -0.16(16)  \\ \hline
\multicolumn{4}{c}{$\gamma=0.453(11)$,  $\chi^2/$dof=2.00, dof=41}
\end{tabular}
\end{ruledtabular}
\end{minipage}
\begin{minipage}{0.5\hsize}
\begin{ruledtabular}
\begin{tabular}{cccc}
fit B3 & $c_0$  & $c_1$ & $c_2$ \\ \hline
$\xi_\pi$  & -0.074(33) & 2.30(21) &  0.44(24) \\
$\xi_F$  & -0.044(25) & 0.564(54)  & -0.147(62) \\ 
$\xi_\rho$ & -0.153(86) & 2.92(27)  & 0.09(31)  \\ \hline
\multicolumn{4}{c}{$\gamma=0.476(39)$,  $\chi^2/$dof=1.88, dof=41}
\end{tabular}
\end{ruledtabular}
\end{minipage}
\end{tabular}
\end{table}

%% file: text_sections/chpt.tex
%

In order to give a fair comparison whether or not the chiral symmetry is 
spontaneously broken, we carry out the fit based on the 
ChPT hypothesis by using our data.

\subsection{Finite-size dependence  of physical quantities}
\begin{figure}[htbp]
  \begin{center}
\rotatebox{0}{
      \includegraphics[width=7.5cm,clip]{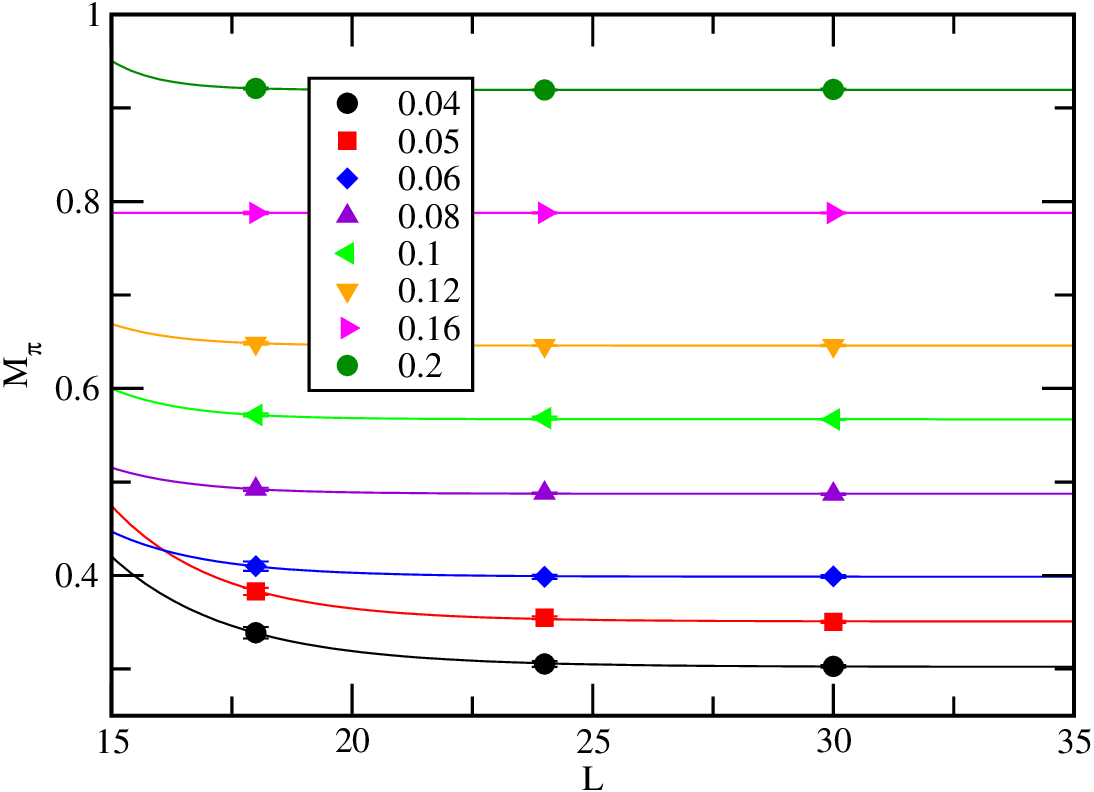} \quad \quad \quad
      \includegraphics[width=7.5cm,clip]{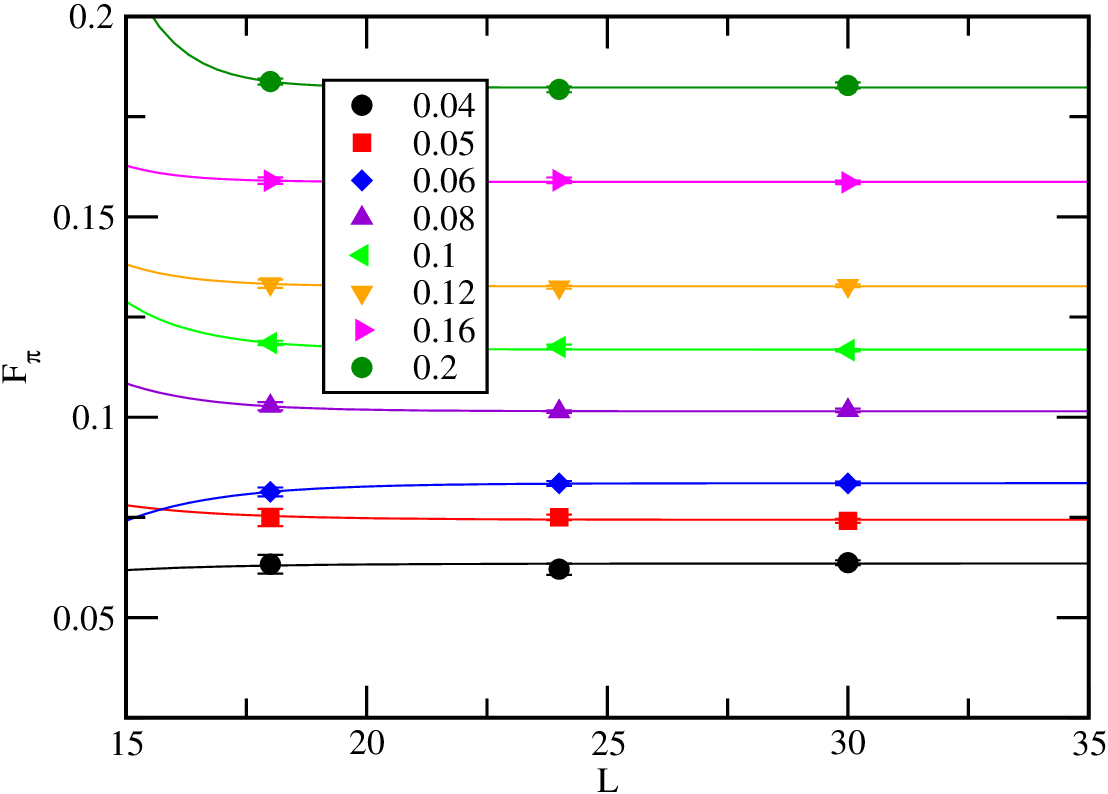}
}
  \end{center}
  \caption{
  The results of the finite-volume scaling fit for $M_\pi$ and $F_\pi$ at $\beta=3.7$
  using the infinite volume extrapolation formula in Eqs.~(\ref{eq:FV}) and (\ref{eq:FV2}).}
   \label{fig:finite}
\end{figure}

\begin{figure}[htbp]
  \begin{center}
\rotatebox{0}{
      \includegraphics[width=7.5cm,clip]{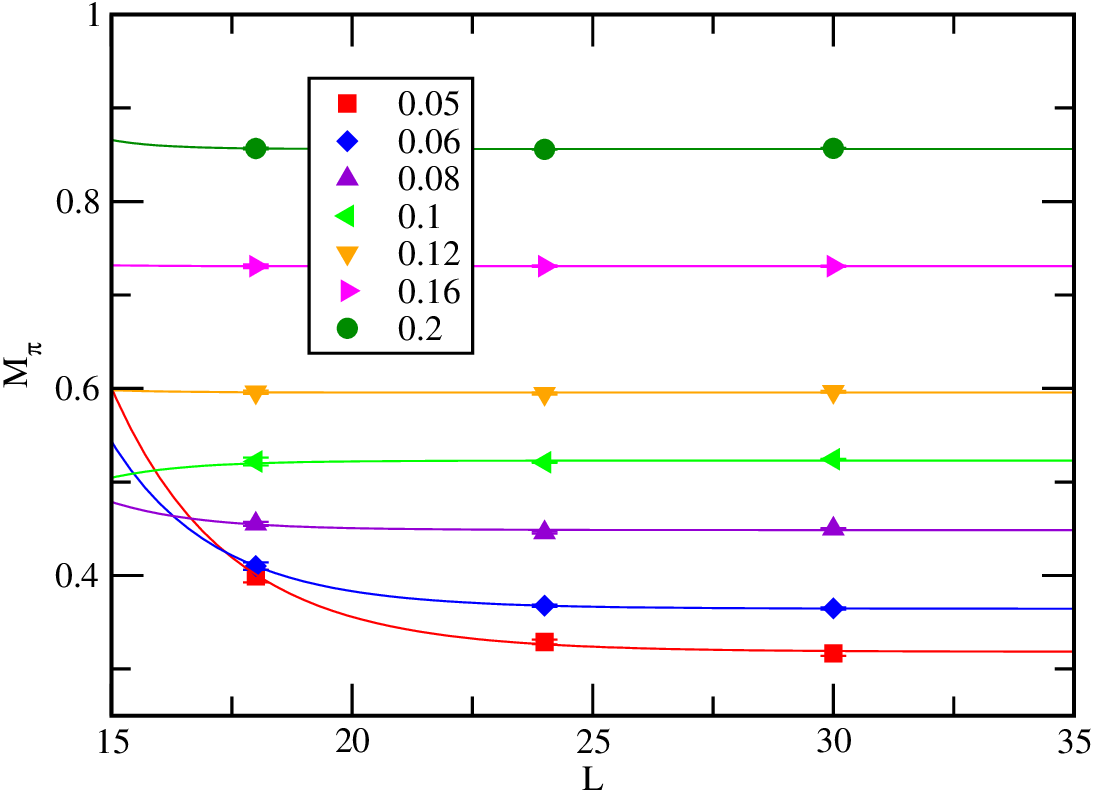} \quad \quad \quad
      \includegraphics[width=7.5cm,clip]{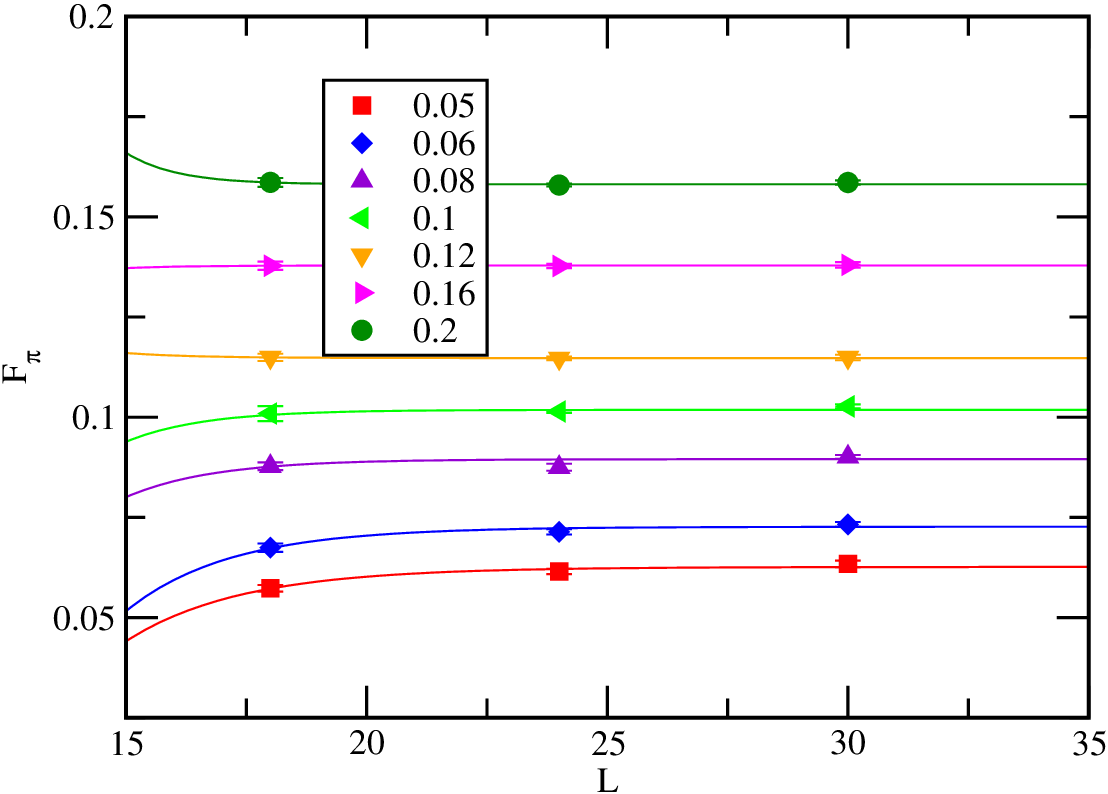}
}
  \end{center}
  \caption{
  The results of the finite volume scaling fit for $M_\pi$ and $F_\pi$ at $\beta=4$
  using the infinite volume extrapolation formula 
  in Eqs.~(\ref{eq:FV}) and (\ref{eq:FV2}).}
   \label{fig:finite_b4}
\end{figure}

Here, we study the finite volume effects for $M_\pi$ and $F_\pi$. 
To obtain the values of $M_\pi$ and $F_\pi$ in the infinite volume limit, 
we use the following ChPT-inspired~\cite{Gasser:1987zq, Luscher:1985dn}
fit functions: 
\begin{eqnarray} \label{eq:FV}
M_\pi(L) - M_\pi &=&
c_{M_\pi} \frac{e^{-LM_\pi} }{(LM_\pi)^{3/2}}, \\
\label{eq:FV2}
F_\pi(L) - F_\pi &=&
c_{F_\pi} \frac{e^{-LM_\pi} }{(LM_\pi)^{3/2}}, 
\end{eqnarray}
where $c_{M_\pi}$ and $c_{F_\pi}$ are the fit parameters.
We carry out the simultaneous fit for each fermion mass 
by using three data points on $L=18, 24$, and $30$ at $\beta=3.7$ and $\beta=4$, 
where we assume for simplicity that possible correlation 
between the data of $M_\pi$ and $F_\pi$ is absent.
The fit results are shown in Figs.~\ref{fig:finite} and \ref{fig:finite_b4}. 
As a result, in the entire fermion mass region, our data is well-fitted, 
and the value of $\chi^2/$dof for each parameter is $\mathcal{O}(1)$. 
Also, one can find that the difference between the value of $L=30$ data and  
that in the infinite volume limit is negligible.   
Thus,  we use the data on $L=30$ 
to analyze the chiral behaviors of both $M_\pi$ and $F_\pi$ 
hereafter.

\subsection{ChPT fit analysis}

We analyze the fit using the second-order polynomial function as 
\begin{equation}  \label{eq:action}
h(m_f)=
c_0 + c_1 m_f + c_2 m_f^2 \\
\end{equation}
where $c_{0,1,2}$ are fit parameters.
Using this simple polynomial function, we carry out the fits 
for $M_\pi^2$ and $F_\pi$ by the function $h(m_f)$,
varying the fit range of the fermion mass 
from  $m_f=0.04$ to $m_f=0.12$.
We denote the fit range of fermion mass as $[m_f^{\rm min}, m_f^{\rm max}]$.
As the current data set of $\beta=3.7$ simulation has physically lighter
quark mass, the property of the chiral limit is captured better for 
$\beta = 3.7$ data than for $\beta=4$ data. 
Therefore, we focus on the former here.

\begin{figure}[h]
  \begin{center}
\rotatebox{0}{
      \includegraphics[width=7.5cm,clip]{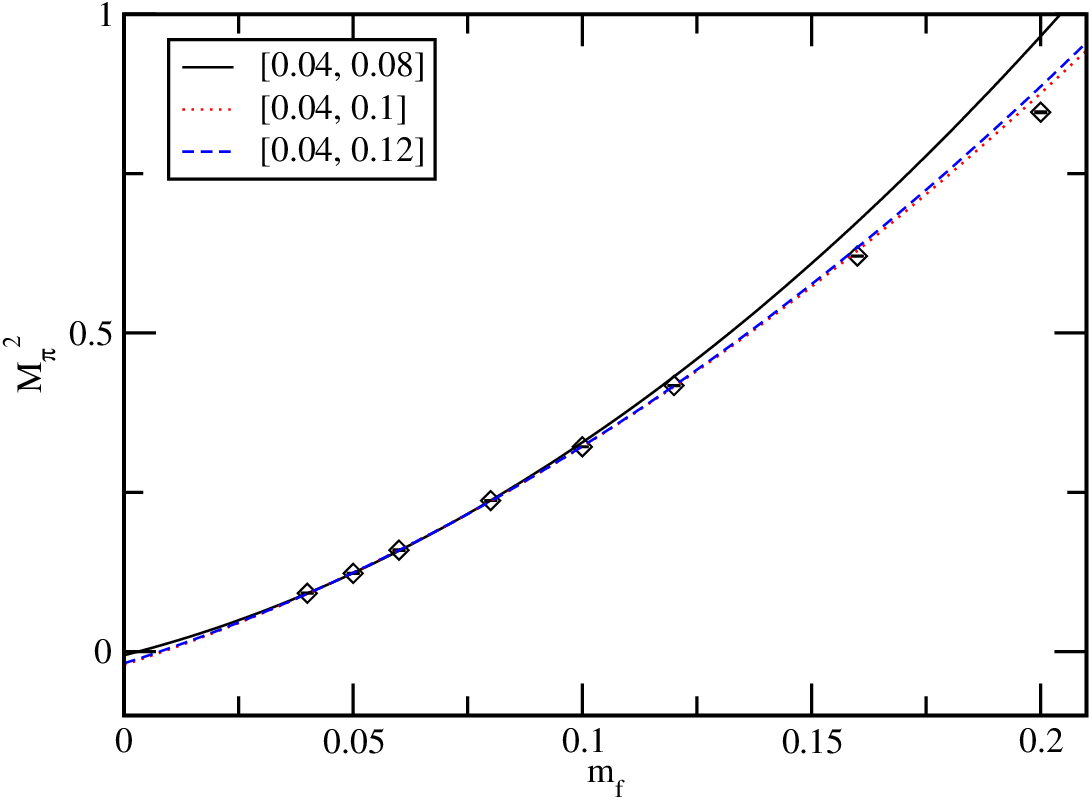} \quad \quad \quad
}
  \end{center}
  \caption{
  The several fit results for $M_\pi^2$ at  $\beta=3.7$ using the data on $L=30$.}
   \label{fig:chpt}
\end{figure}

The fit results for $M_\pi^2$ are shown in Fig.~\ref{fig:chpt} 
and Table \ref{tab:chpt}. 
Among these fits, the best-fit values of intercept $c_0$ 
is consistent with zero for the fit with $m_f^{max}=0.08$, 
while we obtained negative values of $c_0$ for others.
In Table \ref{tab:chpt},  
we also show the fit results with $c_0$ being fixed to zero.
As we expect, the values of $\chi^2/$dof become much larger than
unity except for the case with $m_f^{\rm max} =0.08$. 
Furthermore, the contribution of the higher-order term with $c_2$ is not small enough 
even in the fit result using the data with the smallest mass range.
It may indicate that the fermion mass in the data we have is 
too heavy to take a reliable chiral extrapolation with ChPT-like formula.

\begin{figure}[h]
  \begin{center}
\rotatebox{0}{
      \includegraphics[width=7.5cm,clip]{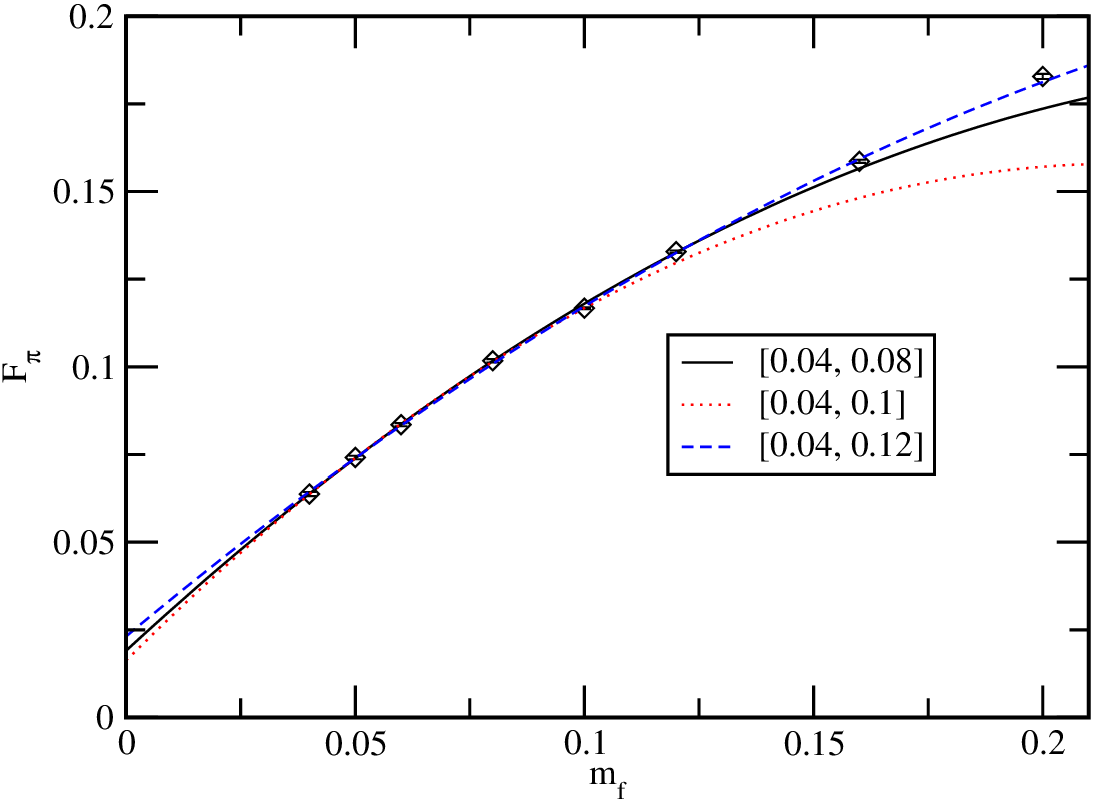} \quad \quad \quad
}
  \end{center}
  \caption{
  The several fit results for $F_\pi$ at  $\beta=3.7$ using the data on $L=30$.}
   \label{fig:chpt_fpi}
\end{figure}

In Fig.~\ref{fig:chpt_fpi} and Table~\ref{tab:chpt_fpi}, 
we show the fit results for the case of $F_\pi$. 
The same fit function and fit ranges as in the case of $M_\pi^2$ 
are used for the fits.
Nonzero value of $F_\pi$ in the chiral limit is an important signal of 
the S$\chi$SB. 
Results show $\mathcal{O}(1)$ value of $\chi^2/$dof for 
the fit with smaller mass range, and we obtain tiny but nonzero 
value of $F_\pi$ in the chiral limit.

If we adopt the fit result with the mass range $[0.04, 0.08]$ at $\beta=3.7$,
which is expected to be the most reliable result among those fits,
it might look like they are consistent with the hypothesis of S$\chi$SB 
scenario because 
the result shows that the pseudoscalar mass squared is going to zero along the
tree-level ChPT, while the decay constant remain nonzero (though it is tiny) 
in the chiral limit. However, to be able to conclude that it is consistent 
with the ChPT fit, one has to check that the expansion parameter of the 
perturbation is in the legitimate region. 
Using the value of $F_\pi$ in the chiral limit, $F_\pi=0.0190(52)$, 
the value of the expansion parameter is estimated as $\mathcal{X} \simeq 39$.
Here, we evaluated it using the data with the lightest mass, $m_f=0.04$, 
on $L=30$ lattice, where $M_\pi=0.3028(14)$.
Since the expansion parameter is much larger than one 
even in the lightest mass region of the fit range, 
we are not in the position to judge if the decay constant is really
nonzero in the chiral limit from the extrapolation performed here.
Therefore, 
investigation with larger volume and smaller masses
is needed to draw more definite conclusions regarding the 
test of the ChPT scenario.


\begin{table}[h]
\caption{The fit results on $M_\pi^2$ 
at $\beta=3.7$ using the data on $L=30$.
The values in the bracket mean the inputs in the fit.
}  
\label{tab:chpt}
\begin{ruledtabular}
\begin{tabular}{l ccc c c}
\multicolumn{1}{c}{fit range} &  $c_0$  & $c_1$ & $c_2$ &  
$\chi^2/$dof & dof \\ \hline 
$[0.04, 0.08]$   & 
-0.0057(91) & 1.82(32) & 15.2(2.6) &  1.35 & 1 \\
 & 
 [0] & 1.62(3) & 16.76(45) &  0.88 & 2 \\
$[0.04, 0.1]$   & 
-0.0209(48) & 2.37(15) &10.6(1.1) &  2.59 & 2\\
 & 
 [0] & 1.729(21) & 14.99(25) &  8.33 & 3 \\
$[0.04, 0.12]$  &
-0.0183(31) & 2.28(87) & 11.21(55) &  1.90 & 3\\
 & 
 [0] & 1.780(17) & 14.28(17) &  10.29 & 4 \\
\end{tabular}
\end{ruledtabular}
\end{table}

\begin{table}[h]
\caption{The fit results on $F_\pi$ at $\beta=3.7$ using the data on $L=30$.}  
\label{tab:chpt_fpi}
\begin{ruledtabular}
\begin{tabular}{l ccc cc}
\multicolumn{1}{c}{fit range} &  $c_0$  & $c_1$ & $c_2$ & 
$\chi^2/$dof & dof \\ \hline 
$[0.04, 0.08]$   & 
0.0190(52) & 1.21(18) & -2.2(1.5) &  0.29 & 1 \\
$[0.04, 0.1]$   & 
0.0162(30) & 1.31(85) & -3.01(58) &  0.37 & 2 \\
$[0.04, 0.12]$  &
0.0231(18) & 1.093(48) & -1.51(29) &  3.30 & 3\\
\end{tabular}
\end{ruledtabular}
\end{table}